\documentclass[11pt]{article}
\RequirePackage{amsthm,amsmath}
\RequirePackage{natbib}

\usepackage{amsgen,amsmath,amstext,amsbsy,amsopn,amssymb,latexsym,float}
\usepackage{array}
\usepackage{cite}
\usepackage{multirow}%
\usepackage{graphicx}%
\usepackage{caption}
\usepackage{subcaption}
\usepackage{amssymb}
\usepackage{hyperref,xr}
\usepackage{mathtools}
\usepackage{booktabs}
\usepackage{comment}
\usepackage{authblk}
\usepackage{algorithm}
 \usepackage{algpseudocode}
 \usepackage{algorithmicx}
 \algdef{SE}[DOWHILE]{Do}{doWhile}{\algorithmicdo}[1]{\algorithmicwhile\ #1}%

\algrenewcommand\algorithmicrequire{\textbf{Input:}}
\algrenewcommand\algorithmicensure{\textbf{Output:}}

 \usepackage[dvipsnames,table,dvipsnames*, svgnames*]{xcolor}

\newcommand{\E}{\mathbb{E}}
\newcommand{\prob}{\mathbb{P}}
\newcommand{\I}{\mathbb{I}}
\newcommand{\Var}{{\rm Var}}

\newcommand{\Cov}{{\rm Cov}}
\newcommand{\SE}{{\rm SE}}

\newcommand{\tr}{^{\mkern-1.5mu\mathsf{T}}}
\newcommand{\pto}{\xrightarrow{p}}
\newcommand{\dto}{\xrightarrow{d}}

\newcommand{\given}{\;\middle|\;}
\newcommand{\what}{\widehat{w}}
\newcommand{\vecl}{{\rm vecl}}
\newcommand{\Int}{{\rm Int}}
\DeclareMathOperator*{\argmax}{arg\,max}
\DeclareMathOperator*{\argmin}{arg\,min}
\usepackage{bm}
\newcommand{\tauhat}{\widehat{\tau}}
\newcommand{\taubar}{\bar{\tau}}
\newcommand{\tauhatSC}{\widehat{\tau}^{\rm SC}}
\newcommand{\tauhatm}{\widehat{\tau}^{[m]}}
\newcommand{\muhat}{\widehat{\mu}}
\newcommand{\muhatm}{\widehat{\mu}^{[m]}}
\newcommand{\muhatmstar}{\widehat{\mu}^{[m^*]}}
\newcommand{\Vhat}{\widehat{\mathbf{V}}}
\newcommand{\VhatSigma}{\Vhat_{\Sigma}}
\newcommand{\Vhatgamma}{\Vhat_{\gamma}}
\newcommand{\Vhatmu}{\Vhat_{\mu}}
\newcommand{\VhatY}{\widehat{{\rm V}}_Y}
\newcommand{\VSigma}{\mathbf{V}_{\Sigma}}
\newcommand{\Vgamma}{\mathbf{V}_{\gamma}}
\newcommand{\Vmu}{\mathbf{V}_{\mu}}
\newcommand{\VY}{{\rm V}_Y}
\newcommand{\Sigmahat}{\widehat{\Sigma}}
\newcommand{\Sigmahatm}{\widehat{\Sigma}^{[m]}}
\newcommand{\Sigmahatmstar}{\Sigmahat^{[m^*]}}
\newcommand{\gammahat}{\widehat{\gamma}}
\newcommand{\gammahatm}{\widehat{\gamma}^{[m]}}
\newcommand{\gammahatmstar}{\gammahat^{[m^*]}}
\newcommand{\Gammahat}{\widehat{\Gamma}}

\newcommand{\Gammahatmstar}{\Gammahat^{[m^*]}}
\newcommand{\deltahat}{\widehat{\delta}}

\newcommand{\deltahatmstar}{\deltahat^{[m^*]}}
\newcommand{\betahat}{\widehat{\beta}}
\newcommand{\betahatSC}{\betahat^{\rm SC}}
\newcommand{\betahatm}{\betahat^{[m]}}

\newcommand{\betahatmstar}{\betahat^{[m^*]}}

\newcommand{\tildebetamstar}{\tilde{\beta}^{[m^*]}}

\newcommand{\barbetamstar}{\bar{\beta}^{[m^*]}}

\newcommand{\err}{{\rm err}(M)}
\newcommand{\ghat}{\widehat{g}}
\newcommand{\ghatm}{\ghat^{[m]}}
\newcommand{\ghatmstar}{\ghat^{[m^*]}}
\newcommand{\fhat}{\widehat{f}}
\newcommand{\fhatm}{\fhat^{[m]}}
\newcommand{\fhatmstar}{\fhat^{[m^*]}}
\newcommand{\Omegahat}{\widehat{\Omega}}
\newcommand{\Omegahatm}{\Omegahat^{[m]}}
\newcommand{\Omegahatmstar}{\Omegahat^{[m^*]}}
\newtheorem{Theorem}{Theorem}

\newtheorem{proposition}{Proposition}
\newtheorem{Lemma}{Lemma}

\newtheorem{remark}{Remark}

\newtheorem{Assumption}{Assumption}

\title{Synthetic Control with Weight Uncertainty: Robust Identification and Statistical Inference}
\author[1]{Taehyeon Koo}
\author[2]{Zijian Guo\thanks{Correspondence to Zijian Guo (\texttt{zijguo@zju.edu.cn}).}}
\date{\today}

\affil[1]{Columbia University Mailman School of Public Health}
\affil[2]{Center for Data Science, Zhejiang University}

\begin{document}

\maketitle

\begin{abstract}
    The synthetic control method estimates causal effects by comparing a treated unit with weighted controls matched on its pre-treatment trajectory. However, validity can be compromised when highly correlated controls leave the synthetic weights weakly determined or when treated–control relationships shift after treatment. We propose a new estimand, the weight-robust treatment effect, defined as the optimizer of a worst-case optimization problem over an uncertainty class of synthetic weights compatible with the pre-treatment fit. We establish its connection to sensitivity analysis: the uncertainty class induces an interval of plausible treatment effects, and the proposed estimand is the {point in} the interval closest to zero. Under the classical identification conditions, the estimand coincides with the true treatment effect. When these conditions fail, the estimand remains point identified; if the uncertainty class contains the true post-treatment weight, it provides a conservative sign-preserving bound on the effect. The estimator of this target {may have a non-normal limiting distribution,} and we propose a novel perturbation-based method for constructing valid confidence intervals. Our proposal may be of independent interest for partial identification and sensitivity analysis, where estimators defined through constrained optimization can have non-standard limiting distributions.
\end{abstract} 

\vfill
\noindent%
\vfill

\newpage

\section{Introduction}\label{sec: introduction}

The synthetic control (SC) method
\citep{abadie2003economic, abadie2010synthetic}
is increasingly used in the social sciences because of its transparency and
interpretability. The method estimates the counterfactual outcome by
constructing a weighted average of control units that closely matches the
treated unit's pre-treatment trajectory and approximates its potential outcome
under control. The SC framework has inspired a wide range of methodological
developments, including linear prediction models
\citep{li2020statistical, chernozhukov2021exact, cattaneo2021prediction}, factor models
\citep{xu2017generalized, hahn2017synthetic, ben2021augmented}, quantile functions
\citep{gunsilius2023distributional}, and matrix completion approaches
\citep{amjad2018robust, athey2021matrix}. 

Despite its widespread use, the SC method has limitations that can compromise
the reliability of the resulting synthetic control. In particular, the synthetic weight might not be identifiable due to 
the following two sources of weight uncertainty.
\begin{itemize}
  \item \textbf{Weight instability:} When the control units are highly
  correlated, distinct weight configurations may yield similar or even
  identical pre-treatment fits while implying different post-treatment
  counterfactuals. 

  \item \textbf{Weight shifts:} The treated--control relationship may change {after treatment,}
  so that weights providing a good
  pre-treatment fit may no longer yield a reliable approximation to the treated
  unit's post-treatment counterfactual.
\end{itemize}

Under either challenge, multiple post-treatment counterfactuals may be
compatible with the observed pre-treatment fit and the allowed weight shifts.
{In this case, the treatment effect targeted by the SC method is not
necessarily point identified. Thus, we propose a new estimand that
coincides with the usual SC treatment effect under the classical identification
conditions. When these conditions fail and the treatment effect is 
partially identified,} the proposed estimand remains point identified and
provides a conservative proxy for the true treatment effect.

\subsection{Our Results and Contributions}

In this paper, we propose a novel causal estimand through the lens of
distributionally robust optimization (DRO). Rather than assuming that the
post-treatment synthetic weight vector is uniquely identified, we allow it to
vary over a class of plausible weights that remain compatible with the
pre-treatment fit up to a prespecified weight-shift tolerance. This class
captures uncertainty arising from shifts in treated--control relationships and
from highly correlated controls that leave the synthetic weights weakly
determined. We define the proposed estimand, the \emph{weight-robust treatment
effect} (WRoTE), as the optimizer of a DRO problem that maximizes the worst-case
reward over this class of plausible weights, where the reward measures the
improvement in post-treatment fit compared {with} a null treatment effect.
The term \emph{weight-robust} reflects the fact that the estimand accounts for
all plausible synthetic weights rather than relying on a uniquely determined
weight vector.

When the SC identification conditions hold and the weight-shift
tolerance is set to zero, the WRoTE coincides with the usual SC-identified
treatment effect. When these conditions fail and the usual treatment effect is
only partially identified, the WRoTE remains point identified. Provided that
the uncertainty class contains the true post-treatment weight vector, the
magnitude of the WRoTE is no greater than that of the true effect, and the
WRoTE cannot have the opposite sign. {Thus, a non-zero WRoTE certifies the
sign of the true effect, while its magnitude gives a conservative measure of
the effect that is robust to weight uncertainty.}

{Although the WRoTE is defined through DRO, we find that it has a useful
interpretation through sensitivity analysis.} The class of plausible
post-treatment weights induces an interval of plausible treatment effects,
which we refer to as the sensitivity interval. If the true post-treatment
weight belongs to this class, the interval contains the true effect and hence
partially identifies it. Theorem~\ref{thm: tau star} shows that the WRoTE is
precisely the point in this sensitivity interval closest to zero, thereby
providing an equivalent sensitivity-analysis characterization of the
DRO-defined estimand. {Hence, the sensitivity-analysis characterization
reveals that the WRoTE selects the most conservative effect compatible with
the specified weight uncertainty.} To the best of our knowledge, this is the
first work {to establish a precise connection} between DRO and sensitivity analysis in the SC framework; see
Section~\ref{subsec: id of DRoSC} for {a detailed comparison}.

Statistical inference for our proposed estimand is challenging since its
estimator, obtained from a constrained optimization problem, need not be
asymptotically normal, rendering conventional Wald inference unreliable.
{This non-regularity can arise when the optimizer is non-unique
or lies near the boundary of the constraint set
\citep[see, e.g.,][]{zhang2023proximal,bennett2026inference,andrews1999estimation}.}
To quantify this uncertainty, we
generate a collection of perturbed optimization problems {at the scale of the sampling error}, show that at least one nearly recovers the population optimization problem, and aggregate the
resulting intervals into a valid confidence interval (CI) for the WRoTE.
{Since the same non-regularity may arise for
optimization-defined bounds in sensitivity analysis and partially identified
models, the method may also be useful beyond SC.}

We evaluate our proposal across diverse data-generating processes, including
highly correlated controls and post-treatment weight shifts. In regimes with
non-regular asymptotics, standard Wald CIs suffer from undercoverage, whereas
our perturbation-based CIs attain nominal coverage; see
Section~\ref{sec: simulation studies}. We further illustrate the practical utility of
our method through a reanalysis of the Basque Country case study
\citep{abadie2003economic}.

\begin{figure}[htp!]
    \centering
    \includegraphics[width=.9\textwidth]{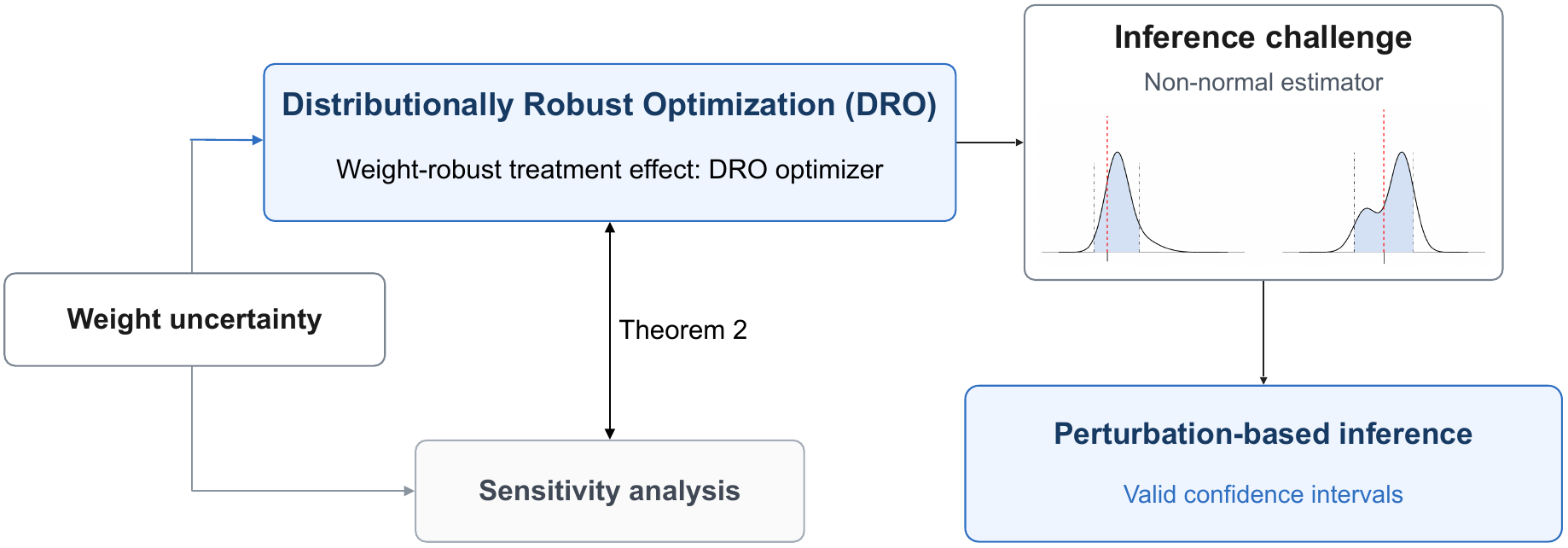}
    \caption{Overview of the two main contributions.}
    \label{fig: contribution map}
\end{figure}

We summarize our main contributions as follows, {with their relationship
further illustrated in Figure~\ref{fig: contribution map}:}
\begin{itemize}
\item[(1)] We propose the WRoTE, a DRO-defined causal estimand that equals the
usual SC treatment effect under the classical identification conditions and
remains point identified when the latter is only partially identified.

\item[(2)] The WRoTE estimator may have a nonstandard limiting distribution.
To address this non-regularity, we propose a perturbation-based procedure that
yields valid CIs. Because the same non-regularity may arise when bounds in
sensitivity analysis and partial identification are defined through
constrained optimization, the procedure is of independent interest beyond SC.
\end{itemize}

\vspace{-0mm}

\subsection{Other Related Works}\label{subsec: related works}

\noindent{\bf Non-unique SC weights.} A relevant work on non-unique synthetic weights is \citet{abadie2021penalized}, who consider multiple treated units and propose a penalized SC method to promote uniqueness.
Their approach {selects a unique solution by penalizing discrepancies between the treated and control units' characteristics}. In contrast, rather than mitigating non-uniqueness through {penalized selection}, we address non-uniqueness by introducing an uncertainty class that {accounts for all synthetic weights compatible with the pre-treatment fit}.

\noindent {\bf Sensitivity analysis in SC}.
{Recent work examines the sensitivity of SC conclusions to violations of its identifying assumptions.}
\citet{zeitler2023non} derive a conservative upper bound on bias caused by shifts in unobserved proxies, rather than the full range of effects implied by a specified weight class. { \citet{ferguson2020assessing} study how the SC estimate changes over a class of alternative weights. Their baseline class consists of weights close to the fitted SC weights, with the allowable distance determined using placebo prediction errors for the control units; their more general framework allows other notions of distance and non-unique SC solutions. Our uncertainty class is different: it is defined at the population level by requiring the pre-treatment residual control moments to be small. Thus, their weight class provides a sample-specific way to assess sensitivity to weight misspecification, whereas ours describes the population weights compatible with the pre-treatment relationship. 
{Most importantly, \citet{ferguson2020assessing} report a sample-specific range for the usual treatment effect and explicitly do not interpret it as sampling-based inference; they do not define a new point-valued population estimand. We use DRO to define the WRoTE as the point in the population sensitivity interval closest to zero and develop inference for this population target, accounting for estimation of the weight class and the resulting non-regularity. See Remark~\ref{rmk: sens dro} for further details.}
}

\noindent{\bf Inference in partially identified models.}
Sensitivity analysis asks how causal conclusions change under prespecified departures from baseline identifying assumptions \citep[e.g.,][]{rosenbaum1987sensitivity}. Such departures often replace point identification with an interval of possible causal effects \citep[e.g.,][]{huang2025variance}. If the sensitivity model contains the true departure, the causal effect lies within this interval. Inference may then aim to cover either the unknown effect \citep{imbens2004confidence} or the entire identified interval. Our target differs: the WRoTE is the point in the population sensitivity interval closest to zero. When the interval excludes zero, the WRoTE equals its nearest endpoint, connecting our problem to inference for optimization-defined bounds. Whether an endpoint estimator is asymptotically normal depends on the local stability of the optimization problem. If the optimizer and its first-order response remain stable under small sampling errors, a normal approximation may be valid \citep[e.g.,][]{tudball2023sample}. If sampling errors instead cause the solution to switch among optimizers or binding constraints with different first-order responses, the endpoint behaves like the minimum or maximum of several approximately Gaussian errors. Its limit is then generally non-Gaussian, and the standard bootstrap may fail \citep{fang2019inference}. Related methods provide complementary solutions for inference on broader set-valued targets. \citet{cho2024simple} develop a computationally simple and uniformly valid procedure based on randomly adjusted linear programs. Their confidence interval is exact for an outer population interval and therefore provides conservative coverage of the original identified interval. \citet{zhao2019sensitivity} combine the percentile bootstrap with a worst-case bounding idea to obtain a computationally tractable interval with {at least nominal coverage of the estimand uniformly over their marginal sensitivity model}. In contrast, we continue to target the same WRoTE defined by the original population sensitivity interval. We perturb the estimated objective and constraint set only when constructing the confidence interval; the population sensitivity interval and the WRoTE remain unchanged. Each perturbed problem yields a candidate interval for the same WRoTE. Because we do not know which perturbation makes the leading non-regular error negligible, we take the union of these candidate intervals.

\noindent {\bf Notation.} 
For $v\in\mathbb{R}^d$, $\|v\|_q$ {denotes} the $\ell_q$ norm and
$\|v\|_\infty=\max_j |v_j|$; $\mathbf{1}_d$ and $\mathbf{0}_d$ denote vectors of ones and zeros. 
For a matrix $M$, $M\tr$ is its transpose, $\|M\|_{\max}=\max_{i,j}|M_{i,j}|$,
and $\|M\|_2$ is its largest singular value; for symmetric $M$, $\lambda_{\min}(M)$ is its smallest eigenvalue; $\mathbf{I}$ denotes the identity matrix. We use $\pto$ and $\dto$
for convergence in probability and distribution. For $a_n>0$, $X_n=O_p(a_n)$ means that $\|X_n\|_\infty/a_n$ is bounded in probability; we write $a_n\lesssim b_n$
when $a_n\le Cb_n$ for a constant $C>0$.

\section{Synthetic Control: Essential Assumptions and Challenges}\label{sec: SC assumptions and challenges}
\begin{figure}[htp!]
    \centering
    \includegraphics[width=0.8\textwidth]{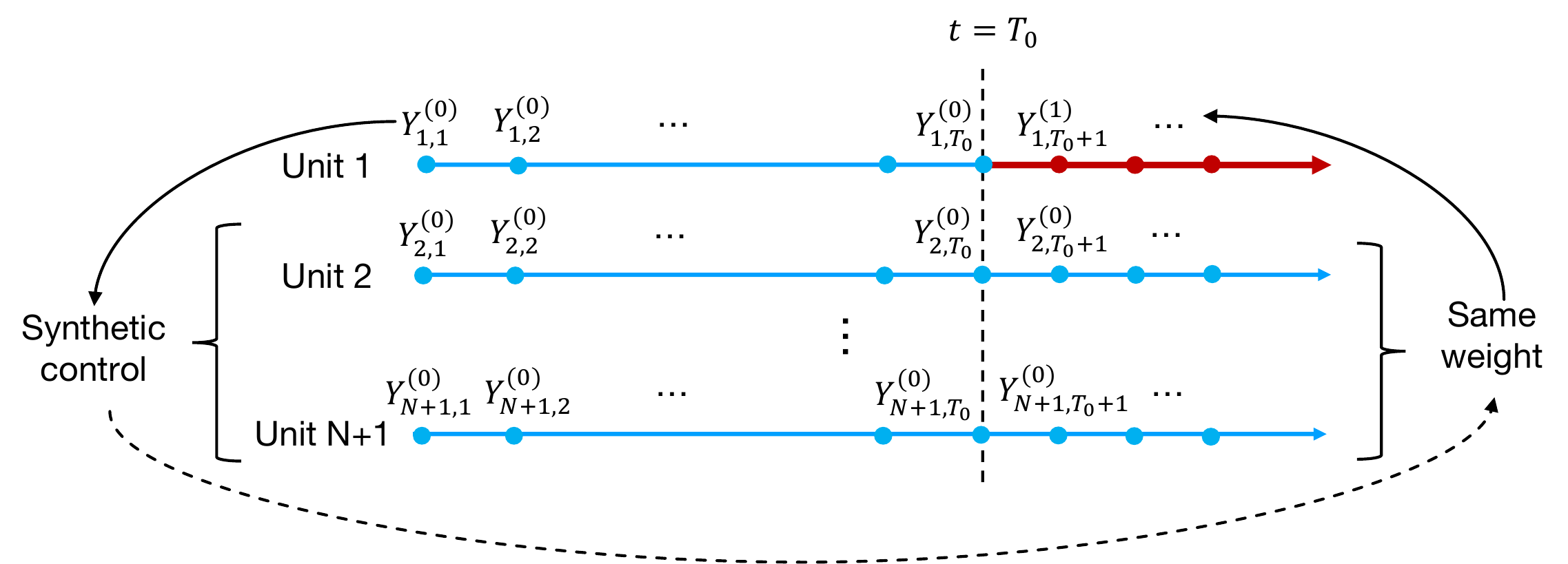}
    \caption{The synthetic control setup: Unit~1 is treated after $T_0$ (red), while the control units $2,\cdots, N+1$ remain untreated. We further illustrate the identification conditions defined below in Section~\ref{subsec: SC}.}
    \label{fig: SC E1E2}
\end{figure}

We review the SC setup with $N+1$ units observed over $T$ time periods; see Figure~\ref{fig: SC E1E2} for an illustration. 
With $T_0\leq T-1$, we refer to $t=1,\dots,T_0$ as the pre-treatment period and $t=T_0+1,\dots,T$ as the post-treatment period. Unit 1 starts receiving treatment from time $T_0+1$  while units $2,\dots,N+1$ do not receive treatment. Let $Y_{j,t}^{(0)}$ and $Y_{j,t}^{(1)}$ denote the potential outcomes of unit $j$ at time $t$ under control and treatment, respectively. Thus, the observed outcomes satisfy $Y_{1,t}=Y_{1,t}^{(0)}$ for $t\le T_0$, $Y_{1,t}=Y_{1,t}^{(1)}$ for $t>T_0$, and $Y_{j,t}=Y_{j,t}^{(0)}$ for $2\le j\le N+1$ and $1\le t\le T$.  

We define the average treatment effect on the treated (ATT) at time $t$ \citep[e.g.,][]{park2025single} as
$\tau_t = \mathbb{E}[Y_{1,t}^{(1)} - Y_{1,t}^{(0)}]$, where the expectation is taken over the randomness of $Y_{1,t}^{(1)}$ and $Y_{1,t}^{(0)}$. 
We further define the time-averaged ATT:
\begin{align} \label{eq: tau bar}
    \taubar = \frac{1}{T_1} \sum_{t=T_0+1}^{T} \tau_t \quad \text{with}\quad T_1=T-T_0.
\end{align}

Our primary focus is on inference for $\bar{\tau}$ \citep{arkhangelsky2021synthetic, liu2024proximal}, or its conservative proxy  
when $\bar{\tau}$ is not identifiable. 
We do not restrict to constant treatment effects but allow for non-constant effects $\tau_t$.  
Inference for a single $\tau_t$ is more challenging since only a single treated unit is observed at $t$, requiring additional assumptions such as a static effect \citep[e.g.,][]{abadie2010synthetic} or a parametric model for $\tau_t$ \citep{park2025single}.
Without such assumptions, researchers turn to constructing a prediction interval for $Y_{1,t}^{(1)} - Y_{1,t}^{(0)}$ \citep[e.g.,][]{cattaneo2021prediction}.
In contrast, focusing on $\bar{\tau}$ enables valid inference without imposing such additional assumptions, or resorting to prediction intervals.

\subsection{Identification Assumptions for the Synthetic Control Method}\label{subsec: SC}
We discuss the key assumptions under which the SC method identifies $\taubar$. 
Throughout, we write $X_t = (Y_{2,t}, \dots, Y_{N+1,t})\tr$ and consider the following models between unit 1 and controls \citep[e.g.,][]{chernozhukov2021exact, shen2023same}: 
\begin{align}\label{eq: outcome model}
    Y_{1,t}^{(0)} = 
    \begin{cases}
        X_t\tr \beta^{(0)} + u_t^{(0)}\ & \text{for } t=1, \dots, T_0, \\
        X_t\tr \beta^{(1)} + u_t^{(1)}\  & \text{for } t=T_0+1, \dots, T, 
    \end{cases}
    \quad \text{with}\quad \beta^{(0)},\beta^{(1)} \in \Delta^{N},
\end{align}
where $\Delta^{N} = \{ \beta : \beta_j \geq 0, \ \mathbf{1}_N\tr \beta = 1 \}$, and $\{u_t^{(0)}\}_{t=1}^{T_0}$ and $\{u_t^{(1)}\}_{t=T_0+1}^{T}$ are sequences of mean-zero error terms satisfying $\mathbb{E}[X_t u_t^{(0)}] =\mathbb{E}[X_t u_t^{(1)}]= \mathbf{0}_N$.

We state the two critical SC identification assumptions and illustrate in Figure~\ref{fig: SC E1E2} how they are used to identify $\taubar$ under model \eqref{eq: outcome model}. 
\begin{enumerate}
\item[(E1)] $\beta^{(0)}$ is the unique minimizer of the following constrained least squares:
\begin{align}\label{eq: beta0}
      \beta^{(0)}=\underset{\beta \in \Delta^{N}}{\argmin}\; \frac{1}{T_0}\sum_{t=1}^{T_0}\E\left[\left(Y_{1,t}-X_t\tr\beta\right)^2\right].
 \end{align} 
\item[(E2)] There is no weight shift before and after the treatment: $\beta^{(0)} = \beta^{(1)}.$
\end{enumerate}

Under conditions (E1) and (E2), the SC method identifies ${\beta}^{(0)}$ via \eqref{eq: beta0} and $\bar{\tau}$ as $T_1^{-1}\sum_{t=T_0+1}^{T}\E[Y_{1,t}-X_t\tr\beta^{(0)}]$,  motivating the SC estimators of $\beta^{(1)}$ and $\taubar$: 
\begin{align}\label{eq: SC estimators}
    \betahatSC=\underset{\beta \in \Delta^{N}}{\argmin}\; \frac{1}{T_0}\sum_{t=1}^{T_0}\left(Y_{1,t}-X_t\tr\beta\right)^2,\quad \tauhatSC = \frac{1}{T_1}\sum_{t=T_0+1}^{T}\left(Y_{1,t}-X_t\tr\betahatSC\right).
\end{align}

\subsection{Identification Challenges: Non-uniqueness and Weight Shift}\label{subsec: challenges}
\begin{figure}[ht]
    \centering
    \includegraphics[width=.85\textwidth]{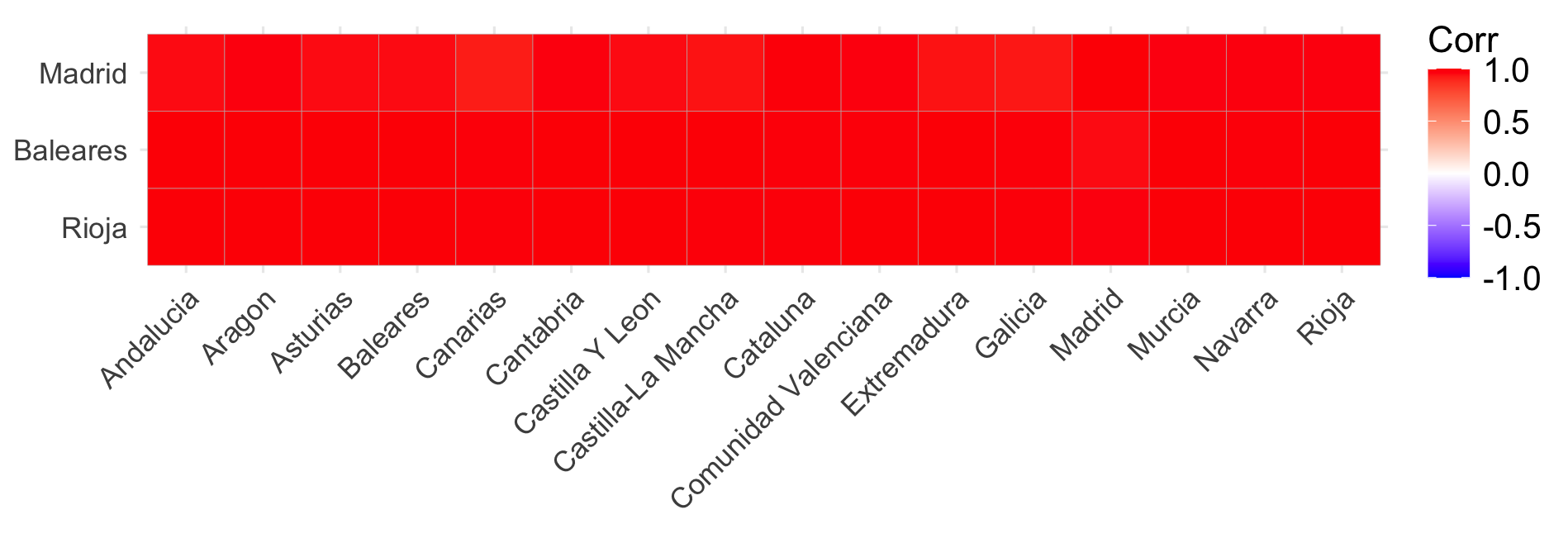}
    \caption{Correlation plot from the Basque study.  The vertical axis denotes control units selected from SC, and the horizontal axis denotes all control units.}
    \label{fig: corr of basque}
\end{figure}

We discuss how the identification conditions (E1) and (E2) may fail to hold in practice. Firstly, the minimizer in \eqref{eq: beta0} may not be unique when control units' pre-treatment outcomes are highly correlated. For example, in the Basque study \citep{abadie2003economic}, most correlations between selected and all control units are near one (see Figure~\ref{fig: corr of basque}).

\begin{figure}[ht]
    \centering
    \includegraphics[width=0.8\textwidth]{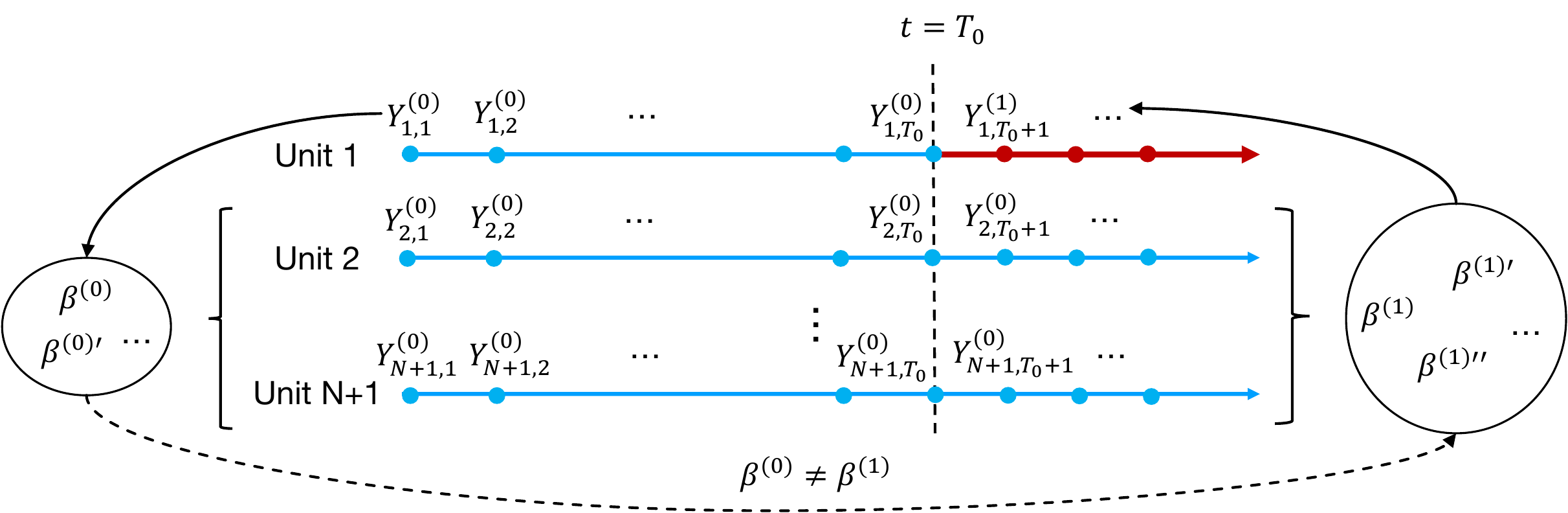}
    \caption{The two sources of weight uncertainty, in contrast to Figure~\ref{fig: SC E1E2}. The violation of (E1) makes the pre-treatment weight non-unique (left circle), while the violation of (E2) induces a weight shift, $\beta^{(0)}\neq\beta^{(1)}$. Consequently, the post-treatment weight is only known to lie in a set of plausible values (right circle) and is not point identified.}
    \label{fig: SC weight shift}
\end{figure}

Secondly, the treated-control relationship may not remain stable, as the treatment itself can alter it, producing weight shifts that violate (E2).
Weight shifts can also be understood through model misspecification.
If the simplex constraint in \eqref{eq: outcome model} were dropped, $\beta^{(0)}$ and $\beta^{(1)}$ would correspond to the population least-squares projections of $Y_{1,t}^{(0)}$ onto the controls $X_t$ during the pre- and post-treatment periods, respectively. These projection coefficients depend on the joint distribution of $(Y_{1,t}^{(0)},X_t)$, in particular on the distribution of $X_t$. Consequently, even if the conditional relationship $\E[Y_{1,t}^{(0)}\mid X_t]$ remains unchanged across periods, a shift in the distribution of the controls generally changes the projection, leading to weight shifts $\beta^{(1)}\neq\beta^{(0)}$. {Figure~\ref{fig: SC weight shift} depicts the two sources of weight uncertainty from the violation of (E1) and (E2).}

{To illustrate how the SC estimator can be unreliable when either (E1) or (E2) fails, we conduct semi-real data analyses using the Basque study \citep{abadie2003economic}. We observed that small perturbations of the pre-treatment data substantially change which control units are selected, illustrating weight instability when (E1) fails. Furthermore, the bias of $\tauhatSC$ increases with the magnitude of the post-treatment weight shift, illustrating the consequence of violating (E2). Implementation details and full results are provided in Appendix~\ref{sec: semi-real simulation}.}

\section{Robust Identification under Weight Uncertainty}
\label{sec: DRoSC}

As discussed in the previous section, violations of the key SC identification
conditions may leave multiple post-treatment synthetic weights compatible with
the available information. This weight uncertainty does not by itself imply
that the time-averaged ATT $\bar{\tau}$ is unidentified: the effect remains
point identified if all plausible weights imply the same time-averaged
counterfactual outcome. The identification problem arises only when the
plausible weights imply different time-averaged counterfactual outcomes and
hence different treatment effects. 

To address this problem, we propose a new
causal estimand through distributionally robust optimization (DRO). We first
introduce an uncertainty class for the plausible post-treatment synthetic
weights in
Section~\ref{subsec: uncertainty class}. We then define the weight-robust
treatment effect directly through DRO in Section~\ref{subsec: def of DRoSC},
before establishing its connection with sensitivity analysis in
Section~\ref{subsec: id of DRoSC}.

\subsection{Uncertainty Class for Weight Uncertainty}
\label{subsec: uncertainty class}

Under oracle knowledge of the true post-treatment synthetic weight
$\beta^{(1)}$, the time-averaged ATT is
$\bar{\tau}=\tau(\beta^{(1)})$, where, for any synthetic weight $\beta$,
\begin{align}
\label{eq: def of tau beta}
\tau(\beta)
    = \mu_Y-\mu\tr\beta,
\qquad
\mu_Y
    = \frac{1}{T_1}\sum_{t=T_0+1}^{T}\E[Y_{1,t}],
\qquad
\mu
    = \frac{1}{T_1}\sum_{t=T_0+1}^{T}\E[X_t].
\end{align}
Thus, identifying the post-treatment weight $\beta^{(1)}$ is sufficient for
identifying $\bar{\tau}$, but it is not necessary. In particular, even if
several post-treatment weights are compatible with the available information,
the treatment effect remains point identified when all such weights imply the
same value of $\tau(\beta)$. Weight uncertainty leads to effect
non-identification only when the plausible weights produce different
post-treatment counterfactuals and hence different values of $\tau(\beta)$.

We focus on settings in which conditions (E1) and (E2) fail and therefore no
longer guarantee a uniquely determined post-treatment weight or treatment
effect. Rather than selecting one weight among the plausible alternatives, we
represent the remaining uncertainty through a class of plausible
post-treatment synthetic weights.

To construct this uncertainty class, we begin by reviewing the moment condition
for the pre-treatment weight $\beta^{(0)}$. By the outcome model
\eqref{eq: outcome model}, the pre-treatment residual
$Y_{1,t}-X_t\tr\beta^{(0)}$ is uncorrelated with the controls $X_t$ for
$t=1,\ldots,T_0$. Therefore, $\beta^{(0)}$ is a solution of
\begin{equation}
\label{eq: beta0 identification}
\frac{1}{T_0}\sum_{t=1}^{T_0}
\E\!\left[
X_t\left(Y_{1,t}-X_t\tr\beta\right)
\right]
=0.
\end{equation}
When the controls are perfectly or highly correlated, this equation may have
multiple or nearly indistinguishable solutions, of which $\beta^{(0)}$ is one.

Moreover, because the treated-control relationship may shift after treatment,
the true post-treatment weight $\beta^{(1)}$ need not satisfy the
pre-treatment moment condition \eqref{eq: beta0 identification} exactly. To
allow for such a shift, we introduce a tolerance parameter $\lambda\geq 0$ and
define the following uncertainty class: %
\begin{equation}
\label{def: Omega}
\begin{aligned}
\Omega(\lambda)
\coloneqq
\left\{
\beta\in\Delta^N:
\left\|
\frac{1}{T_0}\sum_{t=1}^{T_0}
\E\!\left[
X_t\left(Y_{1,t}-X_t\tr\beta\right)
\right]
\right\|_{\infty}
\leq\lambda
\right\}.
\end{aligned}
\end{equation}
The parameter $\lambda$ controls the extent to which a plausible
post-treatment weight may violate the pre-treatment moment condition. When
$\lambda=0$, the uncertainty class contains all weights satisfying the
pre-treatment moment condition exactly; this class may still contain multiple
weights when the controls are perfectly correlated. Increasing $\lambda$
allows larger departures from the pre-treatment relationship.
When there is no ambiguity, we abbreviate $\Omega(\lambda)$ as $\Omega$.

The uncertainty class admits an equivalent representation as a
covariance-scaled neighbourhood of $\beta^{(0)}$. Define
$
\Sigma
=
T_0^{-1}\sum_{t=1}^{T_0}\E[X_tX_t\tr]$ and 
$\gamma
=
T_0^{-1}\sum_{t=1}^{T_0}\E[X_tY_{1,t}].$
Because the pre-treatment moment condition implies
$\gamma=\Sigma\beta^{(0)}$, we can write
\begin{align}
\label{eq: Omega as Sigma ball}
\Omega(\lambda)
=
\left\{
\beta\in\Delta^N:
\|\gamma-\Sigma\beta\|_{\infty}\leq\lambda
\right\}
=
\left\{
\beta\in\Delta^N:
\|\Sigma(\beta-\beta^{(0)})\|_{\infty}\leq\lambda
\right\}.
\end{align}
A main advantage of this covariance-scaled formulation is that it does not
require us to select a unique pre-treatment weight or invert $\Sigma$.
Indeed, when the controls are perfectly correlated, $\Sigma$ is singular and
$\beta^{(0)}$ need not be identifiable. Nevertheless, the moment discrepancy
$\gamma-\Sigma\beta$ remains well defined and directly measures how well a
candidate weight reproduces the pre-treatment treated--control relationship.

\subsection{Causal Estimand via Distributionally Robust Optimization}
\label{subsec: def of DRoSC}

We motivate the weight-robust treatment effect by expressing the population
SC estimand as the solution to an optimization problem. For any synthetic
weight $\beta$ and candidate treatment effect $\tau$, define the reward
function
\begin{equation}
\label{eq: reward}
R_{\beta}(\tau)
\coloneqq
\frac{1}{T_1}\sum_{t=T_0+1}^{T}
\E\left[
\left(Y_{1,t}-X_t\tr\beta\right)^2
-
\left(Y_{1,t}-X_t\tr\beta-\tau\right)^2
\right].
\end{equation}
The first term is the post-treatment prediction error under the null effect
$\tau=0$, whereas the second is the prediction error after allowing for the
candidate effect $\tau$. Thus, $R_{\beta}(\tau)$ measures the improvement in
post-treatment fit obtained by introducing $\tau$. For a fixed $\beta$, this
reward can be written as 
$R_{\beta}(\tau)=2\tau\cdot\tau(\beta)-\tau^2$
with $\tau(\beta)$ in \eqref{eq: def of tau beta}, and is uniquely maximized at $\tau=\tau(\beta)$. With oracle knowledge of the
true post-treatment weight $\beta^{(1)}$, the time-averaged ATT can
be written as
\begin{equation}
\label{eq: standard SC}
\bar{\tau}
=
\argmax_{\tau\in\mathbb{R}}
R_{\beta^{(1)}}(\tau).
\end{equation}

When $\beta^{(1)}$ is unknown and is only known to belong to $\Omega$, the reward associated with a candidate effect $\tau$ depends on which plausible
post-treatment weight $\beta$ is used. We therefore evaluate each $\tau$ by its
worst-case reward over all $\beta\in\Omega$ and define the
\emph{weight-robust treatment effect} (WRoTE) as
\begin{align}
\label{def: DRoSC tau}
\tau^*(\Omega)
\coloneqq
\argmax_{\tau\in\mathbb{R}}
\left[
\min_{\beta\in\Omega}R_{\beta}(\tau)
\right].
\end{align}

Equation~\eqref{def: DRoSC tau} admits a game-theoretic interpretation
\citep{blackwell1979theory}. The analyst chooses a candidate treatment effect
$\tau$, while nature adversarially selects, from $\Omega$, the plausible
post-treatment weight that gives the smallest reward for that choice. The
analyst then selects $\tau$ to maximize this worst-case reward. The resulting
estimand is therefore robust to all plausible post-treatment weights, rather
than relying on one particular estimated weight vector.

\subsection{Identification and Interpretation}
\label{subsec: id of DRoSC}

In the following, we characterize the optimizer of the DRO problem in
\eqref{def: DRoSC tau} and build its connection to the sensitivity analysis.

\begin{Theorem}
\label{thm: id of DRoSC}
The DRO optimizer $\tau^*(\Omega)$ defined in \eqref{def: DRoSC tau} is uniquely
determined by
\begin{align}
\label{eq: tau star and beta star}
\tau^*(\Omega)
=
\mu_Y-\mu\tr\beta^*(\Omega),
\qquad
\beta^*(\Omega)
\in
\argmin_{\beta\in\Omega}
\left[\mu_Y-\mu\tr\beta\right]^2.
\end{align}
\end{Theorem}

We interpret $\beta^*(\Omega)$ as an adversarial post-treatment weight that
drives the time-averaged ATT closest to zero. Theorem~\ref{thm: id of DRoSC}
also provides a direct way to compute $\tau^*(\Omega)$: we first obtain
$\beta^*(\Omega)$ by solving the quadratic program in
\eqref{eq: tau star and beta star}, and then evaluate
$\tau^*(\Omega)=\mu_Y-\mu\tr\beta^*(\Omega)$. When clear from context, we write
$\tau^*$ and $\beta^*$ for $\tau^*(\Omega)$ and $\beta^*(\Omega)$,
respectively. This characterization does not require the key SC identification
conditions (E1) and (E2).

We shall emphasize that the quadratic program in \eqref{eq: tau star and beta star} is degenerate
because $\mu\mu\tr$ has rank one. Consequently, although $\tau^*$ is unique,
$\beta^*$ need not be unique, which complicates estimation, inference, and
theory.
{Despite this degeneracy, the following sensitivity-analysis characterization allows $\tau^*$ to be estimated at the parametric rate, up to a logarithmic factor; see Theorem~\ref{thm: conv of tauhat}.}

Each plausible weight $\beta\in\Omega$ implies a treatment effect $\tau(\beta)$.
Together, these effects form the interval
\begin{equation}
\label{eq: sensitivity interval}
I(\Omega)
\coloneqq
\{\tau(\beta):\beta\in\Omega\}
=
\left[
\min_{\beta\in\Omega}\tau(\beta),
\,
\max_{\beta\in\Omega}\tau(\beta)
\right],
\end{equation}
which we call the \emph{sensitivity interval}. This interval reports all
treatment effects compatible with the specified weight uncertainty. If $\beta^{(1)}\in\Omega$, then
$
\bar{\tau}
=
\tau(\beta^{(1)})
\in I(\Omega),
$
so the interval contains the true time-averaged ATT.

The interval also clarifies when weight uncertainty affects identification of
the treatment effect. If $\tau(\beta)$ is constant over $\Omega$, then
$I(\Omega)$ is a singleton and $\bar{\tau}$ remains point identified even
though the synthetic weight may not be. If $\tau(\beta)$ varies over
$\Omega$, then $I(\Omega)$ is nondegenerate and describes the resulting
partial identification of $\bar{\tau}$. The following theorem shows that the
DRO-defined WRoTE has an exact characterization in terms of this interval.

\begin{Theorem}
\label{thm: tau star}
The weight-robust treatment effect $\tau^*$ defined in
\eqref{def: DRoSC tau} is the projection of the origin onto the sensitivity
interval $I(\Omega)$ in \eqref{eq: sensitivity interval}; that is,
\begin{align*}
\tau^*
=
\begin{cases}
\displaystyle \min_{\beta\in\Omega}\tau(\beta)
& \text{if } \tau(\beta)>0 \text{ for all } \beta\in\Omega, \\[4pt]
\displaystyle \max_{\beta\in\Omega}\tau(\beta)
& \text{if } \tau(\beta)<0 \text{ for all } \beta\in\Omega, \\[4pt]
0
& \text{if there exists } \beta\in\Omega
  \text{ such that } \tau(\beta)=0,
\end{cases}
\end{align*}
where $\tau(\beta)$ is defined in \eqref{eq: def of tau beta}.
\end{Theorem}
\begin{remark}[Sensitivity analysis and DRO.]\label{rmk: sens dro}
Sensitivity analysis and DRO use the same uncertainty class
$\Omega$ but report different objects. Sensitivity analysis reports the full
interval $I(\Omega)$ without selecting among its elements. DRO instead defines
a single effect through the worst-case reward in \eqref{def: DRoSC tau}.
Theorem~\ref{thm: tau star} connects the two approaches by showing that the DRO
optimizer is exactly the point of $I(\Omega)$ closest to zero. Thus, sensitivity
analysis provides an equivalent interpretation of the WRoTE.
\end{remark}

Theorem~\ref{thm: tau star} also yields the following relationship between
$\tau^*$ and $\bar{\tau}$ when the uncertainty class contains the true
post-treatment weight.

\begin{Theorem}
\label{thm: sign of tau star}
If $\beta^{(1)}\in\Omega$, then $\tau^*$ cannot have the opposite sign to
$\bar{\tau}$ and $|\tau^*|\leq|\bar{\tau}|$.
\end{Theorem}

Theorem~\ref{thm: sign of tau star} shows that $\tau^*$ is a conservative
proxy for $\bar{\tau}$ whenever $\Omega$ contains $\beta^{(1)}$. When (E1) and
(E2) hold, $\bar{\tau}$ is identifiable, and setting $\lambda=0$ gives
$\tau^*=\bar{\tau}$. When either condition fails and the plausible weights
imply different treatment effects, $\bar{\tau}$ is no longer point identified,
but $\tau^*$ remains point identified. Its magnitude is no greater than that
of $\bar{\tau}$, and its sign cannot oppose that of $\bar{\tau}$; hence
$\tau^*$ and $\bar{\tau}$ have the same sign whenever $\tau^*\neq0$.

\section{Estimation procedure}\label{sec: estimation of DRoSC}
We devise a data-dependent estimator of the WRoTE $\tau^*$ based on the identification in Theorem~\ref{thm: id of DRoSC}, beginning with an estimator of the uncertainty class $\Omega$ in \eqref{def: Omega}: 
\begin{align}\label{eq: estimator of omega}
    \Omegahat(\lambda) = \left\{\beta \in \Delta^{N}: \|\widehat{\gamma}-\widehat{\Sigma}\beta\|_{\infty} \leq \lambda+ \rho \right\}, %
\end{align}
where $\widehat{\Sigma} = T_0^{-1}\sum_{t=1}^{T_0}X_tX_t\tr$ and $\widehat{\gamma} = T_0^{-1}\sum_{t=1}^{T_0}X_tY_{1,t}$ and $\rho$ is a tuning parameter accounting for the estimation errors in $\widehat{\Sigma}$ and $\widehat{\gamma}$; its selection is discussed at the end of the section. 
When unambiguous, we denote $\Omegahat(\lambda)$ as $\Omegahat$.

Building on the identification in Theorem~\ref{thm: id of DRoSC}, we estimate $\beta^{*}$ and $\tau^*$ by 
\begin{align}\label{eq: betahat tauhat}
    \betahat(\Omegahat){\in}\argmin_{\beta \in \Omegahat}\left[\muhat_Y-\muhat\tr\beta\right]^2,\quad \widehat{\tau}(\Omegahat) = \muhat_Y - \muhat\tr\betahat(\Omegahat),
\end{align}
where we estimate $\Omega$ by $\widehat{\Omega}$ in \eqref{eq: estimator of omega} and estimate
$\mu_Y$ and $\mu$ by the corresponding sample averages $\muhat_Y = T_1^{-1}\sum_{t=T_0+1}^{T} Y_{1,t}$ and $\muhat = T_1^{-1}\sum_{t=T_0+1}^{T}X_t$.
We denote $\betahat(\Omegahat)$ and $\tauhat(\Omegahat)$ as $\betahat$ and $\tauhat$ when there is no confusion. We summarize our estimation procedure in Algorithm~\ref{alg:DRoSC} in Appendix~\ref{subsec: alg DRoSC}.

 { Since the plug-in estimation of $\gamma$ and $\Sigma$ introduces additional uncertainty, we introduce the tuning parameter $\rho$ in \eqref{eq: estimator of omega} accounting for the estimation uncertainty.} For i.i.d.\ data, our theory suggests a data-dependent choice:
\begin{align}
\rho
&= C\left[\widehat{\sigma}\cdot\max_{2 \leq j \leq N+1} \left(\frac{1}{T_0}\sum_{t=1}^{T_0}Y_{j,t}^2\right)^{\frac{1}{2}}+\lambda\right]\frac{{\log (\max\{T_0,N\})}^{1/2}}{\sqrt{T_0}},
\label{eq: tuning from specifying}
\end{align}
where $\widehat{\sigma}^2 = T_0^{-1}\sum_{t=1}^{T_0}(Y_{1,t}-X_t\tr\betahatSC)^2$ and $C>0$ is a constant; see Appendix~\ref{subsec: justify rho} for the justification. To choose $C$, we initialize it at a small value (e.g., $C=0.01$) and increase it by a factor of $1.25$ until the minimization problem in \eqref{eq: betahat tauhat} becomes feasible under $\Omegahat$, yielding the smallest $\rho$ attaining feasibility; the non-i.i.d.\ case is discussed in Appendix~\ref{subsec: non iid}.

\section{Perturbation-based inference}\label{sec: inference}

\subsection{Inference Challenge: Non-regularity and Instability}\label{subsec:inference challenges}
\begin{figure}[htp!]
\centering
    \includegraphics[width=.75\textwidth]{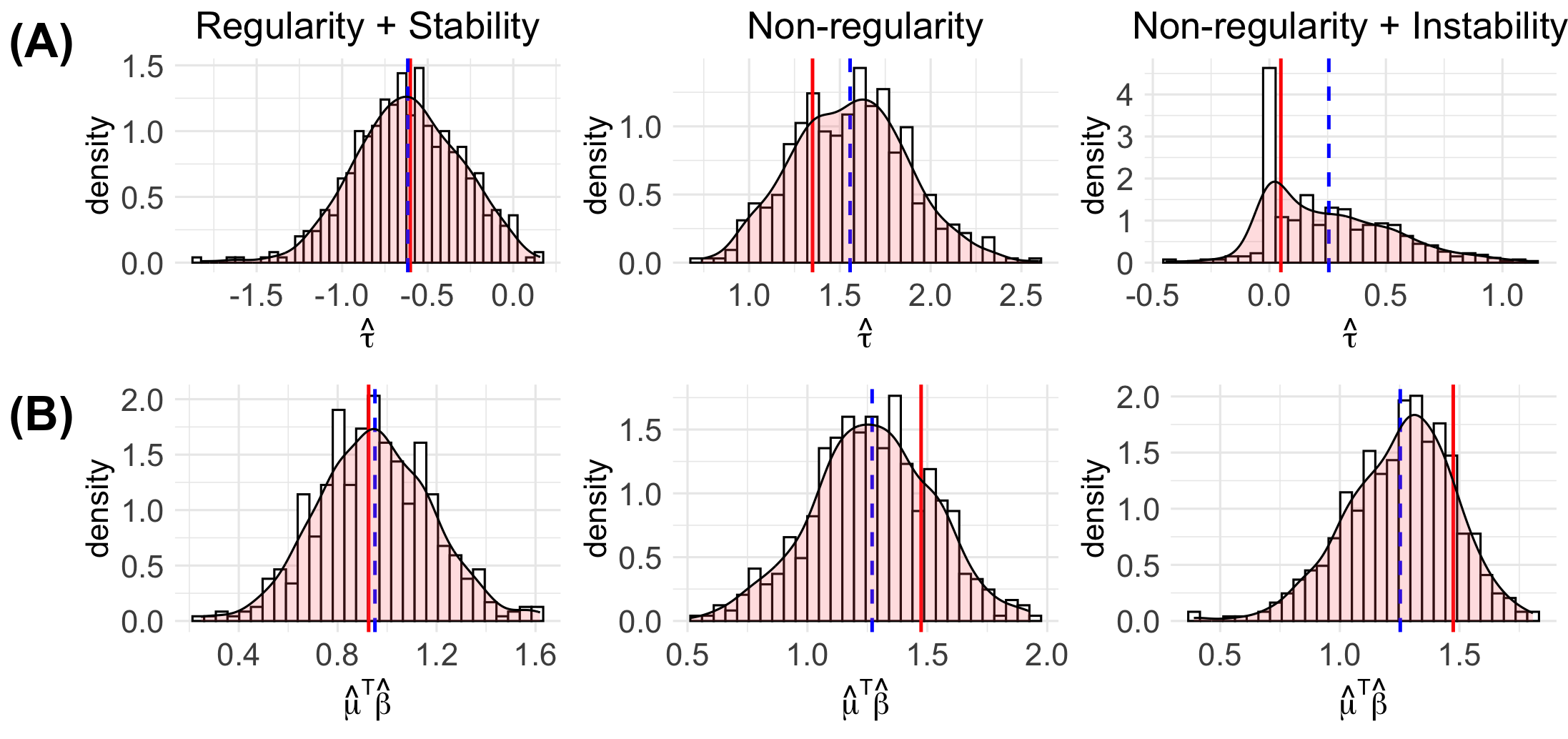}

    \caption{Histograms of $\tauhat$ (panel (A)) and $\muhat\tr\betahat$ (panel (B)) over 500 simulations from setting (S2) in Section~\ref{sec: simulation studies}, with $T_0=T_1=25$; columns correspond to $\tau^*\approx-0.6,\,1.35,\,0.05$. Red solid lines mark $\tau^*$ in the panel (A) and $\mu\tr\beta^*$ in the panel (B); blue dashed lines mark the sample averages.}
    \label{fig:hist}
\end{figure}

 The inference challenge arises since the estimator $\widehat{\tau}$ in \eqref{eq: betahat tauhat} may not admit a standard limiting distribution. The estimation error decomposes as $\tauhat-\tau^* =\muhat_Y-\mu_Y-(\muhat\tr\betahat-\mu\tr\beta^*).$ While $\muhat_Y-\mu_Y$ is asymptotically normal by applying the central limit theorem, {Figure~\ref{fig:hist} shows that} $\muhat\tr\betahat-\mu\tr\beta^*$ {may behave non-regularly} due to the boundary constraint on $\beta^*$ and the highly correlated controls. %
 In Figure~\ref{fig:hist}, we plot histograms of $\tauhat$ (panel (A)) and $\muhat\tr\betahat$ (panel (B)) over 500 simulations of data generated following setting (S2) in Section~\ref{sec: simulation studies}. The left panel exhibits near-normal behavior,
whereas the other panels {deviate from normality, causing} undercoverage of normality-based CIs.%

The non-standard behavior in Figure~\ref{fig:hist} arises from two related but distinct features. First, when the synthetic weight $\beta^*$ lies on or near the boundary, small sampling changes can alter which constraints are active. This can produce a non-normal, often mixture, limiting distribution; we refer to this as non-regularity \citep[e.g.,][]{self1987asymptotic,andrews1999estimation,drton2009likelihood}. Second, highly correlated controls make $\Omega$ nearly flat in some directions, so small errors in estimating $\gamma$ or $\Sigma$ can produce large changes in $\widehat{\Omega}$ and $\widehat{\beta}$; we refer to this sensitivity to small estimation errors as instability. Instability can amplify changes in the active constraints, so the two phenomena may occur together. 
\subsection{Perturbation-based inference}\label{subsec:perturbation}
We propose a perturbation method to address the inferential challenge in Section~\ref{subsec:inference challenges}, beginning with the intuition and recalling the identification result from Theorem~\ref{thm: id of DRoSC}: 
\begin{align}\label{eq: beta star with Omega}
    \beta^*{\in}\argmin_{\beta \in \Omega}\left(\mu_Y-\mu\tr\beta\right)^2\quad\text{with}\quad\Omega = \left\{\beta \in \Delta^N:\|\gamma-\Sigma\beta\|_{\infty}\leq \lambda\right\}.
\end{align}
The estimator $\widehat{\beta}$ presented in \eqref{eq: betahat tauhat} replaces $\{\Sigma,\gamma,\mu_Y,\mu\}$ in \eqref{eq: beta star with Omega} with their sample analogs $\{\Sigmahat,\gammahat,\muhat_Y,\muhat\}$. Our main idea is to add perturbation to the sample-based optimization problem \eqref{eq: betahat tauhat} and create a collection of perturbed optimization problems, with the hope that one of these perturbed problems almost recovers \eqref{eq: beta star with Omega}.

We generate $M$ perturbed quantities $\{\Sigmahatm,\gammahatm,\muhatm_Y,\muhatm\}_{m=1}^M$  by adding perturbations to the sample estimator $\{\Sigmahat,\gammahat,\muhat_Y,\muhat\}$. We use each set of perturbed quantities to define a perturbed optimization problem, and solve for the corresponding weight vector $\betahatm$ as
\begin{align}\label{eq: m-th betahat}
    \betahatm {\in}\argmin_{\beta \in \Omegahatm}\left[\muhat_Y^{[m]}-(\muhat^{[m]})\tr\beta\right]^2,
\end{align}
with the perturbed uncertainty class $\Omegahatm$ defined in the following \eqref{eq: m-th omega}. 
{For each $\betahatm$, we then construct the $m$-th candidate estimator of $\tau^*$ as
\begin{equation}\label{eq: m-th tauhat}
\tauhatm = \muhat_Y - (\muhatm)\tr\betahatm.
\end{equation}
Thus, relative to $\tauhat$ in \eqref{eq: betahat tauhat}, we retain $\muhat_Y$ but replace $\muhat\tr\betahat$ with $(\muhatm)\tr\betahatm$.}

\begin{figure}[htp!]
    \centering
    \includegraphics[width=.85\textwidth]{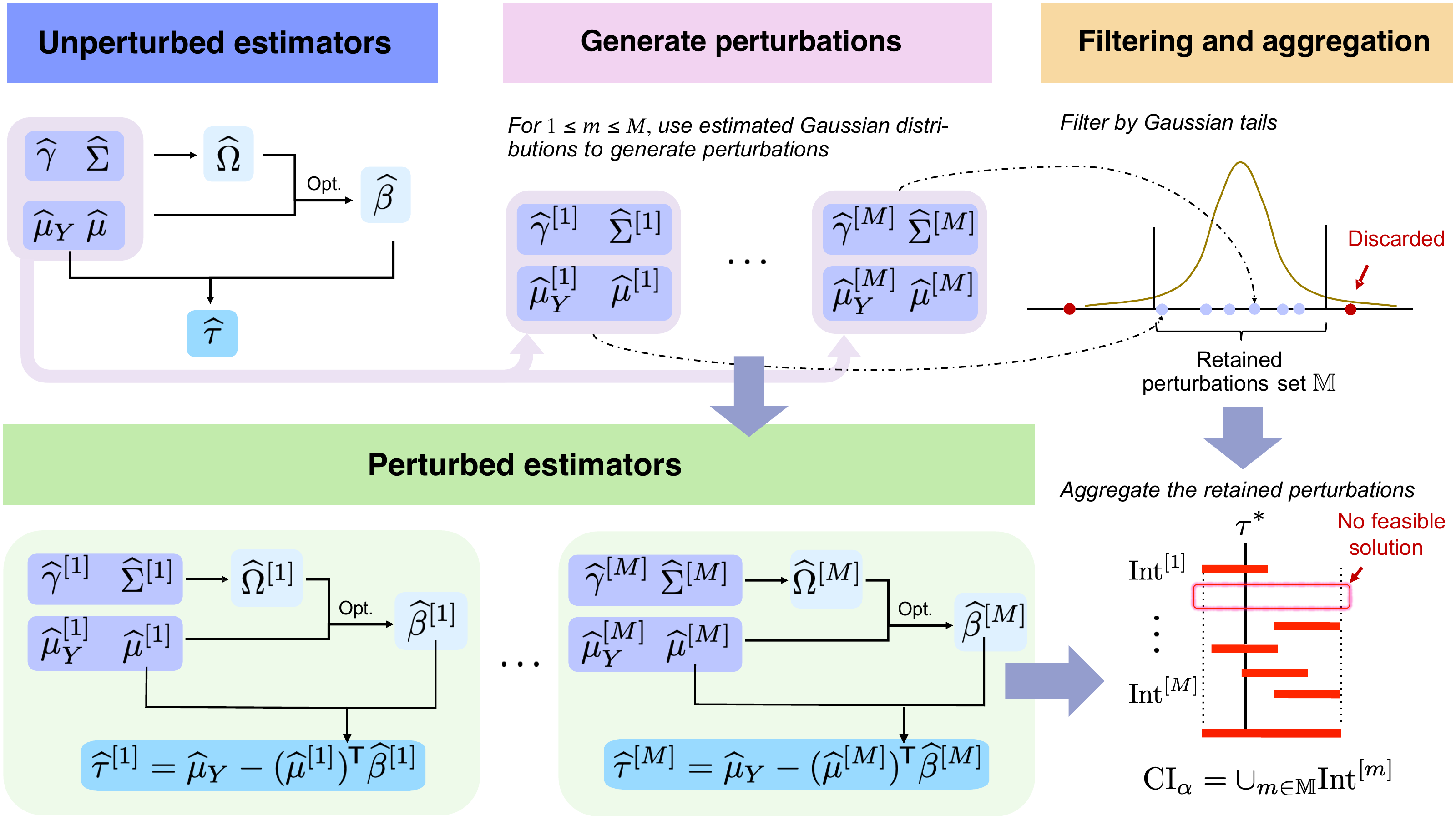}
    \caption{Workflow of the perturbation-based inference procedure.}
    \label{fig: workflow}
\end{figure}

We show that there exists $m^*$ such that \eqref{eq: m-th betahat} with $m=m^*$ nearly recovers \eqref{eq: beta star with Omega} and hence $(\muhatmstar)\tr\betahatmstar$ is nearly the same as $\mu\tr\beta^*$; see Theorem~\ref{thm: negligible} in the Appendix. 
{For $m=m^*$, the remaining uncertainty in $\tauhat^{[m^*]}$ lies primarily in estimating $\mu_Y$. We therefore use a standard normal approximation for $\muhat_Y$ to construct a candidate interval for each perturbation. The interval associated with $m^*$ provides asymptotically valid coverage; however, since $m^*$ is unknown, we retain a set $\mathbb{M}$ of plausible perturbations and aggregate their candidate intervals to obtain a CI for $\tau^*$.
Figure~\ref{fig: workflow} summarizes this procedure; the formal construction is provided below.}

We now provide full details of our proposal. { Under mild conditions, the estimators $\Sigmahat$, $\gammahat$ in \eqref{eq: estimator of omega}, and $\muhat_Y$, $\muhat$ in \eqref{eq: betahat tauhat} admit the asymptotic approximations: 
$\vecl(\Sigmahat-\Sigma) \overset{d}{\approx} \mathcal{N}(0,\VhatSigma)$,  $\gammahat-\gamma \overset{d}{\approx} \mathcal{N}(0,\Vhatgamma)$, $\muhat_Y-\mu_Y\overset{d}{\approx}\mathcal{N}(0,\VhatY)$, $\muhat-\mu \overset{d}{\approx} \mathcal{N}(0,\Vhatmu)$}, %
where $\vecl(\Sigmahat-\Sigma)$ stacks the columns of the lower triangle of $\Sigmahat-\Sigma$, $\overset{d}{\approx}$ denotes approximate equality in distribution,  and $\VhatSigma$, $\Vhatgamma$, $\VhatY$, $\Vhatmu$ are scaled estimated covariance matrices, {with detailed construction provided in \eqref{eq: Vhat} in Appendix~\ref{subsec: choice cov est}. }%
We now detail our two-step proposal: perturbation and aggregation. 

\noindent {\bf Step 1: Perturbation.} We begin by generating perturbed quantities related to the uncertainty class $\Omega$ in \eqref{def: Omega} and objective function $\tau(\beta)$ in \eqref{eq: def of tau beta}. 
Specifically, given the observed data, we generate i.i.d.\ samples $\{\widehat{\Sigma}^{[m]},\widehat{\gamma}^{[m]}\}_{m=1}^M$, following 
\begin{equation}
\vecl(\widehat{\Sigma}^{[m]}) \sim \mathcal{N}\left(\vecl(\widehat{\Sigma}),\VhatSigma+\|\VhatSigma\|_{\max}\mathbf{I}\right), \; \widehat{\gamma}^{[m]}\sim \mathcal{N}\left(\widehat{\gamma},\Vhatgamma+\|\Vhatgamma\|_{\max}\mathbf{I}\right).\label{eq: sampling of sigma and gamma}
 \end{equation}
 To ensure symmetry, we fill the upper triangle of each perturbed matrix $\widehat{\Sigma}^{[m]}$ by setting $\widehat{\Sigma}^{[m]}_{k,l} = \widehat{\Sigma}^{[m]}_{l,k}$ for $1 \leq l < k \leq N$. In addition, we generate i.i.d.\ samples $\{\muhat^{[m]},\muhat_Y^{[m]}\}_{m=1}^M$ as follows:
 \begin{align}
        &\muhat_Y^{[m]} \sim \mathcal{N}\left(\muhat_Y,\VhatY\right),\quad \muhat^{[m]} \sim \mathcal{N}\left(\muhat,\Vhatmu+\|\Vhatmu\|_{\max}\mathbf{I}\right).\label{eq: sampling of mu Y and mu 1}
 \end{align}
Conditional on the data, all four perturbations are independent across the four perturbed quantities and across $m$; the diagonal additions (e.g., $\VhatSigma+\|\VhatSigma\|_{\max}\mathbf{I}$) in \eqref{eq: sampling of sigma and gamma} and \eqref{eq: sampling of mu Y and mu 1} ensure positive-definite generating covariances and mitigate near-singularity when $N$ is large relative to $T_0$ or $T_1$. %

{Throughout, we use $p=1+N(N+5)/2$ to denote the total dimension of $\vecl(\Sigma)$, $\gamma$, $\mu_Y$, and $\mu$ in the population problem \eqref{eq: beta star with Omega}. For $1\leq m\leq M$, we replace $\Sigmahat$ and $\gammahat$ in \eqref{eq: estimator of omega} with $\Sigmahatm$ and $\gammahatm$} and construct the perturbed uncertainty class $\Omegahatm(\lambda)$ as
\begin{align}\label{eq: m-th omega}
    \Omegahatm(\lambda) = \left\{\beta \in \Delta^N: \|\gammahatm - \Sigmahatm\beta\|_{\infty}\leq \lambda+\rho_M\right\},
\end{align}
where $\rho_M \asymp\left[\log(\min\{T_0,T_1\})/M\right]^{1/p}/\sqrt{T_0}$ is a tuning parameter. %
We provide the details for the data-dependent selection of $\rho_M$ after providing the full procedure of our proposed method. When there is no confusion, we denote $\Omegahatm(\lambda)$ as $\Omegahatm$. 

{With the perturbed quantities and $\Omegahatm$ now defined, equations~\eqref{eq: m-th betahat} and \eqref{eq: m-th tauhat} complete the construction of $\betahatm$ and $\tauhatm$.
The estimation error decomposes as
\begin{align}\label{eq: decomposition of tauhatm}
\tauhatm - \tau^* &= (\muhat_Y-\mu_Y)-\left[(\muhatm)\tr\betahatm-\mu\tr\beta^*\right].
\end{align}
{For $m=m^*$, the second term is negligible, motivating the normality-based candidate intervals in Step 2.} %
}

\noindent {\bf Step 2: Filtering and Aggregation.} {Although we do not know which index is $m^*$, we aim to retain it while excluding the indices unlikely to be $m^*$.} Motivated by this, we screen out a small proportion of perturbations if they appear on the tails of the distributions in \eqref{eq: sampling of sigma and gamma} and \eqref{eq: sampling of mu Y and mu 1}.  
We define normalized perturbed statistics as $\widehat{T}^{[m]}= \widehat{S}\widehat{U}^{[m]}$ for $m=1,\ldots,M$ where $\widehat{U}^{[m]}=(\muhatm_Y-\muhat_Y, (\muhatm-\muhat)\tr,(\vecl(\Sigmahatm-\Sigmahat))\tr,(\gammahatm-\gammahat)\tr)\tr$, $\widehat{S}= \text{diag}(\VhatY^{-1/2},D(\Vhatmu) ,D(\VhatSigma),D(\Vhatgamma))$. Here, $D(A)$ denotes a vector of length equal to the diagonal dimension of $A$, with $[D(A)]_j=(A_{j,j}+\|A\|_{\max})^{-1/2}$.
The vector $\widehat{U}^{[m]}$ collects the centered $m$-th perturbed quantities generated from the distributions in \eqref{eq: sampling of sigma and gamma} and \eqref{eq: sampling of mu Y and mu 1}, and
{$\widehat{S}$ is the corresponding diagonal scaling matrix.}
Thus, $\widehat{T}^{[m]}$ represents the vector of normalized deviations between the perturbed quantities and their associated estimators. With $\widehat{T}^{[m]}$, we introduce the following index set $\mathbb{M}$ as
\begin{align}\label{eq: index set}
    &\mathbb{M}=\left\{1\leq m\leq M: \lambda_{\min}(\Sigmahatm)\geq 0, \quad  \left\|\widehat{T}^{[m]}\right\|_{\infty}
    \leq 1.1z_{\alpha_0/(2p)}\right\},
\end{align}
where $z_{q}$ is the upper $q$ quantile of the standard normal distribution, {$\alpha_0 \in (0,0.01]$ is a prespecified tail probability}, and the factor 1.1 adjusts for estimation error (any value greater than 1 {suffices}).
{The threshold $z_{\alpha_0/(2p)}$ is Bonferroni-corrected across the $p$ comparisons to exclude extreme tail perturbations from \eqref{eq: sampling of sigma and gamma} and \eqref{eq: sampling of mu Y and mu 1}, and the eigenvalue condition discards $\Sigmahatm$ with a negative eigenvalue since $\lambda_{\min}(\Sigma)\geq 0$.}

{For significance level $\alpha > \alpha_0$ and each $m \in \mathbb{M}$,} we construct the $m$-th interval as
\begin{align}\label{eq: sampled interval}
    \Int^{[m]} = \left[\tauhatm-z_{\alpha'/2}\VhatY^{1/2},\;\tauhatm+z_{\alpha'/2}\VhatY^{1/2}\right],
\end{align}
where $\alpha' = \alpha - \alpha_0$, $\tauhatm$ is defined in \eqref{eq: m-th tauhat}, and $\VhatY$ denotes the variance estimator of $\muhat_Y$. If no feasible solution exists for \eqref{eq: m-th betahat}, we set $\Int^{[m]} = \varnothing.$ {Each $\Int^{[m]}$ quantifies the uncertainty of $\muhat_Y$ by a standard normal interval, treating $(\muhatm)\tr\betahatm$ as fixed.
}

{Since $m^*$ is unknown}, we take the union to construct the CI for $\tau^*$:
\begin{align}\label{eq: aggregated CI}
    {\rm CI}_{\alpha} = \bigcup_{m\in \mathbb{M}}\Int^{[m]},
\end{align}
with $\Int^{[m]}$ in \eqref{eq: sampled interval}.
{We call ${\rm CI}_{\alpha}$ a confidence interval, although the union need not be an interval.}
We summarize our proposal in Algorithm~\ref{alg:inference} in Appendix~\ref{subsec: alg DRoSC Infer}.

{The tuning parameter $\rho_M$ in \eqref{eq: m-th omega} accounts for the error from replacing $\gamma$ and $\Sigma$ with $\gammahatm$ and $\Sigmahatm$. Proposition~\ref{thm: m star} in the Appendix suggests $\rho_M = C_1[\log(\min\{T_0,T_1\})/M]^{1/p}/\sqrt{T_0}$ for a constant $C_1>0$. As when choosing $C$ for $\rho$ in \eqref{eq: tuning from specifying}, $C_1$ is unknown in practice; we initialize it at a small value and increase it by a factor of $1.25$ until at least a prespecified proportion of the $M$ perturbed problems are feasible (10\% by default). Appendix~\ref{subsec: sens to prop} shows the results are robust to this proportion, with 20\% or 30\% giving similar CI coverage and length.}

{
\begin{remark}[Relation to non-regular inference]
Since Wald CIs and the standard bootstrap may fail under non-regularity \citep{andrews2000inconsistency}, several general alternatives have been proposed \citep[e.g.,][]{wasserman2020universal,kuchibhotla2024hulc,guo2024statistical,xie2024repro}. Our perturbation method is inspired by \citet{xie2024repro} and \citet{guo2024statistical}, but differs from both. \citet{xie2024repro} emphasize non-regular problems involving discrete parameters, whereas our target parameter is continuous. The closer comparison is with \citet{guo2024statistical}, who considers a pre-specified constraint set for a strongly convex optimization problem with a unique minimizer. In contrast, our constraint set is itself data-dependent, and our theory covers the more challenging case of a convex problem with potentially multiple minimizers. More broadly, the methods cited above do not investigate non-regular inference in a partial-identification setting, which is the main focus of our proposal.
\end{remark}
}

\section{Theoretical Justification}\label{sec: theoretical justification}
{We now provide theoretical justification for our methods and focus on the regime of growing $T_0$ and $T_1$ but fixed-$N$ throughout the paper.
We present results for $\lambda>0$, deferring the $\lambda=0$ case to Appendix~\ref{subsec: additional theory}.} We introduce assumptions on pre-treatment data and error terms. {We define $W_t=X_tX_t\tr-\E X_tX_t\tr$ for $t=1,\ldots,T_0$.} %

\begin{Assumption}\label{assumption: pre}
For the pre-treatment control units' outcomes and errors $\{X_{t},u_t^{(0)}\}_{t=1}^{T_0}$ {in \eqref{eq: outcome model}}, there exist constants $C_0,b>0$, %
such that as $T_0\to\infty$, 
\begin{align*}
    \prob\left(\sup_{\beta\in\Delta^N}\left\|\frac{1}{T_0}\sum_{t=1}^{T_0}\left[X_tu_t^{(0)}+W_t(\beta^{(0)}-\beta)\right]\right\|_{\infty}\leq \frac{C_0[\log(\max\{T_0,N\})]^{\frac{1+b}{2b}}}{\sqrt{T_0}}\right)\to 1.
\end{align*}
\end{Assumption}

{Assumption~\ref{assumption: pre} controls, uniformly over $\beta\in\Delta^N$, the empirical fluctuations of $X_tu_t^{(0)}$ and of the sample covariance about its expectation; since $\E X_tu_t^{(0)}=0$ and $\E W_t=0$, these terms vanish as $T_0$ grows, so that $\Omegahat$ behaves like $\Omega$.
Under mild conditions, Assumption \ref{assumption: pre} holds for all $b>0$ under i.i.d.\ data with fixed $N$, and for $\beta$-mixing data with exponential decay with $b$ the mixing order \citep[][Lemma H.8]{chernozhukov2021exact}; similar conditions appear in the SC literature \citep[e.g.,][]{chernozhukov2021exact,ben2021augmented}.}

We now move to the assumption on the post-treatment period. {We define $\epsilon_t= (\nu_{t}\tr,v_t,u_t^{(1)})\tr$ for $t=T_0+1,\ldots,T$ where $\nu_t = X_t - \mathbb{E}X_t$, $v_t=Y_{1,t}^{(1)}-Y_{1,t}^{(0)}-\E[Y_{1,t}^{(1)}-Y_{1,t}^{(0)}]$, and $u_t^{(1)}$ is defined in \eqref{eq: outcome model}.}

\begin{Assumption}\label{assumption: post}
     As $T_1\to\infty$, $\mu=T_1^{-1}\sum_{t=T_0+1}^{T}\E X_t$ and $\mu_Y=T_1^{-1}\sum_{t=T_0+1}^{T}\E Y_{1,t}$ are bounded, and the error terms $\{ \epsilon_t\}_{t=T_0+1}^{T}$ satisfy $T_1^{-1/2}\sum_{t=T_0+1}^{T}\epsilon_{t} = O_p(1)$. %
\end{Assumption}

{Assumption~\ref{assumption: post} holds for the i.i.d.\ post-treatment data, and more generally under dependent structures such as strong mixing \citep[e.g.,][Theorem 27.4]{billingsley2017probability}; a similar condition is used in the SC literature \citep[e.g.,][]{li2020statistical}.}

The following theorem establishes the convergence rate of the estimator $\tauhat$ in \eqref{eq: betahat tauhat}. %
\begin{Theorem}\label{thm: conv of tauhat}
  
  Suppose Assumptions~\ref{assumption: pre} and \ref{assumption: post} hold, and the tuning parameter $\rho$ used in \eqref{eq: estimator of omega} satisfies $\rho =C[\log(\max\{T_0,N\})]^{\frac{1+b}{2b}}/{\sqrt{T_0}}$ for some positive constant $C\geq C_0$ with $C_0$ and $b$ in Assumption~\ref{assumption: pre}. Then, for $\lambda>0$, $\tauhat$ in \eqref{eq: betahat tauhat}  satisfies the following:  
      \begin{align}
      {|\tauhat-\tau^*|=O_p\left(\frac{[\log (\max\{T_0,N\})]^{\frac{1+b}{2b}}}{\sqrt{T_0}\cdot \lambda} + \frac{1}{\sqrt{T_1}}\right).}\label{eq: conv rate of tau hat}
    \end{align}
\end{Theorem}

The convergence rate \eqref{eq: conv rate of tau hat} of $|\tauhat-\tau^*|$ in Theorem~\ref{thm: conv of tauhat} has two components: the first term captures the error from estimating $\Omega$ by $\Omegahat$ and the second term reflects the error in estimating $\tau(\beta)$ by $\muhat_Y-\muhat\tr\beta$.
The rate also applies as $\lambda\to 0$, but consistency of $\tauhat$ requires $\lambda$ large enough so that $[\log(\max\{T_0,N\})]^{\frac{1+b}{2b}}/(\sqrt{T_0}\cdot\lambda)\to 0$ as $T_0,T_1\to\infty$.
{ For fixed $\lambda>0$, the rate attains the parametric rate
$1/\sqrt{\min\{T_0,T_1\}}$, up to a logarithmic factor, despite the degeneracy noted after Theorem~\ref{thm: id of DRoSC}.
 While $\tau^*$ is identified via the optimizer of the degenerate quadratic program in \eqref{eq: tau star and beta star}, Theorem~\ref{thm: tau star} characterizes it as the projection of zero onto the sensitivity interval, whose endpoints are the optimal values of two linear programs. {The errors in these optimal values, and hence in their projection, can be
bounded directly in terms of the errors in estimating the linear objective function and
uncertainty class, despite the degeneracy of the quadratic program and the
possible non-uniqueness of its optimizer.}
}

Finally, we theoretically justify our perturbation-based inference method. We impose the following assumption on the limiting distributions of $\vecl(\Sigmahat)$, $\gammahat$, $\muhat_Y$, and $\muhat$, as well as on the consistency of the corresponding covariance estimators.

\begin{Assumption}\label{assumption: clt and cons}
 $\vecl(\Sigmahat)$, $\gammahat$, $\muhat_Y$, and $\muhat$ admit the following asymptotic distributions: $\VSigma^{-1/2}\left(\vecl(\Sigmahat)-\vecl(\Sigma)\right)\dto\mathcal{N}(0,\mathbf{I})$, $\Vgamma^{-1/2}\left(\gammahat-\gamma\right)\dto\mathcal{N}(0,\mathbf{I})$ as $T_0\to\infty$, and $\VY^{-1/2}\left(\muhat_Y-\mu_Y\right)\dto\mathcal{N}(0,1)$, $\Vmu^{-1/2}\left(\muhat-\mu\right)\dto\mathcal{N}(0,\mathbf{I})$ as $T_1\to\infty$
for some positive scalar $\VY$ and positive definite matrices $\VSigma$, $\Vgamma$, and $\Vmu$, where $T_1\VY$ and the diagonal elements of $T_0\VSigma$, $T_0\Vgamma$, and $T_1\Vmu$ are uniformly bounded above and away from zero. 
The covariance estimators $\VhatSigma$, $\Vhatgamma$, $\VhatY$, and $\Vhatmu$ used in \eqref{eq: sampling of sigma and gamma} and \eqref{eq: sampling of mu Y and mu 1} are positive semidefinite and consistent: $\|T_0(\VhatSigma-\VSigma)\|_2\pto0$ and $ \|T_0(\Vhatgamma-\Vgamma)\|_2\pto0$ as $T_0\to\infty$, and $|T_1(\VhatY-\VY)|\pto0$ and $\|T_1(\Vhatmu-\Vmu)\|_2\pto0$ as $T_1\to\infty$.
\end{Assumption}

Assumption~\ref{assumption: clt and cons}
holds under i.i.d.\ data with fixed $N$, and more generally under $\alpha$-mixing stationarity \citep[e.g.,][Theorem 27.4]{billingsley2017probability}, provided the covariance estimators match the data regime: sample covariances under i.i.d.\ sampling, and HAC estimators under weak dependence \citep{newey1987simple,andrews1991heteroskedasticity}; {concrete formulas for the covariance estimators are in \eqref{eq: Vhat} and \eqref{eq: Vhat HAC} of Appendix~\ref{subsec: choice cov est}.}

The following theorem establishes the coverage property of the proposed CI.
\begin{Theorem}\label{thm: coverage}
Suppose Assumptions~\ref{assumption: post} and \ref{assumption: clt and cons} hold, %
and the tuning parameter $\rho_M$ used in \eqref{eq: m-th omega}, satisfies $\rho_M = C_1[\log(\min\{T_0,T_1\})/M]^{1/p}/\sqrt{T_0}$ for some positive constant $C_1\geq (2/c^*(\alpha_0))^{1/p}$ with the constant $c^*(\alpha_0)$ used in \eqref{def: err} in Appendix~\ref{subsec: additional theory} and $\alpha_0\in(0,0.01]$. Then, for $\alpha\in(\alpha_0,1)$ and $\lambda>0$, ${\rm CI}_{\alpha}$ in \eqref{eq: aggregated CI} satisfies  $\liminf_{T_0,T_1\rightarrow\infty}\liminf_{M\to\infty}\prob(\tau^* \in {\rm CI}_{\alpha})\geq 1-\alpha.$
\end{Theorem}
Theorem~\ref{thm: coverage} establishes a lower bound on the coverage and the CI may be conservative. We investigate this by studying the length of a refined version of the proposed CI.
{
We refine the definition of $\mathbb{M}$ in \eqref{eq: index set} as
\begin{align}
    \tilde{\mathbb{M}}
    = \mathbb{M}\cap\left\{1\leq m\leq M:\ \text{$\exists$  $\beta\in\Delta^N$ such that }\left\|\gammahatm-\Sigmahatm\beta\right\|_{\infty}\leq\rho_M\right\}.
    \label{eq: add filter non-empty}
\end{align}
We show in Proposition~\ref{thm: m star} in the Appendix that with high probability, there exists an index $m^*\in\mathbb{M}$ such that  $\gammahatmstar$ and $\Sigmahatmstar$ nearly recover $\gamma$ and $\Sigma$.
Since $\gamma-\Sigma\beta^{(0)}=0$, the choice of $\rho_M$ ensures that $\beta=\beta^{(0)}$ satisfies $\|\gammahatmstar-\Sigmahatmstar\beta\|_{\infty}\leq\rho_M$. Hence $m^*\in\tilde{\mathbb{M}}$, and replacing $\mathbb{M}$ with $\tilde{\mathbb{M}}$ in \eqref{eq: aggregated CI} preserves the coverage guarantee in Theorem~\ref{thm: coverage}.} The following theorem establishes the length of this refined CI.

\begin{Theorem}\label{thm: precision}
Under the conditions of Theorems~\ref{thm: conv of tauhat} and
\ref{thm: coverage}, construct ${\rm CI}_{\alpha}$ in
\eqref{eq: aggregated CI} using $\tilde{\mathbb{M}}$. Then, for
$\lambda>0$ and $b>0$ defined in Assumption~\ref{assumption: pre}, we have 
\begin{align*}
\liminf_{T_0,T_1\to\infty}\liminf_{M\to\infty}
\mathbb{P}\left\{
\operatorname{Length}({\rm CI}_{\alpha})\lesssim
{\frac{[\log (\max\{T_0,N\})]^{\frac{1+b}{2b}}}{\sqrt{T_0}\cdot\lambda}
+\frac{1}{\sqrt{T_1}}}
\right\}\geq 1-\alpha_0.
\end{align*}
\end{Theorem}

The length of this refined CI contracts at the same rate as the estimator $\tauhat$ in
Theorem~\ref{thm: conv of tauhat}. Regarding the finite-sample performance, we observe that its length is comparable to, although longer than, the oracle benchmark; see Figure~\ref{fig:infer pre25 post25}.

\section{Simulation Studies}\label{sec: simulation studies}

{We assess the finite-sample performance of our inference method via numerical studies; estimation results are reported in Appendix~\ref{subsec: sim est}. We generate the outcomes as follows.} Control units' outcomes $X_t$ are generated via a stationary AR(1) model with coefficient $\phi \in [0,1)$. Pre-treatment outcomes $\{X_t\}_{t=1}^{T_0}$ have mean $\mu_0=(0.8,1.2,\ldots,0.8,1.2)\tr$ and equi-correlation covariance $\Sigma_0=(1 - \rho_0)\mathbf{I}_N + \rho_0 \mathbf{1}_N \mathbf{1}_N\tr$, while post-treatment outcomes $\{X_t\}_{t = T_0 + 1}^{T}$ have mean $\mu$ and covariance $\mathbf{I}_N$; see Appendix~\ref{subsec: dgp} for details. The treated unit's potential outcomes follow \eqref{eq: outcome model} with $\beta^{(0)} = (1/3\cdot \mathbf{1}_3\tr, \mathbf{0}_{N-3}\tr)\tr$ and i.i.d.\ errors $u_t^{(0)}, u_t^{(1)} \sim \mathcal{N}(0, 1)$. We set $Y_{1,t}^{(1)} - Y_{1,t}^{(0)} = \tau + v_t$ with $v_t \overset{i.i.d.}{\sim} \mathcal{N}(0, 0.25^2)$, thus $\taubar = \tau$.

We consider three settings by varying $\mu$, $\rho_0$, and $\beta^{(1)}$. {We present (S2), where high correlations challenge (E1) and a small weight shift violates (E2):}
\begin{enumerate}
\item [(S2)] $\rho_0 = 0.95$, $\mu = \mu_0 + (0.6, 0.4, 0.2, \mathbf{0}_{N-3})\tr$, and $\beta^{(1)} = \beta^{(0)} + 0.05 \cdot (-1, \mathbf{0}_{N-2}\tr, 1)\tr$. 
\end{enumerate}
The additional two settings (S1) and (S3) are detailed in Appendix~\ref{subsec: dgp}.

We vary $\taubar \in \{-1.5,-1.4,\ldots,1.5\}$, $(T_0, T_1) \in \{25, 50\}\times\{25, 50\}$, and $\phi \in \{0, 0.5\}$, with fixed $N = 10$.
 In the main paper, we present results with $\phi = 0$ (i.i.d.\ data) and $T_0 = T_1 = 25$. 
 Additional results are provided in Appendix~\ref{sec: additional simulation} and are qualitatively similar to those in the main paper.

We evaluate the 95\% CIs produced by the perturbation-based method over 500 simulations. For cases sharing the same $\tau^*$ (e.g., $\tau^*=0$ when $\taubar \in\{-0.4,\ldots,0.1\}$), we report the minimum coverage and maximum CI length of each method. %
Our procedure {uses} $M = 500$, and we denote our proposed CI in \eqref{eq: aggregated CI} as {Proposed}.

We compare the {Proposed} CI with two oracle normal-approximation benchmarks for $\tauhat$ in \eqref{eq: betahat tauhat}. The first, denoted Normality, ignores bias and takes {$[\tauhat \pm z_{\alpha/2}\widehat{\SE}(\tauhat)]$, where $\widehat{\SE}(\tauhat)$} is the empirical standard error of $\tauhat$ across the 500 simulation replicates. The second is the oracle bias-aware (OBA) CI of \citet{armstrong2023bias}, {$[\tauhat \pm \chi^*_{\alpha}]$ with $\chi^*_{\alpha} = \widehat{\SE}(\tauhat)\,[\mathrm{cv}_{\alpha}(|\widehat{\E}\tauhat - \tau^*|^2/\widehat{\SE}(\tauhat)^2)]^{1/2}$ and $\mathrm{cv}_{\alpha}(B^2)$ the $1-\alpha$ quantile of a non-central $\chi^2(1)$ with non-centrality $B^2$;} it uses oracle knowledge of the bias and standard error of $\tauhat$ and is therefore infeasible, but represents the best achievable normal-approximation benchmark.

\begin{figure}[htp!]
    \centering
    \includegraphics[width=.75\textwidth]{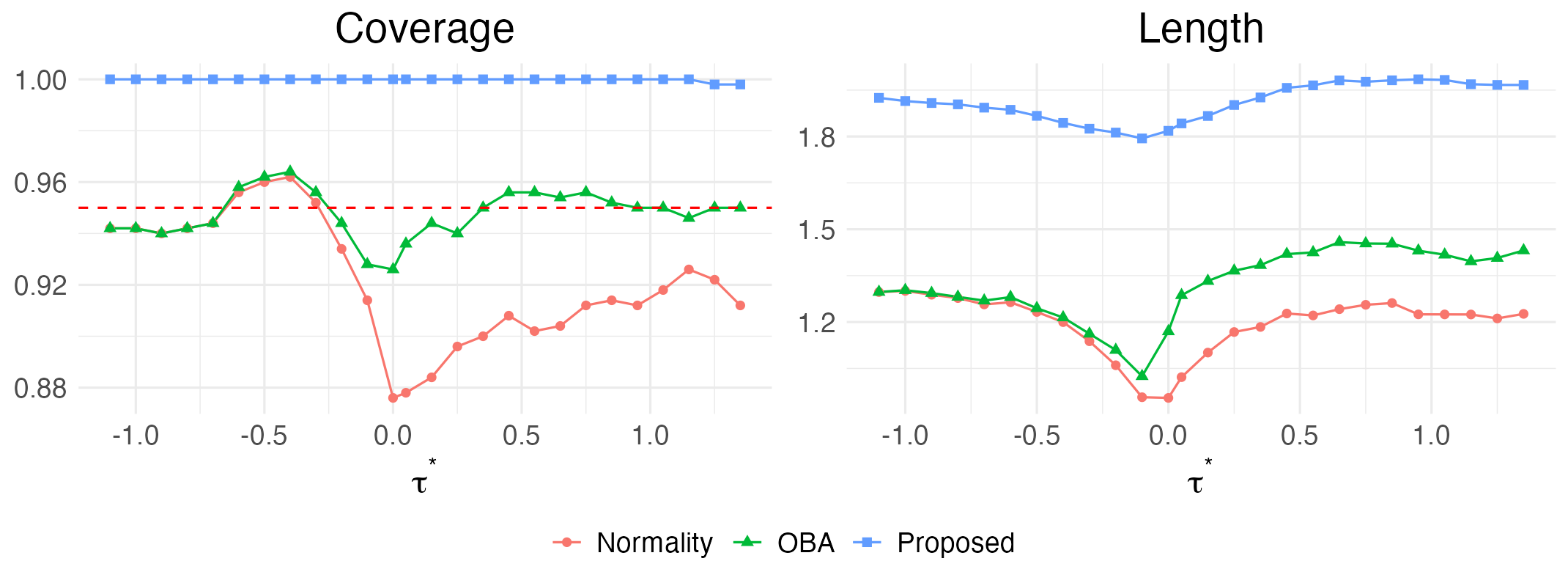}
    \caption{Empirical coverages and interval lengths for setting (S2). %
    $x$-axis plots the values of $\tau^*$. \texttt{Normality} refers to the oracle normality-based CI, \texttt{OBA} denotes the oracle bias-aware CI, and {\texttt{Proposed}} refers to the perturbation-based CI in \eqref{eq: aggregated CI}.}
    \label{fig:infer pre25 post25}
\end{figure}

We present inference results with empirical coverages and mean lengths of CIs in Figure~\ref{fig:infer pre25 post25}.
The Normality CI's coverage is near 0.95 when $\tau^*<0$ but drops to $0.9$ for some positive $\tau^*$ and below $0.9$ near zero, reflecting the non-regularity discussed in Section~\ref{subsec:inference challenges}. However, the OBA and {Proposed} CIs improve coverage substantially; the Proposed CI {attains at least nominal coverage across the configurations considered} and is conservative but comparable in length to the OBA.

\section{Real Data Application}\label{sec: real data}

We reanalyze the Basque Country case study of \citet{abadie2003economic}, which examines the economic impact of terrorism on per capita GDP. The Basque Country, affected by terrorism, is treated, and the remaining $N=16$ Spanish regions are controls, observed over 1955–1997 ($T=43$), with pre-treatment period 1955–1969 ($T_0=15$) and post-treatment period 1970–1997 ($T_1=28$). {While the original analysis matched on both pre-treatment outcomes and additional covariates, we use pre-treatment outcomes only.}

As discussed in Section~\ref{subsec: challenges}, highly correlated controls and weight shifts threaten SC's stability and causal conclusions. We address these challenges by estimating $\tau^*$ and constructing its 95\% CIs using $M=500$. {We vary $\lambda\in\{0,0.001,\ldots,0.06\}$ for the point estimate and report the CIs at $\lambda\in\{0,0.015,\ldots,0.06\}$;} for comparison, we also report the outcome-only SC point estimate \eqref{eq: SC estimators}.

\begin{figure}[htp!]
\centering
\includegraphics[width=.8\textwidth]{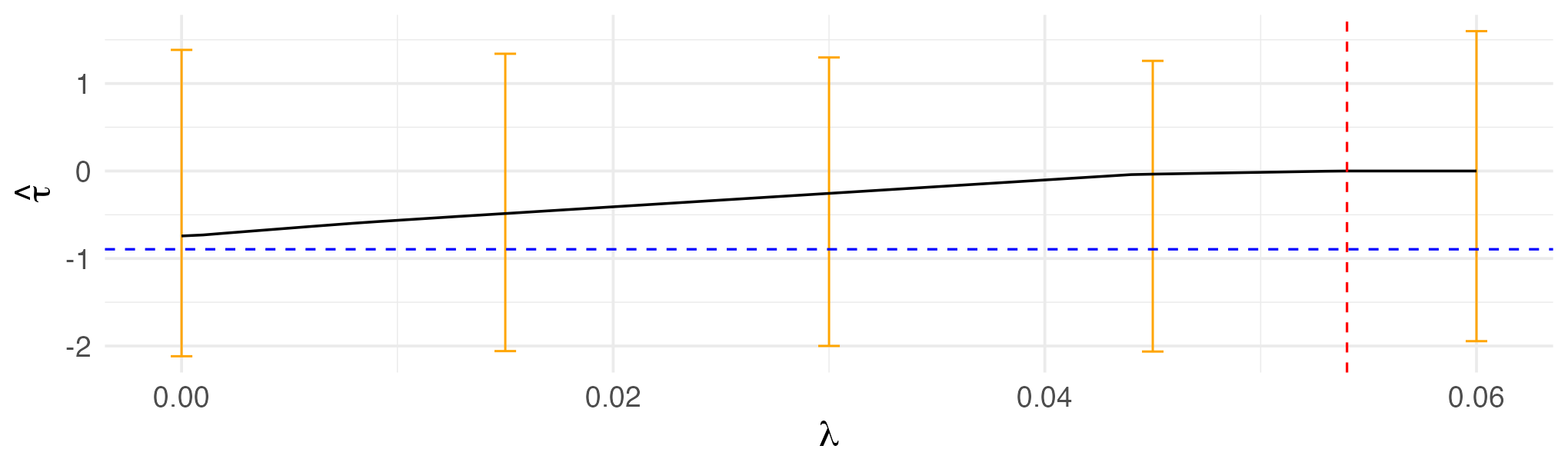}
\caption{Reanalysis of the Basque study. The black line shows $\tauhat$ in \eqref{eq: betahat tauhat} and the orange intervals are 95\% CIs in \eqref{eq: aggregated CI} for $\lambda\in\{0,0.015,\dots,0.06\}$; the blue dashed line shows $\tauhatSC$, and the red dashed line marks 0.054, where $\tauhat$ first reaches 0.
}
\label{fig: basque lambda CI}
\end{figure}

Figure~\ref{fig: basque lambda CI} presents the results. At $\lambda=0$, our estimate is $\tauhat\approx -0.74$, compared to the SC estimate $\tauhatSC\approx -0.89$, yielding a more conservative effect even without weight shifts. As $\lambda$ increases, $\tauhat$ moves toward zero, reaching and remaining at zero for $\lambda\geq 0.054$. 
 {Thus, the difference from SC at $\lambda=0$ reflects the method’s adjustment for weight instability, whereas the movement toward zero occurs because larger values of $\lambda$
admit a wider range of plausible post-treatment weights. 
{At every reported value of $\lambda$, the CI contains zero; hence, by Theorem~\ref{thm: tau star}, we do not reject that $0 \in I\{\Omega(\lambda)\}$.} The relatively wide CIs may reflect the short time periods
and the high correlations (Figure~\ref{fig: corr of basque}), which inflate the estimated covariance matrices} used in \eqref{eq: sampling of sigma and gamma} and \eqref{eq: sampling of mu Y and mu 1}, yielding more dispersed perturbations and wider aggregated CIs in \eqref{eq: aggregated CI}.

\section{Conclusion and Discussion}\label{sec: conclusion}
{The SC method is widely used for estimating causal effects, but its validity relies on key identification conditions. To remain informative when these conditions fail, we introduce a DRO-based causal estimand $\tau^*$ that equals the time-averaged ATT $\taubar$ when $\taubar$ is identifiable, and otherwise serves as a conservative proxy for it, namely the least-favourable point of the sensitivity interval, bridging sensitivity analysis and DRO in the SC setting. To address the possibly non-standard limiting distribution of the WRoTE estimator, we propose a novel perturbation-based method for valid inference. Beyond SC, the perturbation method is of independent interest for handling non-regular inference in other partial identification inference problems.}

Several avenues for future research remain. One direction is to extend the framework to incorporate covariates, {which can improve counterfactual construction \citep{abadie2026}, 
by enlarging the uncertainty class so that weights balance both outcomes and covariates.}
Another direction is to extend the framework to staggered adoption settings, where treatment timing varies across units \citep[e.g.,][]{athey2022design, ben2022synthetic,cattaneo2025uncertainty}; {the same identification challenges persist, and the DRO target could be generalized by defining a reward function that averages across adoption times.}

\section*{Data Availability Statement}
The code for reproducing the simulation studies and empirical analysis is available at 
\href{https://github.com/taehyeonkoo/DRoSC}{https://github.com/taehyeonkoo/DRoSC}. The Basque Country data used in
the empirical analysis are freely available as the \texttt{basque} dataset
distributed with the R package \texttt{Synth} \citep{abadie2011synth}.

\section*{Acknowledgments}
The authors thank Mengchu Zheng for producing Figures~\ref{fig: SC E1E2}, \ref{fig: SC weight shift}, and \ref{fig: workflow}.

\bibliographystyle{apalike}
\bibliography{ref}

\newpage
\appendix

\renewcommand{\thefigure}{S\arabic{figure}}
\setcounter{figure}{0}
\renewcommand{\thetable}{S\arabic{table}}
\setcounter{table}{0}
\setcounter{page}{1}

\newpage
\begin{center}
{\bf \Large Appendix for ``Synthetic Control with Weight Uncertainty: Robust Identification and Statistical Inference''}
\end{center}
These supplements are organized as follows:

\begin{enumerate}
    \item Section~\ref{sec: alg} provides algorithms that summarize the estimation and perturbation-based inference procedures.

    \item Section~\ref{sec: add meth theo} mainly introduces additional methods and theoretical results omitted from the main paper.

    \item Section~\ref{sec: pf thm prop} contains the proofs of the results in the main paper, and Section~\ref{sec: pf other main} includes the proofs of the other main results not proved in  Section~\ref{sec: pf thm prop}. Section~\ref{sec: pf lem} presents the proofs of lemmas used in Sections~\ref{sec: pf thm prop} and \ref{sec: pf other main}. 

    \item Section~\ref{sec: semi-real simulation} details the implementation of the semi-real data analysis based on the Basque study introduced in Section~\ref{subsec: challenges} of the main paper.

    \item Section~\ref{sec: additional simulation} presents additional simulation results.
\end{enumerate}

\section{Algorithms}\label{sec: alg}
In this section, we provide algorithms for the estimation procedure of the weight-robust treatment effect (WRoTE) and the perturbation-based inference procedure described in Sections~\ref{sec: estimation of DRoSC} and \ref{subsec:perturbation} of the main paper. Section~\ref{subsec: alg DRoSC} presents the estimation procedure, and Section~\ref{subsec: alg DRoSC Infer} presents perturbation-based inference.
\subsection{Estimation procedure}\label{subsec: alg DRoSC}
Here, we present the estimation procedure described in Section~\ref{sec: estimation of DRoSC}.
\begin{algorithm}[H]   
\caption{Estimation procedure}\label{alg:DRoSC}
    \begin{algorithmic}[1]
        \Require{Data $\{Y_{1,t},X_t\}_{t=1}^{T}$;  Weight shift parameter $\lambda \geq 0$; Tuning parameter $\rho \geq 0$.}
        \Ensure{ Point estimator $\betahat$ of $\beta^*$; Point estimator $\widehat{\tau}$ of $\tau^*$. }
        \State Construct the uncertainty class $\Omegahat$ as in \eqref{eq: estimator of omega}; 
        \State Construct $
        \muhat_{Y}=T_1^{-1}\sum_{t=T_0+1}^{T}Y_{1,t}$ and $\muhat=T_1^{-1}\sum_{t=T_0+1}^{T}X_{t}$;
        \State Construct $\betahat$ as in \eqref{eq: betahat tauhat} and $\widehat{\tau}$ as in \eqref{eq: betahat tauhat}. \Comment{WRoTE estimator}
    \end{algorithmic}
\end{algorithm}

\subsection{Perturbation-based inference}\label{subsec: alg DRoSC Infer}
Here, we present the perturbation-based inference procedure described in Section~\ref{subsec:perturbation}.
\begin{algorithm}[H]   
\caption{Perturbation-based inference}\label{alg:inference}
    \begin{algorithmic}[1]
        \Require{Pre-treatment data $\{Y_{1,t},X_t\}_{t=1}^{T_0}$; Post-treatment data $\{Y_{1,t},X_t\}_{t=T_0+1}^{T}$; Weight shift parameter $\lambda\geq 0$; Sampling size $M=500$; Significance level $\alpha \in (0,1)$; Pre-specified constant $\alpha_0\in(0,0.01]$; Tuning parameter $\rho_M\geq0$.}
        \Ensure{Confidence interval $ {\rm CI}_{\alpha}$. }
        \State Construct $\widehat{\Sigma}$ and $\widehat{\gamma}$ as in \eqref{eq: estimator of omega}, and $ \muhat_{Y}=T_1^{-1}\sum_{t=T_0+1}^{T}Y_{1,t}$ and $\muhat=T_1^{-1}\sum_{t=T_0+1}^{T}X_{t}$;
        \For{$m \gets 1,...,M$}
        \State Sample $\widehat{\Sigma}^{[m]}$ and $\widehat{\gamma}^{[m]}$ as in \eqref{eq: sampling of sigma and gamma} with $\VhatSigma$, $\Vhatgamma$ as in \eqref{eq: Vhat};
        \State  Sample $\muhat_Y^{[m]}$ and $\muhat^{[m]}$ as in \eqref{eq: sampling of mu Y and mu 1} with $\widehat{{\rm V}}_{Y}$ and $\Vhatmu$ as in \eqref{eq: Vhat};
        \State Construct $\Omegahatm$ as in \eqref{eq: m-th omega};
        \State Construct $\betahatm$ as in \eqref{eq: m-th betahat} and $\tauhatm$ as in \eqref{eq: m-th tauhat};
        \State Construct $\text{Int}^{[m]}$ as in \eqref{eq: sampled interval};
        \EndFor \Comment{Perturbation}
        \State Construct the index set $\mathbb{M}$ as in \eqref{eq: index set}; \Comment{Filtering}
        \State Construct ${\rm CI}_{\alpha}$ as in \eqref{eq: aggregated CI}; 
        \Comment{Aggregation}
    \end{algorithmic}
\end{algorithm}

\section{Additional Methods and Theories}\label{sec: add meth theo} 
In this section, we provide additional methods and theoretical results omitted from the main text. 
Section~\ref{subsec: non iid} introduces and discusses tuning parameter selection procedure for estimation with non-i.i.d.\;data. Section~\ref{subsec: choice cov est} discusses the choice of the covariance estimators used in the perturbation generating distributions \eqref{eq: sampling of sigma and gamma} and \eqref{eq: sampling of mu Y and mu 1}. Lastly, Section~\ref{subsec: additional theory} provides theoretical results omitted from the main paper.

\subsection{Tuning Parameter Selection for Estimation with Non-i.i.d. 
Data}\label{subsec: non iid}
In this section, we discuss selecting the tuning parameter $\rho$, which is used to define $\Omegahat$ in \eqref{eq: estimator of omega}, in the context of non-i.i.d.\;data.

While we specify $\rho$ to be of order $[\log(\max\{T_0, N\})]^{1/2} / \sqrt{T_0}$ in \eqref{eq: tuning from specifying} with i.i.d.\;data, our theoretical analysis (Assumption~\ref{assumption: pre}) establishes that temporal dependence requires a more general form: $\rho = C[\log(\max\{T_0, N\})]^a / \sqrt{T_0}$, for some positive constant $C>0$. Here, the exponent $a \geq 1/2$ reflects the temporal dependence structure of the pre-treatment data, where larger values of $a$ accommodate stronger temporal dependence.  Thus, while the default specification of $a = 1/2$ suffices when the pre-treatment data are i.i.d., theory suggests that $a > 1/2$ is necessary to ensure valid estimation in the presence of serial dependence.

In practice, however, we select the constant $C$ in \eqref{eq: tuning from specifying} using a data-dependent procedure: we initialize $C$ at a small value (e.g., $C=0.01$) and incrementally increase it by a factor of 1.25 until a feasible solution to \eqref{eq: betahat tauhat} is obtained. Because this procedure adapts the magnitude of $\rho$ to ensure feasibility, the specific choice of the exponent $a$ does not strongly affect the final value of $\rho$. Large-scale numerical studies confirm that the estimator remains reliable across varying choices of $a$; see Section~\ref{subsec: sens rho}.

\subsection{Choice of the Covariance Estimators for Inference}\label{subsec: choice cov est}
In this section, we discuss the choice of the covariance estimators used in the perturbation generating distribution.

We remark that the choice of the covariance estimator $\VhatSigma$, $\Vhatgamma$, $\VhatY$, and $\Vhatmu$ used in the perturbation procedure \eqref{eq: sampling of sigma and gamma} and \eqref{eq: sampling of mu Y and mu 1} depends on the underlying data-generating mechanism.  For i.i.d.\ pre-treatment and post-treatment data with fixed $N$, we use the following estimators:
\begin{equation}\label{eq: Vhat}
    \begin{aligned}
        &\VhatSigma = \frac{1}{T_0(T_0-1)}\sum_{t=1}^{T_0}\left(\vecl(X_tX_t\tr)-\vecl(\widehat{\Sigma})\right)\left(\vecl(X_tX_t\tr)-\vecl(\widehat{\Sigma})\right)\tr, \\
    &\Vhatgamma=\frac{1}{T_0(T_0-1)}\sum_{t=1}^{T_0}\left(X_tY_{1,t}-\widehat{\gamma}\right)\left(X_tY_{1,t}-\widehat{\gamma}\right)\tr,\\
    &\VhatY = \frac{1}{T_1(T_1-1)}\sum_{t=T_0+1}^T\left(Y_{1,t}-\muhat_Y\right)^2,\quad
    \Vhatmu = \frac{1}{T_1(T_1-1)}\sum_{t=T_0+1}^T\left(X_t-\muhat\right)\left(X_t-\muhat\right)\tr. 
    \end{aligned}
\end{equation}

For weakly dependent and stationary processes,  we instead use the Heteroskedasticity and Autocorrelation Consistent (HAC) covariance estimators \citep{newey1987simple,andrews1991heteroskedasticity} for consistency. For example, we define the HAC covariance estimators for $\VSigma$ as
\begin{equation}\label{eq: Vhat HAC}
    \begin{aligned}
        &\VhatSigma^{\rm HAC} = \frac{1}{T_0}\left[\mathbf{\Gammahat}(0)+\sum_{l=1}^{L_0}w_l(\mathbf{\Gammahat}(l)+\mathbf{\Gammahat}(l)\tr)\right],\quad\text{with}\\
        & \mathbf{\Gammahat}(l)= \frac{1}{T_0}\sum_{t=l+1}^{T_0}\left(\vecl(X_tX_t\tr)-\vecl(\widehat{\Sigma})\right)\left(\vecl(X_{t-l}X_{t-l}\tr)-\vecl(\widehat{\Sigma})\right)\tr,
    \end{aligned}
\end{equation}
where $w_l=w(l,L_0)$ are kernel weights and $L_0$ is the bandwidth \citep[see, e.g.,][for choices of weights and bandwidth]{newey1987simple}. HAC covariance estimators for $\Vgamma$, $\VY$, and $\Vmu$ are defined analogously.

In more general settings, however, obtaining consistent covariance estimators can be difficult without knowledge of the data's dependence structure. While Assumption~\ref{assumption: clt and cons} indicates that consistent covariance estimators are sufficient to justify our method, our simulation studies in Section~\ref{subsec:inconsistency} demonstrate that even inconsistent estimators (specifically, using \eqref{eq: Vhat} under an AR(1) process) can yield valid perturbation inference, provided that a larger number of perturbations $M$ is used and the covariance estimator employed in the perturbation-generating distribution is sufficiently large to ensure near recovery of the true quantities.

\subsection{Additional Theoretical Results}\label{subsec: additional theory}

In this section, we present the theoretical results that were omitted from the main paper. The following proposition quantifies the difference between $\tau^*$ and $\taubar$.
\begin{proposition}\label{prop: magnitude of lower bound}
  If $\beta^{(1)}\in\Omega$ and $\lambda_{\min}(\Sigma)>0$, then $\left|\tau^*-\taubar\right| \leq 2[\lambda_{\min}(\Sigma)]^{-1}\|\mu\|_1\sqrt{N}\lambda.$
\end{proposition}

When (E1) and (E2) hold, we can take $\lambda = 0$, yielding $\tau^* = \taubar$. More generally, $\tau^*$ recovers $\taubar$ closely when $\lambda \to 0$ sufficiently fast such that the upper bound of $|\tau^*-\taubar|$ in Proposition~\ref{prop: magnitude of lower bound} converges to zero. For example, if $\{X_t\}_{t=T_0+1}^{T}$ is stationary (so $\|\mu\|_1$ is bounded) and $\lambda_{\min}(\Sigma)$ is bounded away from zero with fixed $N$, then $\tau^*\to\taubar$ as $\lambda \to 0$. 

Next, we provide theoretical results for our proposed estimator and inference procedures including $\lambda=0$ case. Throughout this section, we additionally assume that $\lambda_{\min}(\Sigma)>0$ for $\lambda = 0$. Consequently, the uncertainty class $\Omega$ defined in \eqref{def: Omega} reduces to a singleton containing the true pre-treatment weight, $\{\beta^{(0)}\}$. Provided that $\beta^{(1)}\in\Omega(0)$, the key identification conditions (E1) and (E2) (introduced in Section~\ref{subsec: SC} of the main paper) are satisfied; thus, $\tau^*=\taubar$. 

First, we present the theorem about the convergence rate of the estimator $\tauhat$ for $\lambda=0$, whose proof is provided in Section~\ref{pf: conv of tauhat lambda0}. 
\begin{Theorem}\label{thm: conv of tauhat lambda0}
  Under the same assumptions in Theorem~\ref{thm: conv of tauhat}, together with the additional assumption that $\lambda_{\min}(\Sigma)>0$, \eqref{eq: betahat tauhat} for $\lambda=0$ satisfies the following:  
      \begin{align}
      {|\tauhat-\tau^*|=O_p\left(\frac{[\log (\max\{T_0,N\})]^{\frac{1+b}{2b}}}{\sqrt{T_0}\cdot \lambda_{\min}(\Sigma)/\sqrt{N}} + \frac{1}{\sqrt{T_1}}\right),}\label{eq: conv rate of tau hat lambda0}
    \end{align}
    where $b>0$ is defined in Assumption~\ref{assumption: pre}.
\end{Theorem}
For $\lambda=0$, the additional condition $\lambda_{\min}(\Sigma)>0$ in Theorem~\ref{thm: conv of tauhat lambda0} is required to ensure that $\Omega$ contains a unique weight; this uniqueness is necessary for the convergence of $\Omegahat$ to $\Omega$ when $\lambda = 0$. 
Thus, the theorem precludes non-unique SC weights when no weight shift is assumed by $\lambda = 0$. 
Theorem~\ref{thm: conv of tauhat lambda0} also covers the limiting case where $\lambda_{\min}(\Sigma) \to 0$ holds, but it requires $\lambda_{\min}(\Sigma)$ to satisfy $[\log(\max\{T_0,N\})]^{\frac{1+b}{2b}}/({\sqrt{T_0}}\cdot\lambda_{\min}(\Sigma)/\sqrt{N}) \to 0$, similar to the case of $\lambda>0$.

Next, we present the results of our inference procedure. We introduce the function ${\rm err}(\cdot)$ to quantify the effect of the perturbation step,
\begin{align}\label{def: err}
    \err = \frac{1}{2}\left[\frac{2\log(\min\{T_0,T_1\})}{c^*(\alpha_0)\cdot M}\right]^{1/p},
\end{align}
where $\alpha_0 \in (0,0.01]$ is a pre-specified constant used to construct $\mathbb{M}$ in \eqref{eq: index set}, and {$c^*(\alpha_0)$ is the fixed positive constant defined in \eqref{eq: c star alpha 0}.} Thus, $\err$ is of order $[\log(\min\{T_0,T_1\})/M]^{1/p}$ and vanishes as $M$ goes to infinity.
We define the following events to control $\min_{m\in\mathbb{M}}|(\muhatm)\tr\betahatm-\mu\tr\beta^*|$: %
\begin{equation}\label{event: negligible}
    \begin{aligned}
        \text{For $\lambda>0$, }&\mathcal{E}^* =\left\{\min_{m\in\mathbb{M}}\left|(\muhatm)\tr\betahatm-\mu\tr\beta^*\right|\lesssim \left[\frac{\err}{\min\{\lambda\sqrt{T_0},\sqrt{T_1}\}}\right]^{\frac{1}{2}}\right\},\\
             \text{For $\lambda=0$, }&\mathcal{E}^* =\left\{\min_{m\in\mathbb{M}}\left|(\muhatm)\tr\betahatm-\mu\tr\beta^*\right|\lesssim \left[\frac{\err}{\min\{\lambda_{\min}(\Sigma)\sqrt{T_0/N},\sqrt{T_1}\}}\right]^{\frac{1}{2}}\right\}.
    \end{aligned}
\end{equation}
With the above event, we introduce the following theorem, which establishes critical perturbation property and the theoretical foundation of our proposed CI.

\begin{Theorem}\label{thm: negligible} 
Suppose Assumptions~\ref{assumption: post} and \ref{assumption: clt and cons} hold, and the tuning parameter $\rho_M$ used in \eqref{eq: m-th omega} satisfies $\rho_M = C_1[\log(\min\{T_0,T_1\})/M]^{1/p}/\sqrt{T_0}$ for some positive constant $C_1\geq (2/c^*(\alpha_0))^{1/p}$, {where $c^*(\alpha_0)$ is defined in \eqref{eq: c star alpha 0} and $\alpha_0\in(0,0.01]$ is the pre-specified constant used to define $\mathbb{M}$ in \eqref{eq: index set}.} Then, the following holds:
\begin{enumerate}
    \item[(i)] Case of $\lambda > 0 $: For $\mathcal{E}^*$ defined in \eqref{event: negligible}, 
    \begin{align}
\liminf_{T_0,T_1\rightarrow\infty}\liminf_{M\rightarrow\infty}\prob\left(\mathcal{E}^*\right)\geq 1-\alpha_0.\label{eq: E star prob}
\end{align}
\item[(ii)] Case of $\lambda = 0$: Under the additional assumption of $\lambda_{\min}(\Sigma)>0$, \eqref{eq: E star prob} still holds.
\end{enumerate}
\end{Theorem}
The proof of Theorem~\ref{thm: negligible} is provided in Section~\ref{pf: negligible}.

Theorem~\ref{thm: negligible} guarantees that the event $\mathcal{E}^*$ in \eqref{event: negligible} occurs with high probability when the number of perturbations $M$ is sufficiently large. This result formalizes the intuition behind our perturbation-based inference method: for some $m = m^*$, it ensures that the quantity $(\muhatm)\tr \betahatm - \mu\tr \beta^*$ whose uncertainty exhibits non-regular behavior can be rendered negligible. 
The pre-specified constant $\alpha_0 \in (0, 0.01]$, used to define the index set $\mathbb{M}$ in \eqref{eq: index set}, represents a small probability reserved for excluding the possibility that resampled quantities from \eqref{eq: sampling of sigma and gamma} and \eqref{eq: sampling of mu Y and mu 1} lie in the extreme tails of the generating distributions. We note that Theorem~\ref{thm: negligible} ensures only the existence of such an index; it does not identify which specific index satisfies \eqref{event: negligible}.

Using Theorem~\ref{thm: negligible}, Theorem~\ref{thm: coverage} and the following theorem establish the coverage properties for $\lambda>0$ and $\lambda=0$, respectively: 
\begin{Theorem}\label{thm: coverage lambda0}
 Under the same assumptions as in Theorem~\ref{thm: coverage}, and with the additional assumption that $\lambda_{\min}(\Sigma)>0$, for $\alpha\in(\alpha_0,1)$ with $\alpha_0\in(0,0.01]$ and $\lambda=0$, ${\rm CI}_{\alpha}$ defined in \eqref{eq: aggregated CI} satisfies  $\liminf_{T_0,T_1\rightarrow\infty}\liminf_{M\to\infty}\prob(\tau^* \in {\rm CI}_{\alpha})\geq 1-\alpha.$
\end{Theorem}
The proof is deferred to Section~\ref{pf: coverage lambda0}.

The following theorem establishes the precision property of our proposed CI in \eqref{eq: aggregated CI} for $\lambda=0$.
\begin{Theorem}\label{thm: precision lambda0}
For $\alpha\in(\alpha_0,1)$ with $\alpha_0\in(0,0.01]$, suppose ${\rm CI}_{\alpha}$ in \eqref{eq: aggregated CI} is constructed using the refined index set $\tilde{\mathbb{M}}$ in \eqref{eq: add filter non-empty}. Under the same assumptions as in Theorem~\ref{thm: precision}, together with the additional assumption that $\lambda_{\min}(\Sigma)>0$, for $\lambda=0$, $\operatorname{Length}({\rm CI}_{\alpha})$ satisfies the following precision property:
$$\liminf_{T_0,T_1\to\infty}\liminf_{M\to\infty}\mathbb{P}\left(\operatorname{Length}\left({\rm CI}_{\alpha}\right)\lesssim \frac{[\log (\max\{T_0,N\})]^{\frac{1+b}{2b}}}{\sqrt{T_0}\cdot \lambda_{\min}(\Sigma)/\sqrt{N}} + \frac{1}{\sqrt{T_1}}\right)\geq 1-\alpha_0,$$
where $b>0$ is defined in Assumption~\ref{assumption: pre}.
\end{Theorem}

The proof is deferred to Section~\ref{pf: precision lambda0}.

Similar to Theorem~\ref{thm: conv of tauhat lambda0}, the above theorems cover the limiting case where $\lambda_{\min}(\Sigma) \to 0$, but they require $\lambda_{\min}(\Sigma)$ to satisfy $[\log(\max\{T_0,N\})]^{\frac{1+b}{2b}}/({\sqrt{T_0}}\cdot\lambda_{\min}(\Sigma)/\sqrt{N}) \to 0$ for a meaningful result.

\section{Proofs of Theorems and Propositions}\label{sec: pf thm prop}
In this section, we provide the proofs of the results in the main paper. The proofs of the supporting theorem and lemmas used in this section are deferred to Sections~\ref{sec: pf other main} and \ref{sec: pf lem}, respectively. 
\subsection{Proof of Theorem~\ref{thm: id of DRoSC}}\label{pf: identification}
We recall the definition of $R_{\beta}(\tau)$:
\begin{align*}
    R_{\beta}(\tau) &= \frac{1}{T_1}\sum_{t=T_0+1}^{T}\E\left[ \left(Y_{1,t}-X_t\tr\beta\right)^2 - \left(Y_{1,t}-X_t\tr\beta-\tau\right)^2\right].\\
    &=\frac{1}{T_1}\sum_{t=T_0+1}^{T}\E\left[ 2\tau\left(Y_{1,t}-X_t\tr\beta\right)-\tau^2\right]=2\tau\left(\mu_Y-\mu\tr\beta\right) -\tau^2,
\end{align*}
where $\mu_Y = T_1^{-1}\sum_{t=T_0+1}^{T} \E Y_{1,t}$, $\mu = T_1^{-1}\sum_{t=T_0+1}^{T}\E X_t$ as in \eqref{eq: def of tau beta}. We also recall the definition of $\beta^*$ in \eqref{eq: tau star and beta star}: $$\beta^* {\in} \argmin_{\beta \in \Omega}\left(\mu_Y-\mu\tr\beta\right)^2,$$
where $\Omega$ is defined in \eqref{def: Omega}.

For any $\nu \in (0,1)$ and $\tilde{\beta}\in \Omega$, $\nu\tilde{\beta}+(1-\nu)\beta^*\in \Omega$, since $\Omega$ is a convex set. Since $\left(\mu_Y-\mu\tr\beta^*\right)^2 \leq \left(\mu_Y-\mu\tr\beta\right)^2$ for all $\beta \in \Omega$, we have
\begin{align*}
&\left(\mu_Y -\mu\tr\beta^*\right)^2 \leq 
    \left(\mu_Y -\mu\tr\left(\nu\tilde{\beta}+(1-\nu)\beta^*)\right)\right)^2= \left[\nu\left(\mu_Y -\mu\tr\tilde{\beta}\right)+(1-\nu)\left(\mu_Y -\mu\tr{\beta}^*\right)\right]^2\\
    &=\nu^2\left(\mu_Y -\mu\tr\tilde{\beta}\right)^2+(1-\nu)^2\left(\mu_Y -\mu\tr{\beta}^*\right)^2+2\nu(1-\nu)\left(\mu_Y -\mu\tr\tilde{\beta}\right)\left(\mu_Y -\mu\tr{\beta}^*\right),
\end{align*}
and this leads to 
\begin{align*}
    0&\leq \nu^2\left(\mu_Y -\mu\tr\tilde{\beta}\right)^2+(-2\nu+\nu^2)\left(\mu_Y -\mu\tr{\beta}^*\right)^2+2\nu(1-\nu)\left(\mu_Y -\mu\tr\tilde{\beta}\right)\left(\mu_Y -\mu\tr{\beta}^*\right).
\end{align*}
By dividing by $\nu$ and taking $\underset{\nu\rightarrow 0}{\lim}$ both sides, we establish
\begin{align*}
  \left(\mu_Y -\mu\tr\beta^*\right)^2\leq \left(\mu_Y -\mu\tr\beta^*\right)\left(\mu_Y -\mu\tr\tilde{\beta}\right),
\end{align*}
for any $\tilde{\beta} \in \Omega$. By taking $\tau =\mu_Y -\mu\tr\beta^*$, we have
\begin{align*}
    \max_{\tau \in \mathbb{R}}\min_{\beta \in \Omega}\left\{2\tau\left(\mu_Y -\mu\tr\beta\right)-\tau^2\right\}&\geq \min_{\beta\in \Omega}\left\{2\left(\mu_Y -\mu\tr\beta^*\right)\left(\mu_Y -\mu\tr\beta\right)-\left(\mu_Y -\mu\tr\beta^*\right)^2\right\}\\
    &\geq \left(\mu_Y -\mu\tr\beta^*\right)^2.
\end{align*}

On the other hand, by taking $\beta = \beta^*$,

\begin{align*}
    &\max_{\tau \in \mathbb{R}}\min_{\beta \in \Omega}\left\{2\tau\left(\mu_Y -\mu\tr\beta\right)-\tau^2\right\}\leq \max_{\tau \in \mathbb{R}}\left\{2\tau\left(\mu_Y -\mu\tr\beta^*\right)-\tau^2\right\}= \left(\mu_Y -\mu\tr\beta^*\right)^2.
\end{align*}
By matching the two bounds above, we have
\begin{align*}
    \max_{\tau \in \mathbb{R}}\min_{\beta \in \Omega}\left\{2\tau\left(\mu_Y -\mu\tr\beta\right)-\tau^2\right\}=\left(\mu_Y -\mu\tr\beta^*\right)^2.
\end{align*}
Thus,
\begin{align*}
    \argmax_{\tau \in \mathbb{R}}\min_{\beta \in \Omega}\left\{2\tau\left(\mu_Y -\mu\tr\beta\right)-\tau^2\right\} = \mu_Y -\mu\tr\beta^*.
\end{align*}

For the uniqueness, we note that $R_{\beta}(\tau)$ can be written as
\begin{align*}
    R_{\beta}(\tau) = -\tau^2+2\tau\left(\mu_Y-\mu\tr\beta\right),
\end{align*}
which is a twice-differentiable function of $\tau$ with a second derivative $\frac{d^2}{d\tau^2}R_{\beta}(\tau) = -2<0$ for all $\tau$ and $\beta$. Thus, $R_{\beta}(\tau)$ is a strictly concave function. That is, for $s\in (0,1)$ and $\tau,\tilde{\tau}\in \mathbb{R}$ with $\tau\neq\tilde{\tau}$, by strict concavity,
\begin{align*}
&R_{\beta}(s\tau+(1-s)\tilde{\tau})>sR_{\beta}(\tau)+(1-s)R_{\beta}(\tilde{\tau}).
\end{align*}
Since $\Omega$ is compact set and $R_{\beta}(\tau)$ is continuous function, $\min_{\beta\in\Omega}R_{\beta}(\tau)$ exists for all $\tau$. 
Combining the above, we have
\begin{align*}
    \min_{\beta\in\Omega}R_{\beta}(s\tau+(1-s)\tilde{\tau})>\min_{\beta\in\Omega}\left\{sR_{\beta}(\tau)+(1-s)R_{\beta}(\tilde{\tau})\right\}\geq s\min_{\beta\in\Omega}R_{\beta}(\tau)+(1-s)\min_{\beta\in\Omega}R_{\beta}(\tilde{\tau}).
\end{align*}
Thus, $\min_{\beta\in\Omega}R_{\beta}(\tau)$ is also strictly concave, which leads to the optimizer $\tau^*(\Omega)$ being unique. This concludes the proof.

\subsection{Proof of Theorem~\ref{thm: tau star} and Theorem~\ref{thm: sign of tau star}}
We first provide the proof of Theorem~\ref{thm: tau star}. Since $\Omega$ in \eqref{def: Omega} is a convex set, which was proved in Section~\ref{pf: identification}, $\Omega$ is a connected set. Since $\tau(\beta)$, which is defined in \eqref{eq: def of tau beta}, is a continuous function, the image of the connected set $\Omega$ under the continuous function $\tau(\beta)$  is also a connected set. The image $\tau(\Omega)$ is connected subset in $\mathbb{R}$, so we can classify $\tau(\Omega)$ into three categories: $$\text{(i) $\tau(\Omega) \subset (-\infty,0)$, \quad (ii) $\tau(\Omega) \subset (0,\infty)$, \quad (iii) $0 \in \tau(\Omega)$}. $$ 
Since $\beta^* \in \argmin_{\beta \in \Omega} [\tau(\beta)]^2$ by Theorem~\ref{thm: id of DRoSC}, $(\tau^*)^2 = [\tau(\beta^*)]^2 = \min_{\beta\in\Omega} [\tau(\beta)]^2$ so  $\tau(\beta^*) = \max_{\beta\in\Omega}\tau(\beta)$ for (i). Analogously, $\tau^*(\Omega) = \tau(\beta^*) =\min_{\beta\in\Omega}\tau(\beta)$ for (ii). If $0 \in \tau(\Omega)$, then there exists $\beta^* \in \Omega$ such that $[\tau(\beta^*)]^2=0$, so $\tau(\beta^*) = 0$.

We now prove Theorem~\ref{thm: sign of tau star}. Since $Y_{1,t}^{(1)}=Y_{1,t}$ for $t=T_0+1,...,T$ and by the definition of $\tau_t$, we have 
\begin{align*}
    \tau(\beta) = \taubar+\frac{1}{T_1}\sum_{t=T_0+1}^{T}\E\left[ Y_{1,t}^{(0)}-X_t\tr\beta\right].
\end{align*}
For $\Omega$ containing $\beta^{(1)}$, we have $(\tau^*)^2 = [\tau(\beta^*)]^2 \leq [\tau(\beta^{(1)})]^2 = (\taubar)^2$ since $\E[Y_{1,t}^{(0)}] = \E[X_t\tr\beta^{(1)}]$ under the model \eqref{eq: outcome model}. For (i), $\tau(\beta^{(1)}) = \taubar<0$ and $\tau(\beta)<0$ for all $\beta\in\Omega$, so $\tau^* = \tau(\beta^*)<0$. Analogously, for (ii), $\tau^*>0$ if $\taubar>0$. Thus for non-zero $\tau^*$, $\tau^*$ and $\taubar$ do not have the opposite sign. This concludes the proofs.

\subsection{Proof of Theorem~\ref{thm: conv of tauhat}}
\label{pf: conv of tauhat}

We prove the result based directly on the sensitivity-interval characterization in Theorem~\ref{thm: tau star}. Recall from Section~\ref{sec: theoretical justification} that $W_t=X_tX_t\tr-\E X_tX_t\tr$, and $\epsilon_t= (\nu_{t}\tr,v_t,u_t^{(1)})\tr$ for $t=T_0+1,\ldots,T$ where $\nu_t = X_t - \mathbb{E}X_t$, $v_t=Y_{1,t}^{(1)}-Y_{1,t}^{(0)}-\E[Y_{1,t}^{(1)}-Y_{1,t}^{(0)}]$, and $u_t^{(1)}$ is defined in \eqref{eq: outcome model}.
{For $K>0$, define the following events:}
\begin{align}
    &\mathcal{G}_1 = \left\{\sup_{\beta\in\Delta^N}\left\|\frac{1}{T_0}\sum_{t=1}^{T_0}\left[X_tu_t^{(0)}+W_t(\beta^{(0)}-\beta)\right]\right\|_{\infty}\leq C_0\frac{\left(\log(\max\{T_0,N\})\right)^{\frac{1+b}{2b}}}{\sqrt{T_0}}\right\},\label{event: pre}\\
    &{\mathcal{G}_2(K) = \left\{\left\|\frac{1}{\sqrt{T_1}}\sum_{t=T_0+1}^{T}\epsilon_t\right\|_{\infty}\leq K\right\}.}\label{event: post}
\end{align}
where $C_0$ and $b$ are defined in Assumption~\ref{assumption: pre}.
{Assumptions~\ref{assumption: pre} and \ref{assumption: post} imply $\prob(\mathcal{G}_1)\to1$ and
$\lim_{K\to\infty}\liminf_{T_1\to\infty}\prob\{\mathcal{G}_2(K)\}=1$, respectively. Constants hidden by $\lesssim$ below may depend on fixed $K$.}

We first control the post-treatment estimation error. Note that
\begin{align*}
    \sqrt{T_1}(\muhat_Y-\mu_Y)
    &=\frac{1}{\sqrt{T_1}}\sum_{t=T_0+1}^{T}\left[\nu_t\tr\beta^{(1)}+u_t^{(1)}+v_t\right]\\
    &={\left((\beta^{(1)})\tr,1,1\right)}
    \left(\frac{1}{\sqrt{T_1}}\sum_{t=T_0+1}^{T}\epsilon_t\right).
\end{align*}
Thus, on the event ${\mathcal{G}_2(K)}$, H\"older inequality gives
\begin{align}
    |\muhat_Y-\mu_Y|\lesssim\frac{1}{\sqrt{T_1}}.
    \label{eq: conv of first term}
\end{align}
Moreover,
\begin{align*}
    \|\muhat-\mu\|_{\infty}
    =\left\|\frac{1}{T_1}\sum_{t=T_0+1}^{T}\nu_t\right\|_{\infty}
    \leq\left\|\frac{1}{T_1}\sum_{t=T_0+1}^{T}\epsilon_t\right\|_{\infty},
\end{align*}
so on the same event,
\begin{align}
    |(\muhat-\mu)\tr\beta|
    \leq\|\muhat-\mu\|_{\infty}\|\beta\|_1
    \lesssim\frac{1}{\sqrt{T_1}}
    \quad\text{for every }\beta\in\Delta^N.
    \label{eq: conv of second term}
\end{align}
Combining the boundedness of $\mu$ in Assumption~\ref{assumption: post} with \eqref{eq: conv of first term} and \eqref{eq: conv of second term}, on ${\mathcal{G}_2(K)}$ we have
\begin{align}
    \|\muhat-\mu\|_{\infty}\lesssim\frac{1}{\sqrt{T_1}},
    \quad |\muhat_Y-\mu_Y|\lesssim\frac{1}{\sqrt{T_1}}.
    \label{eq: mu rate}
\end{align}

To compare the population and sample uncertainty classes, define the convex functions
\begin{align}
    \fhat(\beta)
    &=\left\|\frac{1}{T_0}\sum_{t=1}^{T_0}X_t\left(Y_{1,t}-X_t\tr\beta\right)\right\|_{\infty}-\lambda,
    \label{eq: def of f hat}\\
    f(\beta)
    &=\left\|\frac{1}{T_0}\sum_{t=1}^{T_0}\E\left[X_t\left(Y_{1,t}-X_t\tr\beta\right)\right]\right\|_{\infty}-\lambda.
    \label{eq: def of f}
\end{align}
Then $\beta\in\Omega$ is equivalent to $f(\beta)\leq0$, whereas $\beta\in\Omegahat$ is equivalent to $\fhat(\beta)\leq\rho$. The following lemma gives the required uniform comparison; its proof is provided in Section~\ref{pf: unif conv of f}.
\begin{Lemma}\label{lem: unif conv of f}
On the event $\mathcal{G}_1$ defined in \eqref{event: pre},
\begin{align*}
    \sup_{\beta\in\Delta^N}|\fhat(\beta)-f(\beta)|
    \leq C_0\frac{[\log(\max\{T_0,N\})]^{\frac{1+b}{2b}}}{\sqrt{T_0}}.
\end{align*}
\end{Lemma}

We first compare the population uncertainty class $\Omega(\lambda)$ and its estimator $\Omegahat(\lambda)$. Define
\begin{align}
    r_{T_0}=C_0\frac{[\log(\max\{T_0,N\})]^{\frac{1+b}{2b}}}{\sqrt{T_0}},
    \label{eq: alternative r T0}
\end{align}
where $C_0$ and $b$ are defined in Assumption~\ref{assumption: pre}. By Lemma~\ref{lem: unif conv of f}, on the event $\mathcal{G}_1$,
\begin{align}
    \sup_{\beta\in\Delta^N}|\fhat(\beta)-f(\beta)|\leq r_{T_0}.
    \label{eq: alternative uniform f}
\end{align}
The choice of $\rho$ in Theorem~\ref{thm: conv of tauhat} satisfies $\rho\geq r_{T_0}$. Thus, if $\beta\in\Omega(\lambda)$, then $f(\beta)\leq0$ and
\begin{align*}
    \fhat(\beta)\leq f(\beta)+r_{T_0}\leq r_{T_0}\leq\rho,
\end{align*}
which implies $\beta\in\Omegahat(\lambda)$. Conversely, if $\beta\in\Omegahat(\lambda)$, then $\fhat(\beta)\leq\rho$ and
\begin{align*}
    f(\beta)\leq\fhat(\beta)+r_{T_0}\leq\rho+r_{T_0}.
\end{align*}
Therefore, on $\mathcal{G}_1$,
\begin{align}
    \Omega(\lambda)\subseteq\Omegahat(\lambda)
    \subseteq\Omega(\lambda+\rho+r_{T_0}).
    \label{eq: alternative omega inclusions}
\end{align}

We next show that each weight in the estimated class is close, in terms of its implied treatment effect, to a weight in the population class. Let $s_{T_0}=\rho+r_{T_0}$. For any $\beta\in\Omegahat(\lambda)$, define
\begin{align}
    \beta^{\circ}=\frac{\lambda}{\lambda+s_{T_0}}\beta
    +\frac{s_{T_0}}{\lambda+s_{T_0}}\beta^{(0)}.
    \label{eq: alternative beta circle}
\end{align}
Since $\beta,\beta^{(0)}\in\Delta^N$, we have $\beta^{\circ}\in\Delta^N$. Moreover, $\gamma-\Sigma\beta^{(0)}=0$, and the second inclusion in \eqref{eq: alternative omega inclusions} gives $\|\gamma-\Sigma\beta\|_{\infty}\leq\lambda+s_{T_0}$. It follows that
\begin{align*}
    \|\gamma-\Sigma\beta^{\circ}\|_{\infty}
    &=\frac{\lambda}{\lambda+s_{T_0}}
    \|\gamma-\Sigma\beta\|_{\infty}
    \leq\lambda,
\end{align*}
so $\beta^{\circ}\in\Omega(\lambda)$. By H\"older inequality and $\|\beta-\beta^{(0)}\|_1\leq2$,
\begin{align}
    |\tau(\beta)-\tau(\beta^{\circ})|
    &=|\mu\tr(\beta-\beta^{\circ})|
    \leq\|\mu\|_{\infty}\|\beta-\beta^{\circ}\|_1
    \leq2\|\mu\|_{\infty}\frac{s_{T_0}}{\lambda+s_{T_0}}
    \eqqcolon d_{T_0}.
    \label{eq: alternative effect distance}
\end{align}

We now compare the ranges of treatment effects over the two uncertainty classes. Write the population sensitivity interval in \eqref{eq: sensitivity interval} as $I(\Omega)=[L,U]$, and define
\begin{align*}
    [\widetilde L,\widetilde U]
    &=\left[\min_{\beta\in\Omegahat}\tau(\beta),
    \max_{\beta\in\Omegahat}\tau(\beta)\right],\quad
    [\widehat L,\widehat U]
    =\left[\min_{\beta\in\Omegahat}\widehat\tau(\beta),
    \max_{\beta\in\Omegahat}\widehat\tau(\beta)\right],
\end{align*}
where
\begin{equation}\label{eq: tauhat(beta)}
    \widehat\tau(\beta)=\muhat_Y-\muhat\tr\beta.
\end{equation}
On the event $\mathcal{G}_1$, $\beta^{(0)}\in\Omega\subseteq\Omegahat$ by the first inclusion in \eqref{eq: alternative omega inclusions}, so $\Omegahat$ is non-empty.
These ranges are closed intervals because the corresponding uncertainty classes are compact and convex and $\tau(\beta)$ and $\widehat\tau(\beta)$ are affine in $\beta$.

The first inclusion in \eqref{eq: alternative omega inclusions} implies $\widetilde L\leq L$ and $\widetilde U\geq U$. On the other hand, for every $\beta\in\Omegahat$, the weight $\beta^{\circ}$ in \eqref{eq: alternative beta circle} belongs to $\Omega$ and satisfies \eqref{eq: alternative effect distance}. Hence,
\begin{align*}
    \tau(\beta)\geq\tau(\beta^{\circ})-d_{T_0}\geq L-d_{T_0},
    \quad
    \tau(\beta)\leq\tau(\beta^{\circ})+d_{T_0}\leq U+d_{T_0}.
\end{align*}
Taking the minimum and maximum over $\beta\in\Omegahat$ yields
\begin{align}
    \max\{|\widetilde L-L|,|\widetilde U-U|\}\leq d_{T_0}.
    \label{eq: alternative population endpoints}
\end{align}

To account for estimating the post-treatment objective, define
\begin{align*}
    r_{T_1}=|\muhat_Y-\mu_Y|+\|\muhat-\mu\|_{\infty}.
\end{align*}
Since $\|\beta\|_1=1$ for all $\beta\in\Delta^N$, H\"older inequality gives
\begin{align*}
    \sup_{\beta\in\Delta^N}|\widehat\tau(\beta)-\tau(\beta)|
    \leq r_{T_1}.
\end{align*}
Applying this uniform bound to the minimum and maximum over the same set $\Omegahat$ gives
\begin{align}
    \max\{|\widehat L-\widetilde L|,|\widehat U-\widetilde U|\}
    \leq r_{T_1}.
    \label{eq: alternative estimated endpoints}
\end{align}
Combining \eqref{eq: alternative population endpoints} and \eqref{eq: alternative estimated endpoints}, we obtain
\begin{align}
    \max\{|\widehat L-L|,|\widehat U-U|\}
    \leq r_{T_1}+d_{T_0}.
    \label{eq: alternative endpoint distance}
\end{align}

It remains to translate the endpoint bound into a bound for the WRoTE. By Theorem~\ref{thm: tau star}, $\tau^*$ is the projection of zero onto $[L,U]$; we denote $\tau^* = \operatorname{proj}_{[L,U]}(0)$. By the definition of $\betahat$ in \eqref{eq: betahat tauhat}, minimizing the squared sample objective over $\Omegahat$ is equivalent to projecting zero onto its range, so $\tauhat$ is the projection of zero onto $[\widehat L,\widehat U]$. For any two closed intervals $[a,b]$ and $[c,d]$,
\begin{align}
    |\operatorname{proj}_{[a,b]}(0)-\operatorname{proj}_{[c,d]}(0)|
    \leq\max\{|a-c|,|b-d|\},
    \label{eq: projection interval inequality}
\end{align}
because $\operatorname{proj}_{[a,b]}(0)=\max\{a,\min\{0,b\}\}$ and the minimum and maximum operators are Lipschitz under the maximum norm. Consequently, \eqref{eq: alternative endpoint distance} gives
\begin{align}
    |\tauhat-\tau^*|
    \leq r_{T_1}+d_{T_0}
    =r_{T_1}+2\|\mu\|_{\infty}
    \frac{\rho+r_{T_0}}{\lambda+\rho+r_{T_0}}.
    \label{eq: alternative finite sample rate}
\end{align}
This projection argument covers all three cases in Theorem~\ref{thm: tau star}, including the cases in which exactly one of $\tau^*$ and $\tauhat$ equals zero.

On the event ${\mathcal{G}_2(K)}$, the same arguments as in \eqref{eq: conv of first term}--\eqref{eq: conv of second term} yield $r_{T_1}\lesssim T_1^{-1/2}$. Assumption~\ref{assumption: post} also ensures that $\|\mu\|_{\infty}\lesssim1$. Since $\lambda>0$, \eqref{eq: alternative finite sample rate}, the definition of $r_{T_0}$ in \eqref{eq: alternative r T0}, and the choice of $\rho$ imply that, on ${\mathcal{G}_1\cap\mathcal{G}_2(K)}$,
\begin{align}
    |\tauhat-\tau^*|
    \lesssim
    \frac{1}{\sqrt{T_1}}
    +\frac{[\log(\max\{T_0,N\})]^{\frac{1+b}{2b}}}
    {\lambda\sqrt{T_0}}.
    \label{eq: alternative  rate}
\end{align}
{For every fixed $K$, the preceding bound holds on $\mathcal G_1\cap\mathcal G_2(K)$. The probability limits stated after \eqref{event: post} establish the $O_p$ rate in \eqref{eq: conv rate of tau hat} and prove Theorem~\ref{thm: conv of tauhat}. In particular, if the uniform pre-treatment moment error and $\rho$ are both of order $T_0^{-1/2}$, then $|\tauhat-\tau^*|=O_p(T_0^{-1/2}+T_1^{-1/2})$ for fixed $\lambda>0$.}

\subsection{Proof of Theorem~\ref{thm: coverage}}\label{pf: coverage}
{
In this section, we provide the proof of Theorem~\ref{thm: coverage}. Let $\alpha'=\alpha-\alpha_0$. On the event $\mathcal E^*$ defined in \eqref{event: negligible}, choose an index $m^*\in\mathbb M$ attaining the minimum that defines $\mathcal E^*$. Since ${\rm Int}^{[m^*]}\subseteq {\rm CI}_{\alpha}$ on this event, the definition of ${\rm CI}_{\alpha}$ and the decomposition in \eqref{eq: decomposition of tauhatm} give
\begin{equation*}
\begin{aligned}
\prob(\tau^*\notin {\rm CI}_{\alpha})
&\leq 1-\prob(\mathcal E^*)
+\prob\left(\tau^*\notin {\rm Int}^{[m^*]},\mathcal E^*\right)\\
&=1-\prob(\mathcal E^*)
+\prob\left(\left|\muhat_Y-\mu_Y-\left\{(\muhatmstar)\tr\betahatmstar-\mu\tr\beta^*\right\}\right|
\geq z_{\alpha'/2}\VhatY^{1/2},\mathcal E^*\right)\\
&\leq1-\prob(\mathcal E^*)
+\prob\left(\left\{|\muhat_Y-\mu_Y|+
|(\muhatmstar)\tr\betahatmstar-\mu\tr\beta^*|\geq
z_{\alpha'/2}\VhatY^{1/2}\right\}\cap\mathcal E^*\right).
\end{aligned}
\end{equation*}
For $\lambda>0$, on $\mathcal E^*$ there exists a constant $C>0$ such that
\begin{equation*}
\left|(\muhatmstar)\tr\betahatmstar-\mu\tr\beta^*\right|
\leq\zeta_M,
\quad
\zeta_M^2=\frac{C\err}{\min\{\lambda\sqrt{T_0},\sqrt{T_1}\}}.
\end{equation*}
Consequently,
\begin{equation*}
\begin{aligned}
\prob(\tau^*\notin {\rm CI}_{\alpha})
&\leq
\prob\left(\left|\frac{\muhat_Y-\mu_Y}{\VY^{1/2}}\right|
\geq z_{\alpha'/2}\left(\frac{\VhatY}{\VY}\right)^{1/2}
-\frac{\zeta_M}{\VY^{1/2}}\right)
+1-\prob(\mathcal E^*).
\end{aligned}
\end{equation*}
}

We observe that
\begin{equation*}
    \left|\sqrt{\frac{\VhatY}{\VY}}-1\right|=\left|\frac{\sqrt{\VhatY}-\sqrt{\VY}}{\sqrt{\VY}}\right|=\frac{|\VhatY-\VY|}{\sqrt{\VY}(\sqrt{\VhatY}+\sqrt{\VY})}\leq \frac{|\VhatY-\VY|}{\VY},
\end{equation*}
which implies that
\begin{equation*}
    \sqrt{\frac{\VhatY}{\VY}}\geq 1-\frac{|T_1(\VhatY-\VY)|}{T_1\VY}.
\end{equation*}
By applying the above inequality, we establish that
\begin{align*}
    &\prob\left(\left|\frac{\muhat_Y-\mu_Y}{{\rm V}_Y^{1/2}}\right|\geq z_{\alpha'/2}\left(\frac{\VhatY}{{\rm V}_Y}\right)^{1/2}-\frac{\zeta_M}{\VY^{1/2}}\right)+1-\prob(\mathcal{E}^*)\\
    &\leq \prob\left(\left|\frac{\muhat_Y-\mu_Y}{{\rm V}_Y^{1/2}}\right|\geq z_{\alpha'/2}\left(1-\frac{|T_1(\VhatY-\VY)|}{T_1\VY}\right)-\frac{\zeta_M}{{\rm V}_Y^{1/2}}\right)+1-\prob(\mathcal{E}^*).
\end{align*}
Note that $\zeta_M/{\rm V}_Y^{1/2}\rightarrow 0$ as $M\rightarrow\infty$ for fixed $T_1$ and ${|T_1(\VhatY-\VY)|}/({T_1\VY})\pto 0$ as $T_1\to\infty$ under Assumption~\ref{assumption: clt and cons}. Since ${\rm V}_Y^{-1/2}(\muhat_Y-\mu_Y)\dto N(0,1)$ as $T_1\to\infty$ under Assumption~\ref{assumption: clt and cons}, by the bounded convergence theorem, we establish
\begin{align*}
    &\lim_{T_0,T_1\rightarrow\infty}\lim_{M\rightarrow\infty}\prob\left(\left|\frac{\muhat_Y-\mu_Y}{{\rm V}_Y^{1/2}}\right|\geq z_{\alpha'/2}\left(1-\frac{|T_1(\VhatY-\VY)|}{T_1\VY}\right)-\frac{\zeta_M}{{\rm V}_Y^{1/2}}\right)=\prob(|Z|\geq z_{\alpha'/2})=\alpha',
\end{align*}
where $Z$ denotes the standard normal random variable.
Thus, combining with $$\liminf_{T_0,T_1\rightarrow\infty}\liminf_{M\rightarrow\infty}\prob (\mathcal{E}^*) \geq 1- \alpha_0,$$ by Theorem~\ref{thm: negligible}, we establish
\begin{align*}
    \liminf_{T_0,T_1\rightarrow\infty}\liminf_{M\rightarrow\infty}\prob(\tau^*\in {\rm CI}_{\alpha}) \geq 1-\alpha' -\alpha_0=1-\alpha.
\end{align*}
This concludes the proof.

\subsection{Proof of Theorem~\ref{thm: precision}}\label{pf: precision}

Throughout this proof, ${\rm CI}_{\alpha}$ is constructed using the refined index set $\tilde{\mathbb M}$ in \eqref{eq: add filter non-empty}. We consider the sequential asymptotic regime in the theorem: $M\to\infty$ first for fixed $T_0$ and $T_1$, followed by $T_0,T_1\to\infty$. Thus, for sufficiently large $M$,
\begin{align*}
    \rho_M\lesssim\frac{1}{\sqrt{T_0}}.
\end{align*}
Let $\mathcal E$ be the event in \eqref{event: m star}. On $\mathcal E$, there exists $m^*\in\mathbb M$ such that
\begin{align*}
    \|\gammahatmstar-\gamma\|_\infty
    +\|\vecl(\Sigmahatmstar-\Sigma)\|_\infty
    \leq\frac{2\err}{\sqrt{T_0}}.
\end{align*}
Since $\gamma-\Sigma\beta^{(0)}=0$ and $\|\beta^{(0)}\|_1=1$, it follows that
\begin{align*}
    \|\gammahatmstar-\Sigmahatmstar\beta^{(0)}\|_\infty
    &\leq\|\gammahatmstar-\gamma\|_\infty
    +\|(\Sigmahatmstar-\Sigma)\beta^{(0)}\|_\infty\\
    &\leq\frac{2\err}{\sqrt{T_0}}\leq\rho_M.
\end{align*}
Therefore, $m^*\in\tilde{\mathbb M}$ and $\tilde{\mathbb M}$ is non-empty on $\mathcal E$.

Regarding the length of ${\rm CI}_{\alpha}$, we first have
\begin{align}
    \operatorname{Length}({\rm CI}_{\alpha})
    &\leq \max_{m\in\tilde{\mathbb M}}\tauhatm-
    \min_{m\in\tilde{\mathbb M}}\tauhatm+
    2z_{\alpha'/2}\VhatY^{1/2}\nonumber\\
    &\leq2\left[\max_{m\in\tilde{\mathbb M}}|\tauhatm-\tauhat|
    +z_{\alpha'/2}\VhatY^{1/2}\right],
    \label{eq: precision initial length}
\end{align}
where $\alpha'=\alpha-\alpha_0$. It remains to control the candidate centers and their common half-length.

By Assumption~\ref{assumption: clt and cons}, there exists a positive constant $c_1$ such that
\begin{equation}
    \max\{\|T_0\VSigma\|_{\max},\|T_0\Vgamma\|_{\max},
    \|T_1\Vmu\|_{\max},T_1\VY\}\leq c_1.
    \label{eq: max cov}
\end{equation}
Define
\begin{align}
    \mathcal G_3
    &=\left\{\VhatY\leq\frac{2c_1}{T_1},\quad
    d_\mu\leq\frac{2c_1}{T_1}\right\},\quad
    \mathcal G_4
    =\left\{d_\Sigma\leq\frac{2c_1}{T_0},\quad
    d_\gamma\leq\frac{2c_1}{T_0}\right\},
    \label{event: consistency}
\end{align}
where $d_\mu=\|\Vhatmu\|_{\max}$, $d_\Sigma=\|\VhatSigma\|_{\max}$, and $d_\gamma=\|\Vhatgamma\|_{\max}$.
\begin{Lemma}\label{lem: cons of cov}
Under Assumption~\ref{assumption: clt and cons},
\begin{align*}
    \prob(\mathcal G_3)\to1\quad\text{as }T_1\to\infty,
    \quad
    \prob(\mathcal G_4)\to1\quad\text{as }T_0\to\infty.
\end{align*}
\end{Lemma}
The proof is provided in Section~\ref{pf: cons of cov}. On $\mathcal G_3$, for fixed $\alpha'=\alpha-\alpha_0$,
\begin{align}
    z_{\alpha'/2}\VhatY^{1/2}\lesssim\frac{1}{\sqrt{T_1}}.
    \label{eq: VhatY bound}
\end{align}
Moreover, by the definition of $\mathbb M$ in \eqref{eq: index set},
\begin{align*}
    |\muhatm_Y-\muhat_Y|
    &\leq1.1z_{\alpha_0/(2p)}\sqrt{\VhatY},\\
    \|\muhatm-\muhat\|_\infty
    &\leq1.1z_{\alpha_0/(2p)}\sqrt{2\|\Vhatmu\|_{\max}}
    \quad\text{for every }m\in\mathbb M.
\end{align*}
Since $N$ and $\alpha_0$ are fixed, on $\mathcal G_3$ this yields
\begin{align}
    \max_{m\in\mathbb M}|\muhatm_Y-\muhat_Y|
    \lesssim\frac{1}{\sqrt{T_1}},
    \quad
    \max_{m\in\mathbb M}\|\muhatm-\muhat\|_\infty
    \lesssim\frac{1}{\sqrt{T_1}}.
    \label{eq: muhatm bound on M}
\end{align}

{We first compare each perturbed sample discrepancy with the unperturbed sample discrepancy.} For $\lambda\geq0$, define
\begin{align}
    \fhatm(\beta)=\|\gammahatm-\Sigmahatm\beta\|_\infty-\lambda.
    \label{def: fhatm}
\end{align}
For any matrix $A=(a_1,\ldots,a_m)\tr\in\mathbb R^{m\times N}$ and $\beta\in\Delta^N$,
\begin{align}
    \|A\beta\|_\infty
    =\max_{1\leq j\leq m}|a_j\tr\beta|
    \leq\max_{1\leq j\leq m}\|a_j\|_\infty\|\beta\|_1
    \leq\|A\|_{\max}=\|\vecl(A)\|_\infty.
    \label{eq: vecl norm}
\end{align}
Thus, for every $m\in\mathbb M$,
\begin{align*}
    \sup_{\beta\in\Delta^N}|\fhatm(\beta)-\fhat(\beta)|
    &\leq\|\gammahatm-\gammahat\|_\infty
    +\|\vecl(\Sigmahatm-\Sigmahat)\|_\infty.
\end{align*}
On $\mathcal G_4$, the definition of $\mathbb M$ and \eqref{eq: max cov} therefore give
\begin{align}
    e_{T_0}
    &\coloneqq\max_{m\in\mathbb M}\sup_{\beta\in\Delta^N}
    |\fhatm(\beta)-\fhat(\beta)|\leq\frac{4.4z_{\alpha_0/(2p)}\sqrt{c_1}}{\sqrt{T_0}}
    \lesssim\frac{1}{\sqrt{T_0}}.
    \label{eq: fhatm fhat max m}
\end{align}
{
We now compare every candidate center directly with the unperturbed center $\tauhat$. For $\beta\in\Delta^N$, write
\begin{equation*}
\begin{aligned}
\widetilde\tau^{[m]}(\beta)&=\muhatm_Y-(\muhatm)\tr\beta.
\end{aligned}
\end{equation*}
Moreover, on $\mathcal G_1$, \eqref{eq: alternative uniform f}, $\gamma-\Sigma\beta^{(0)}=0$, and $\rho\geq r_{T_0}$ give
$\fhat(\beta^{(0)})\leq-\lambda+r_{T_0}\leq-\lambda+\rho$.

Fix $m\in\tilde{\mathbb M}$. By the definition of $\tilde{\mathbb M}$, there exists $\beta_0^{[m]}\in\Delta^N$ such that
$\fhatm(\beta_0^{[m]})\leq\rho_M-\lambda$.
If $\beta\in\Omegahatm$, then \eqref{eq: fhatm fhat max m} gives
$\fhat(\beta)\leq\rho_M+e_{T_0}$. By convexity of $\fhat$,
\begin{equation*}
\fhat\left(
\frac{\lambda}{\lambda+\rho_M+e_{T_0}}\beta
+\frac{\rho_M+e_{T_0}}{\lambda+\rho_M+e_{T_0}}\beta^{(0)}
\right)
\leq
\frac{\rho_M+e_{T_0}}{\lambda+\rho_M+e_{T_0}}\rho
\leq\rho.
\end{equation*}
Thus, the displayed convex combination belongs to $\Omegahat$. Since $\beta$ and $\beta^{(0)}$ belong to the simplex, its objective value differs from $\tauhat(\beta)$ in \eqref{eq: tauhat(beta)} by at most
$2\|\muhat\|_\infty(\rho_M+e_{T_0})/(\lambda+\rho_M+e_{T_0})$.

Conversely, if $\beta\in\Omegahat$, then
$\fhatm(\beta)\leq\rho+e_{T_0}$. The same argument gives
\begin{equation*}
\fhatm\left(
\frac{\lambda}{\lambda+\rho+e_{T_0}}\beta
+\frac{\rho+e_{T_0}}{\lambda+\rho+e_{T_0}}\beta_0^{[m]}
\right)
\leq
\frac{\rho+e_{T_0}}{\lambda+\rho+e_{T_0}}\rho_M
\leq\rho_M.
\end{equation*}
Hence, this convex combination belongs to $\Omegahatm$, and its objective value differs from $\tauhat(\beta)$ by at most
$2\|\muhat\|_\infty(\rho+e_{T_0})/(\lambda+\rho+e_{T_0})$.

Assumption~\ref{assumption: post} implies
$\|\muhat-\mu\|_\infty=O_p(T_1^{-1/2})=o_p(1)$. Because $\mu$ is bounded, $\|\muhat\|_\infty$ is bounded with probability tending to one.

Let $[\widehat L,\widehat U]$ be the range of $\tauhat(\beta)$ over $\Omegahat$, and let $[\widehat L^{[m]},\widehat U^{[m]}]$ be the range of $\widetilde\tau^{[m]}(\beta)$ over $\Omegahatm$. Since $\|\beta\|_1=1$ for $\beta\in\Delta^N$, \eqref{eq: muhatm bound on M} gives, on $\mathcal G_3$,
\begin{equation}
\max_{m\in\mathbb M}\sup_{\beta\in\Delta^N}
|\widetilde\tau^{[m]}(\beta)-\tauhat(\beta)|
\lesssim\frac{1}{\sqrt{T_1}}.
\end{equation}
The preceding class comparisons therefore imply, on $\mathcal G_1\cap\mathcal G_3\cap\mathcal G_4$ and the event above, that
\begin{equation}
\begin{aligned}
\max\{|\widehat L^{[m]}-\widehat L|,
|\widehat U^{[m]}-\widehat U|\}
&\lesssim\frac{1}{\sqrt{T_1}}+
\frac{\rho+\rho_M+e_{T_0}}{\lambda}\\
&\lesssim\frac{1}{\sqrt{T_1}}
+\frac{[\log(\max\{T_0,N\})]^{\frac{1+b}{2b}}}
{\lambda\sqrt{T_0}}.
\end{aligned}
\label{eq: direct perturbed endpoint comparison}
\end{equation}

By \eqref{eq: m-th betahat}, $\widetilde\tau^{[m]}(\betahatm)$ is the projection of zero onto $[\widehat L^{[m]},\widehat U^{[m]}]$, while $\tauhat$ is the projection of zero onto $[\widehat L,\widehat U]$. Hence, the projection inequality in \eqref{eq: projection interval inequality}, \eqref{eq: direct perturbed endpoint comparison}, and \eqref{eq: muhatm bound on M} give, uniformly over $m\in\tilde{\mathbb M}$,
\begin{equation}
\begin{aligned}
|\tauhatm-\tauhat|
&\leq|\widetilde\tau^{[m]}(\betahatm)-\tauhat|
+|\muhatm_Y-\muhat_Y|\\
&\lesssim\frac{1}{\sqrt{T_1}}
+\frac{[\log(\max\{T_0,N\})]^{\frac{1+b}{2b}}}
{\lambda\sqrt{T_0}}.
\end{aligned}
\end{equation}
Together with \eqref{eq: precision initial length} and \eqref{eq: VhatY bound}, this gives, on the intersection of the preceding events and $\mathcal E$,
\begin{equation}
\operatorname{Length}({\rm CI}_{\alpha})
\lesssim\frac{1}{\sqrt{T_1}}
+\frac{[\log(\max\{T_0,N\})]^{\frac{1+b}{2b}}}
{\lambda\sqrt{T_0}}.
\label{eq: precision  rate}
\end{equation}
Therefore, for a sufficiently large fixed constant $C$, Lemma~\ref{lem: cons of cov}, Assumptions~\ref{assumption: pre} and \ref{assumption: post}, and Proposition~\ref{thm: m star} imply
\begin{equation*}
\begin{aligned}
&\liminf_{T_0,T_1\to\infty}\liminf_{M\to\infty}
\prob\left\{\operatorname{Length}({\rm CI}_{\alpha})
\leq C\left(\frac{1}{\sqrt{T_1}}
+\frac{[\log(\max\{T_0,N\})]^{\frac{1+b}{2b}}}
{\lambda\sqrt{T_0}}\right)\right\}\\
&\quad\geq
\liminf_{T_0,T_1\to\infty}\liminf_{M\to\infty}
\prob\{\mathcal G_1\cap\mathcal G_3\cap\mathcal G_4
\cap\mathcal E\}
\geq1-\alpha_0.
\end{aligned}
\end{equation*}
This establishes Theorem~\ref{thm: precision} and concludes the proof.
}

\section{Proofs of Other Main Results}\label{sec: pf other main}
In this section, we establish the proof of other main results not proved in Section~\ref{sec: pf thm prop}.

\subsection{Proof of Proposition~\ref{prop: magnitude of lower bound}}
Under the model \eqref{eq: outcome model}, $\Omega$ in \eqref{def: Omega} can be rewritten as
\begin{align*}
    \Omega = \left\{\beta\in\Delta^N:\left\|\Sigma(\beta-\beta^{(0)})\right\|\leq\lambda\right\}.
\end{align*}
When $\beta^{(1)}\in\Omega$, by the above definition of $\Omega$, we have
\begin{align*}
    \left\|\Sigma(\beta^{(1)}-\beta^{(0)})\right\|_{\infty}\leq \lambda.
\end{align*}
The equivalence between $\ell^{2}$-norm and $\ell^{\infty}$-norm and the representation of the minimum of a quadratic form lead to the following inequalities: For any $\beta$, we have
\begin{align}
    \left\|\Sigma(\beta-\beta^{(0)})\right\|_{\infty} \geq \frac{\left\|\Sigma(\beta-\beta^{(0)})\right\|_{2}}{\sqrt{N}}\geq \frac{\lambda_{\min}(\Sigma)\left\|\beta-\beta^{(0)}\right\|_{2}}{\sqrt{N}}\geq \frac{\lambda_{\min}(\Sigma)\left\|\beta-\beta^{(0)}\right\|_{\infty}}{\sqrt{N}}.\label{eq: beta and beta0}
\end{align}
If $\beta\in\Omega$, we ensure 
\begin{align*}
\lambda \geq \left\|\Sigma(\beta-\beta^{(0)})\right\|_{\infty} \geq \frac{\lambda_{\min}(\Sigma)\left\|\beta-\beta^{(0)}\right\|_{\infty}}{\sqrt{N}}.
\end{align*}
For $\lambda_{\min}(\Sigma)>0$, this leads to 
\begin{align*}
    \left\|\beta^{(1)}-\beta^{(0)}\right\|_{\infty} \leq \frac{\sqrt{N}\lambda}{\lambda_{\min}(\Sigma)}.
\end{align*}
Since $\beta^* \in \Omega$, we have the similar inequality $$\left\|\beta^*-\beta^{(0)}\right\|_{\infty} \leq \frac{\sqrt{N}\lambda}{\lambda_{\min}(\Sigma)}.$$
On the other hand, by H\"older's inequality and triangle inequality, we have
\begin{align*}
    \left|\tau^*-\taubar\right| = \left|\mu\tr(\beta^*-\beta^{(1)})\right|\leq \left\|\mu\right\|_{1}\left\|\beta^*-\beta^{(1)}\right\|_{\infty} &\leq \left\|\mu\right\|_{1}\left(\left\|\beta^{(1)}-\beta^{(0)}\right\|_{\infty}+ \left\|\beta^*-\beta^{(0)}\right\|_{\infty}\right)\\
    &\leq 2\left\|\mu\right\|_{1}\frac{\sqrt{N}\lambda}{\lambda_{\min}(\Sigma)},
\end{align*}
and this concludes the proof.

\subsection{Proof of Theorem~\ref{thm: conv of tauhat lambda0}}\label{pf: conv of tauhat lambda0}

Throughout this proof, let $\lambda=0$ and suppose $\lambda_{\min}(\Sigma)>0$. Since $\gamma=\Sigma\beta^{(0)}$, the population class satisfies
\begin{align*}
    \Omega(0)=\{\beta\in\Delta^N:\|\Sigma(\beta-\beta^{(0)})\|_\infty=0\}
    =\{\beta^{(0)}\}.
\end{align*}
Consequently, $\beta^*=\beta^{(0)}$ and $\tau^*=\tau(\beta^{(0)})$.

Recall $r_{T_0}$ in \eqref{eq: alternative r T0}. On the event $\mathcal G_1$, Lemma~\ref{lem: unif conv of f} gives
\begin{align*}
    \sup_{\beta\in\Delta^N}|\fhat(\beta)-f(\beta)|\leq r_{T_0}.
\end{align*}
Since $\rho\geq r_{T_0}$ and $\beta^{(0)}\in\Omega(0)$, \eqref{eq: alternative omega inclusions} gives $\beta^{(0)}\in\Omegahat(0)$ on $\mathcal G_1$, so $\betahat$ is well-defined.
For any $\beta\in\Omegahat(0)$, $\fhat(\beta)\leq\rho$, so
\begin{align}
    \|\Sigma(\beta-\beta^{(0)})\|_\infty
    =f(\beta)\leq\fhat(\beta)+r_{T_0}\leq\rho+r_{T_0}.
    \label{eq: lambda0 sample discrepancy}
\end{align}
Because $\Sigma$ is positive definite,
\begin{align*}
    \lambda_{\min}(\Sigma)\|\beta-\beta^{(0)}\|_2
    &\leq\|\Sigma(\beta-\beta^{(0)})\|_2\\
    &\leq\sqrt N\|\Sigma(\beta-\beta^{(0)})\|_\infty.
\end{align*}
Combining this inequality with \eqref{eq: lambda0 sample discrepancy}, for every $\beta\in\Omegahat(0)$ we obtain
\begin{align}
    \|\beta-\beta^{(0)}\|_2
    \leq\frac{\sqrt N(\rho+r_{T_0})}{\lambda_{\min}(\Sigma)}.
    \label{eq: lambda0 sample weight distance}
\end{align}

Since $\betahat\in\Omegahat(0)$, H\"older inequality and \eqref{eq: lambda0 sample weight distance} give
\begin{align*}
    |\tauhat-\tau^*|
    &=|\muhat_Y-\muhat\tr\betahat-(\mu_Y-\mu\tr\beta^{(0)})|\\
    &\leq|\muhat_Y-\mu_Y|+|(\muhat-\mu)\tr\betahat|
    +|\mu\tr(\betahat-\beta^{(0)})|\\
    &\leq|\muhat_Y-\mu_Y|+\|\muhat-\mu\|_\infty
    +\|\mu\|_2\frac{\sqrt N(\rho+r_{T_0})}{\lambda_{\min}(\Sigma)}.
\end{align*}
On ${\mathcal G_2(K)}$, equations \eqref{eq: conv of first term}--\eqref{eq: mu rate} show that the first two terms are of order $T_1^{-1/2}$ and that $\|\mu\|_2$ is bounded because $N$ is fixed. Therefore, on $\mathcal G_1\cap{\mathcal G_2(K)}$,
\begin{align}
    |\tauhat-\tau^*|
    \lesssim\frac{1}{\sqrt{T_1}}
    +\frac{\sqrt N}{\lambda_{\min}(\Sigma)}
    \frac{[\log(\max\{T_0,N\})]^{\frac{1+b}{2b}}}{\sqrt{T_0}}.
    \label{eq: lambda0  rate}
\end{align}
{For each fixed $K$, this bound holds on $\mathcal G_1\cap\mathcal G_2(K)$. The probability properties of these events established in \eqref{event: pre}--\eqref{event: post} therefore yield the $O_p$ rate in \eqref{eq: conv rate of tau hat lambda0} and prove Theorem~\ref{thm: conv of tauhat lambda0}.}

\subsection{Proof of Theorem~\ref{thm: negligible}}\label{pf: negligible}

In this section, we provide the proof of Theorem~\ref{thm: negligible}. To facilitate the proof, we define the quantities $\widehat{Z}$ and $\widehat{Z}^{[m]}$ as follows:
\begin{align}
     \widehat{Z} = \begin{bmatrix}
        \sqrt{T_1}(\muhat_Y-\mu_Y)\\
        \sqrt{T_1}(\muhat-\mu)\\
        \sqrt{T_0}(\vecl(\Sigmahat-\Sigma))\\
        \sqrt{T_0}(\gammahat-\gamma)
    \end{bmatrix},\quad  \widehat{Z}^{[m]} = \begin{bmatrix}
        \sqrt{T_1}(\muhat_Y-\muhatm_Y)\\
        \sqrt{T_1}(\muhat-\muhatm)\\
        \sqrt{T_0}(\vecl(\Sigmahat-\Sigmahatm))\\
        \sqrt{T_0}(\gammahat-\gammahatm)
    \end{bmatrix}.\label{eq: Z and Zm}
\end{align}
With the above notation, we define the following event:
\begin{align}
    \mathcal{E}=\left\{\min_{m\in\mathbb{M}}\left\|\widehat{Z}-\widehat{Z}^{[m]}\right\|_{\infty}
    \leq \err\right\},\label{event: m star}
\end{align}
where $\err$ is defined in \eqref{def: err}. 
On the event $\mathcal{E}$, since
\begin{align*}
\widehat{Z}-\widehat{Z}^{[m]} = \begin{bmatrix}
        \sqrt{T_1}(\muhatm_Y-\mu_Y)\\
        \sqrt{T_1}(\muhatm-\mu)\\
        \sqrt{T_0}(\vecl(\Sigmahatm-\Sigma))\\
        \sqrt{T_0}(\gammahatm-\gamma)
    \end{bmatrix},
\end{align*}
there exists an index $m = m^* \in \mathbb{M}$ such that the perturbed quantities, $\Sigmahatmstar$, $\gammahatmstar$, $\muhatmstar_Y$, and $\muhatmstar$, which are generated from the distribution \eqref{eq: sampling of sigma and gamma} and \eqref{eq: sampling of mu Y and mu 1}, nearly recover the corresponding population parameters, $\Sigma$, $\gamma$, $\mu_Y$, and $\mu$, respectively. 

We present the following proposition, whose proof is provided in Section~\ref{pf: m star}. 
 \begin{proposition}\label{thm: m star} Suppose Assumption~\ref{assumption: clt and cons} holds. 
 Then, for a prespecified constant $\alpha_0\in(0,0.01]$, $$\liminf_{T_0,T_1\rightarrow\infty}\liminf_{M\rightarrow\infty}\prob\left(\mathcal{E}\right)\geq 1-\alpha_0,$$ where the event $\mathcal{E}$ is defined in \eqref{event: m star}.
 \end{proposition}
Proposition~\ref{thm: m star} ensures that the event $\mathcal{E}$ occurs with high probability for small $\alpha_0$, which provides the theoretical foundation of near equivalence between \eqref{eq: m-th betahat} with $m=m^*$ and the population version of the optimization problem in \eqref{eq: tau star and beta star}.

Now we turn to the main proof of Theorem~\ref{thm: negligible}. 
 Note that
 \begin{equation}\label{eq: mubeta bound}
     \begin{aligned}
         \left|(\muhatm)\tr\betahatm-\mu\tr\beta^{*}\right| &= \left|(\muhatm-\mu)\tr\beta^*+(\muhatm)\tr(\beta^*-\betahatm)\right|\\
    &\leq \left|(\muhatm-\mu)\tr\beta^*\right|+\left|(\muhatm)\tr(\beta^*-\betahatm)\right|\\
    &\leq \left\|\muhatm-\mu\right\|+\left|(\muhatm)\tr(\beta^*-\betahatm)\right|\\
    &\lesssim \max\left\{\left\|\muhatm-\mu\right\|,\left|(\muhatm)\tr(\beta^*-\betahatm)\right|\right\}.
     \end{aligned}
 \end{equation}
For $\lambda\geq 0$, we define the set $\mathcal{E}^{**}$ where 
\begin{equation}\label{event: negligible2}
\begin{aligned}
    \lambda>0:\;&\mathcal{E}^{**} = \left\{\min_{m\in\mathbb{M}}\max\left\{\|\muhatm-\mu\|_{\infty},|(\muhatm)\tr(\beta^*-\betahatm)|\right\}\lesssim \left[\frac{\err}{\min\{\lambda\sqrt{T_0},\sqrt{T_1}\}}\right]^{\frac{1}{2}}\right\}, \\
    \lambda=0:\;&\mathcal{E}^{**} =\left\{\min_{m\in\mathbb{M}}\max\left\{\|\muhatm-\mu\|_{\infty},|(\muhatm)\tr(\beta^*-\betahatm)|\right\}\lesssim \left[\frac{\err}{\min\{\lambda_{\min}(\Sigma)\sqrt{T_0/N},\sqrt{T_1}\}}\right]^{\frac{1}{2}}\right\}.
\end{aligned}
\end{equation}
For given $\lambda\geq0$, we ensure that $\mathcal{E}^{**}\subset \mathcal{E}^*$ by \eqref{eq: mubeta bound}, where the event $\mathcal{E}^*$ is defined in \eqref{event: negligible}.
Specifically, on the event $\mathcal{E}^{**}$ defined in \eqref{event: negligible2}, there exists $m = m^*$ such that we can simultaneously control $\muhatm-\mu$ and $(\muhatm)\tr(\beta^*-\betahatm)$. Thus, the term $(\muhatm)\tr\betahatm-\mu\tr\beta^{*}$ can also be controlled.
 
 By Proposition~\ref{thm: m star}, on the event $\mathcal{E}$, there exists $m^*$ such that $\|\muhatmstar-\mu\|_{\infty}\leq \err/\sqrt{T_1}$. For given $T_0$ and $T_1$, since $\lim_{M\to\infty}\err = 0$, we have
 $$\|\muhatmstar-\mu\|_{\infty}^2\lesssim (\err/\sqrt{T_1})^2 \lesssim \err/\sqrt{T_1}.$$

Now we provide the proof of the control of $(\muhatm)\tr(\beta^*-\betahatm)$. In the following proof, we establish that on the event $\mathcal{E}$, such $m^*$ satisfies $|(\muhatmstar)\tr(\betahatmstar-\beta^*)|$
can be negligible with sufficiently large $M$.
To facilitate the proof, we define the following quantities for $m=m^*$ with slightly abusing the notation:
\begin{align}\label{eq: ghat with Omegahatm m}
    &\tildebetamstar {\in} \argmin_{\beta \in \Omega}\ghatmstar(\beta),\quad\barbetamstar {\in} \argmin_{\beta\in\Omegahatmstar\cap \Omega}\ghatmstar(\beta),\quad\text{with}\quad \ghatmstar(\beta)=\beta\tr\Gammahatmstar\beta-2(\deltahatmstar)\tr\beta,
\end{align}
where $\Omegahatmstar$ is defined in \eqref{eq: m-th omega}, $\Gammahatmstar = \muhatmstar(\muhatmstar)\tr$, and $\deltahatmstar = \muhatmstar\muhatmstar_Y$. We remark that the quantities $\tildebetamstar$ and $\barbetamstar$ defined above are distinct from those in Section~\ref{pf: precision}. 

Since $\beta^{(0)}\in\Omega$, $\Omega$ is non-empty thus $\tildebetamstar$ exists. To show that $\Omegahatmstar\cap \Omega$ is non-empty on the event $\mathcal{E}$ in \eqref{event: m star}, we introduce the following lemma, whose proof is provided in Section~\ref{pf: fhatmstar f}.
\begin{Lemma}\label{lem: fhatmstar f}
    On the event $\mathcal{E}$ defined in \eqref{event: m star}, for the index $m^*$ satisfying \eqref{eq: m star}, we have
    \begin{align}
        &\sup_{\beta\in\Delta^N}|\fhatmstar(\beta)-f(\beta)| \leq \frac{2\err}{\sqrt{T_0}}, \label{eq: fhatm conv}
    \end{align}
    where $\fhatmstar$ is defined in \eqref{def: fhatm} and $\err$ is defined in \eqref{def: err}. If $\rho_M$ used in \eqref{eq: m-th omega} satisfies $\rho_M=C_1[\log(\min\{T_0,T_1\})/M]^{1/p}/\sqrt{T_0}$ for some positive constant $C_1>0$ which makes $\rho_M\geq 2\err/\sqrt{T_0}$, then $\fhatmstar(\beta^{(0)}) \leq -\lambda+ \rho_M$. 
\end{Lemma}
Since $\fhatmstar(\beta^{(0)})\leq-\lambda+\rho_M\leq \rho_M$ on the event $\mathcal{E}$ and $f(\beta^{(0)})\leq 0$, $\beta^{(0)}\in\Omegahatmstar\cap\Omega$ on the event $\mathcal{E}$, which ensures that $\Omegahatmstar\cap\Omega$ is non-empty. Thus, on the same event $\mathcal{E}$, $\barbetamstar$ exists by Lemma~\ref{lem: fhatmstar f} with $\rho_M$ satisfying the conditions in Lemma~\ref{lem: fhatmstar f}.

We decompose $|(\muhatm)\tr(\widehat{\beta}^{[m]}-\beta^*)|$ by triangle inequality as follows:
\begin{align*}
    |(\muhatmstar)\tr(\betahatmstar-\beta^*)| &= |(\muhatmstar)\tr(\tildebetamstar-\beta^*)+(\muhatmstar)\tr(\barbetamstar-\tildebetamstar)+(\muhatmstar)\tr(\betahatmstar-\barbetamstar)|\\
    &\leq \underbrace{|(\muhatmstar)\tr(\tildebetamstar-\beta^*)|}_{(i)}+\underbrace{|(\muhatmstar)\tr(\barbetamstar-\tildebetamstar)|}_{(ii)}+\underbrace{|(\muhatmstar)\tr(\betahatmstar-\barbetamstar)|}_{(iii)}
\end{align*}
In the following, we shall control each of the three terms separately.\\

To control the first term, define the population quadratic objective
\begin{align}
    g(\beta)=\beta\tr\Gamma\beta-2\delta\tr\beta,
    \quad \Gamma=\mu\mu\tr,
    \quad \delta=\mu\mu_Y.
    \label{eq: def of g tilde bar}
\end{align}
Then $\beta^*{\in}\argmin_{\beta\in\Omega}g(\beta)$. We use the following comparison of minimizers under two quadratic objectives; its proof is provided in Section~\ref{pf: w conv}.
\begin{Lemma}\label{lem: w conv}
For a convex set $\mathcal{H}\subset\Delta^N$ and symmetric matrices $A$ and $\widehat A$, define
\begin{align*}
    w^*{\in}\argmin_{w\in\mathcal H}\{w\tr Aw-2a\tr w\},
    \quad
    \widehat w{\in}\argmin_{w\in\mathcal H}\{w\tr\widehat A w-2\widehat a\tr w\}.
\end{align*}
Then
\begin{align*}
    (w^*-\widehat w)\tr A(w^*-\widehat w)
    \leq\|(a-\widehat a)+(\widehat A-A)\widehat w\|_\infty
    \|w^*-\widehat w\|_1.
\end{align*}
\end{Lemma}

\noindent
\underline{\textbf{Control of} $(i)$.} Applying Lemma~\ref{lem: w conv} to the population and perturbed objectives over $\Omega$, we have
\begin{align*}
    \left|(\muhatmstar)\tr(\tildebetamstar-\beta^*)\right|^2 &= \left|(\tildebetamstar-\beta^*)\tr\Gammahatmstar(\tildebetamstar-\beta^*)\right|\\
    &\leq 2(\|\deltahatmstar-\delta\|_{\infty}+\|\Gammahatmstar-\Gamma\|_{\infty}).
\end{align*}
By the triangle inequality, direct calculation shows that
\begin{align*}
    \|\deltahatmstar-\delta\|_{\infty} &= \|\muhatmstar\muhatmstar_Y-\mu\mu_Y\|_{\infty} = \|(\muhatmstar-\mu)\muhatmstar_Y+\mu(\muhatmstar_Y-\mu_Y)\|_{\infty}\\
    &\leq(|\mu_Y|+|\muhatmstar_Y-\mu_Y|)\|\muhatmstar-\mu\|_{\infty}+\|\mu\|_{\infty}|\muhatmstar_Y-\mu_Y|\\
    &\leq(|\mu_Y|+\|\mu\|_{\infty}+|\muhatmstar_Y-\mu_Y|)\max\{\|\muhatmstar-\mu\|_{\infty},|\muhatmstar_Y-\mu_Y|\},\\
     \|\Gammahatmstar-\Gamma\|_{\infty} &= \|\muhatmstar(\muhatmstar)\tr-\mu\mu\tr\|_{\infty} = \|\muhatmstar(\muhatmstar-\mu)\tr+(\muhatmstar-\mu)\mu\tr\|_{\infty}\\
     &\leq(2\|\mu\|_{\infty}+\|\muhatmstar-\mu\|_{\infty})\|\muhatmstar-\mu\|_{\infty},
    \end{align*}
and thus $ \left|(\muhatmstar)\tr(\tildebetamstar-\beta^*)\right|^2$ is further bounded as
\begin{align*}
     &\left|(\muhatmstar)\tr(\tildebetamstar-\beta^*)\right|^2\\
     &\leq 2(|
     \mu_Y|+3\|\mu\|_{\infty}+|\muhatmstar_Y-\mu_Y|+\|\muhatmstar-\mu\|_{\infty})\max\{\|\muhatmstar-\mu\|_{\infty},|\muhatmstar_Y-\mu_Y|\}\\
     &\leq4\left[\max\{\|\muhatmstar-\mu\|_{\infty},|\muhatmstar_Y-\mu_Y|\}\right]^2+2(|
     \mu_Y|+3\|\mu\|_{\infty})\max\{\|\muhatmstar-\mu\|_{\infty},|\muhatmstar_Y-\mu_Y|\}.
\end{align*}

On the event $\mathcal{E}$ defined in \eqref{event: m star}, we denote $m=m^*$ to denote the index of perturbation such that
\begin{align}\label{eq: m star}
    & \left\|\widehat{Z}-\widehat{Z}^{[m^*]}\right\|_{\infty}\leq \err,
\end{align}  
where $\widehat{Z}$ and $\widehat{Z}^{[m]}$ are defined in \eqref{eq: Z and Zm}.
For such $m = m^*$ satisfying \eqref{eq: m star}, we ensure
\begin{align*}
    \max\{\|\muhatmstar-\mu\|_{\infty},|\muhatmstar_Y-\mu_Y|\} \leq \frac{\err}{\sqrt{T_1}}.
\end{align*}
By \eqref{eq: outcome model}, $\mu_Y=\mu\tr\beta^{(1)}+\taubar$, so
$|\mu_Y|\leq\|\mu\|_{\infty}+|\taubar|$ is bounded under
Assumption~\ref{assumption: post}. Hence, on the same event $\mathcal{E}$,
we establish the following:
\begin{align}\label{eq: control i}
    \left|(\muhatmstar)\tr(\tildebetamstar-\beta^*)\right|^2 \lesssim \frac{\err}{\sqrt{T_1}}.
\end{align}
\\

\noindent
\underline{\textbf{Control of} $(ii)$.}
We use the following comparison between a quadratic minimum over a convex set and the minimum over its intersection with another convex set; its proof is provided in Section~\ref{pf: g eq}.
\begin{Lemma}\label{lem: g eq}
For convex sets $\mathcal H$ and $\widehat{\mathcal H}$, a positive semidefinite matrix $A$, and a vector $a$, define
\begin{align*}
    \widetilde w{\in}\argmin_{w\in\mathcal H}h(w),
    \quad
    \overline w{\in}\argmin_{w\in\mathcal H\cap\widehat{\mathcal H}}h(w),
    \quad h(w)=w\tr Aw-2a\tr w.
\end{align*}
Then $(\overline w-\widetilde w)\tr A(\overline w-\widetilde w)\leq h(\overline w)-h(\widetilde w)$.
\end{Lemma}
By taking $\mathcal{H} = \Omega$ in \eqref{def: Omega}, $\widehat{\mathcal{H}} = \Omegahatm$ in \eqref{eq: m-th omega}, and $h = \ghatm$ in \eqref{eq: ghat with Omegahatm m} in Lemma~\ref{lem: g eq}, we ensure 
\begin{align*}
     &|(\muhatmstar)\tr(\barbetamstar-\tildebetamstar)|^2\leq \ghatmstar(\barbetamstar)-\ghatmstar(\tildebetamstar).
\end{align*}

In the following, we consider two separate cases: $(i)$ $\lambda>0$; $(ii)$ $\lambda=0$.
\\

\noindent
\underline{$(i)$ Case of $\lambda>0$.}
To compare quadratic optimal values over the population and perturbed classes, we first use the following convex-set comparison; its proof is provided in Section~\ref{pf: control of obj}.
\begin{Lemma}\label{lem: control of obj}
For convex sets $\mathcal{H}$ and $\mathcal{H}'$ contained in $\Delta^N$ for fixed $N$, a positive semidefinite matrix $A$, and a vector $a$, define
\begin{align*}
    \widetilde w{\in}\argmin_{w\in\mathcal H}h(w),
    \quad
    \overline w{\in}\argmin_{w\in\mathcal H\cap\mathcal H'}h(w),
    \quad h(w)=w\tr Aw-2a\tr w.
\end{align*}
For convex functions $e,e':\Delta^N\to\mathbb R$ and nonnegative constants $c,c'$, suppose
\begin{align*}
    w\in\mathcal H\iff e(w)\leq c,
    \quad
    w\in\mathcal H'\iff e'(w)\leq c',
\end{align*}
and that there exists $w^0\in\mathcal H$ such that $e'(w^0)<c'$. Then
\begin{align*}
    h(\overline w)-h(\widetilde w)
    \leq4(\|A\|_\infty+\|a\|_\infty)
    \frac{\sup_{w\in\Delta^N}|e'(w)-e(w)|+\max\{c-c',0\}}
    {c'-e'(w^0)}.
\end{align*}
\end{Lemma}

Applying this comparison to $\ghatmstar$ gives the following bounds, proved in Section~\ref{pf: ghatm fhatm}.

\begin{Lemma}\label{lem: ghatm fhatm}
Suppose $\lambda>0$ and the tuning parameter $\rho_M = C_1[\log(\min\{T_0,T_1\})/M]^{1/p}/\sqrt{T_0}$ with some positive constant $C_1$ which satisfies $C_1\geq (2/c^*(\alpha_0))^{1/p}$ where $c^*(\alpha_0)$ in \eqref{def: err} with $\alpha_0\in(0,0.01]$, a pre-specified constant used to define $\mathbb{M}$ in \eqref{eq: index set}. On the event $\mathcal{E}$ defined in \eqref{event: m star}, let $m^*$ denote the index of perturbation satisfying \eqref{eq: m star}. Then, the following holds:
    \begin{align}
&\ghatmstar(\barbetamstar)-\ghatmstar(\tildebetamstar) \lesssim \frac{\left(\log(\min\{T_0,T_1\})/{M}\right)^{1/p}}{\lambda\sqrt{T_0}} ,\label{eq: ghatm bar tilde}\\
        & \ghatmstar(\barbetamstar)-\ghatmstar(\betahatmstar) \lesssim \frac{\left(\log(\min\{T_0,T_1\})/{M}\right)^{1/p}}{\lambda\sqrt{T_0}}.\label{eq: ghatm hat bar}
    \end{align}
    where $p = 1+N(N+5)/2$.
\end{Lemma}

On the event $\mathcal{E}$, if we denote $m^*$ as the index satisfying \eqref{eq: m star}, Lemma~\ref{lem: ghatm fhatm} ensures that 
\begin{align}\label{eq: control ii}
    |(\muhatmstar)\tr(\barbetamstar-\tildebetamstar)|^2\leq  \ghatmstar(\barbetamstar)-\ghatmstar(\tildebetamstar)\lesssim \frac{\err}{\lambda\sqrt{T_0}},
\end{align}
since $\err \asymp \left(\log(\min\{T_0,T_1\})/{M}\right)^{1/p}$.\\

\noindent
\underline{$(ii)$ Case of $\lambda=0$.}
When $\lambda=0$, we assume $\lambda_{\min}(\Sigma)>0$. Then $\Omega=\{\beta^{(0)}\}$, so $\tildebetamstar=\barbetamstar=\beta^{(0)}$ and
\begin{align}
    |(\muhatmstar)\tr(\barbetamstar-\tildebetamstar)|^2\leq\ghatmstar(\barbetamstar)-\ghatmstar(\tildebetamstar) =0.\label{eq: control ii lambda0}
\end{align}

\noindent
\underline{\textbf{Control of} $(iii)$.} Similar to Control of $(ii)$, we consider two cases: $(i)$ $\lambda>0$; $(ii)$ $\lambda=0$.
\\
\noindent
\underline{$(i)$ Case of $\lambda>0$.}
Because we can exchange $\Omega$ and $\Omegahatm$ in the use of Lemma~\ref{lem: ghatm fhatm} for the control of $(ii)$, we establish the similar inequality on the event $\mathcal{E}$ as follows: For $m^*$ satisfying \eqref{eq: m star},
\begin{align}\label{eq: control iii}
     |(\muhatmstar)\tr(\barbetamstar-\betahatmstar)|^2\lesssim \frac{\err}{\lambda\sqrt{T_0}}.
\end{align}
\\
\noindent
\underline{$(ii)$ Case of $\lambda=0$.}
To control this case, we use the following bound, whose proof is provided in Section~\ref{pf: control lambda0}.
\begin{Lemma}\label{lem: control lambda0}
Let $\lambda=0$. Suppose $\lambda_{\min}(\Sigma)>0$. For a convex and differentiable function $h:\Delta^N\to\mathbb R$ and $\beta\in\Delta^N$,
\begin{align}
    h(\beta^{(0)})-h(\beta)
    \leq\|\nabla h(\beta^{(0)})\|_1
    \frac{\sqrt N f(\beta)}{\lambda_{\min}(\Sigma)}.
    \label{eq: f lambda 0}
\end{align}
\end{Lemma}
Since $\Omega = \{\beta^{(0)}\}$ when $\lambda=0$ under $\lambda_{\min}(\Sigma)>0$, $\barbetamstar=\beta^* = \beta^{(0)}$. Thus, by Lemma~\ref{lem: control lambda0}, we ensure $$ |(\muhatmstar)\tr(\barbetamstar-\betahatmstar)|^2\leq\ghatmstar(\barbetamstar)-\ghatmstar(\betahatmstar) \leq  \|\nabla \ghatmstar(\beta^{(0)})\|_{1}\frac{\sqrt{N}f(\betahatmstar)}{\lambda_{\min}(\Sigma)}.$$
 On the event $\mathcal{E}$ defined in \eqref{event: m star}, we have $$\| \nabla \ghatmstar(\beta^{(0)})\|_{1} \lesssim \|\Gamma\|_{\infty}+\|\delta\|_{\infty}+\err/\sqrt{T_1}.$$
By Lemma~\ref{lem: fhatmstar f}, on the event $\mathcal{E}$, we have
\begin{align*}
    f(\betahatmstar) \leq \fhatmstar(\betahatmstar)+2\err/\sqrt{T_0}\leq 2\rho_M.
\end{align*}
since $\fhatmstar(\betahatmstar)\leq \rho_M$ and $2\err/\sqrt{T_0}\leq \rho_M$. Thus, on the event $\mathcal{E}$, since $\rho_M\asymp \err/\sqrt{T_0}$, we establish that
\begin{align}
    &|(\muhatmstar)\tr(\barbetamstar-\betahatmstar)|^2\leq \ghatmstar(\beta^{(0)})-\ghatmstar(\betahatmstar) \lesssim \frac{\sqrt{N}\err}{\lambda_{\min}(\Sigma)\sqrt{T_0}},\label{eq: control iii lambda0}
\end{align}
by Lemma~\ref{lem: control lambda0}.

For $\lambda>0$, on the event $\mathcal{E}$ in \eqref{event: m star} with $m^*$ satisfying \eqref{eq: m star}, since
\begin{align*}
     &|(\muhatmstar)\tr(\betahatmstar-\beta^*)| \\
     &\lesssim \max\left\{|(\muhatmstar)\tr(\betahatmstar-\tildebetamstar)|,|(\muhatmstar)\tr(\tildebetamstar-\barbetamstar)|,|(\muhatmstar)\tr(\barbetamstar-\beta^*)|\right\},
\end{align*}
by combining \eqref{eq: control i}, \eqref{eq: control ii}, and \eqref{eq: control iii}, we establish that 
\begin{align*}
    |(\muhatmstar)\tr(\betahatmstar-\beta^*)|
    &\lesssim \left[\frac{\err}{\min\{\lambda\sqrt{T_0},\sqrt{T_1}\}}\right]^{\frac{1}{2}}.
\end{align*}
for sufficiently large $M$, since $a^{-1}+b^{-1}\lesssim (\min\{a,b\})^{-1}$ for $a,b>0$.

For $\lambda=0$ with $\lambda_{\min}(\Sigma)>0$, on the same event $\mathcal{E}$, by combining \eqref{eq: control i}, \eqref{eq: control ii lambda0}, and \eqref{eq: control iii lambda0}, we have
\begin{align*}
    |(\muhatmstar)\tr(\betahatmstar-\beta^*)|
    &\lesssim \left[\frac{\err}{\min\{\lambda_{\min}(\Sigma)\sqrt{T_0/N},\sqrt{T_1}\}}\right]^{\frac{1}{2}}
\end{align*}
for sufficiently large $M$.

To summarize, we showed that $\prob(\mathcal{E}^{**}) \geq \prob(\mathcal{E})$, and this implies that $\prob(\mathcal{E}^{*})\geq \prob(\mathcal{E}^{**}) \geq \prob(\mathcal{E})$. This concludes the proof by Proposition~\ref{thm: m star}.

\subsection{Proof of Proposition~\ref{thm: m star}}\label{pf: m star}
We give the outline of the proof. Using the asymptotic normality, we establish an event $\mathcal{E}$ defined on the observed data, denoted by $\mathcal{O}$, which occurs with high probability. Next, on the event $\mathcal{E}$, we study the behavior of the perturbed quantities. Finally, we combine the results and establish the desired argument.

Now we provide the proof. We recall from \eqref{eq: Z and Zm} that 
\begin{align*}
     \widehat{Z} = \begin{bmatrix}
        \sqrt{T_1}(\muhat_Y-\mu_Y)\\
        \sqrt{T_1}(\muhat-\mu)\\
        \sqrt{T_0}(\vecl(\Sigmahat-\Sigma))\\
        \sqrt{T_0}(\gammahat-\gamma)
    \end{bmatrix},\quad  \widehat{Z}^{[m]} = \begin{bmatrix}
        \sqrt{T_1}(\muhat_Y-\muhatm_Y)\\
        \sqrt{T_1}(\muhat-\muhatm)\\
        \sqrt{T_0}(\vecl(\Sigmahat-\Sigmahatm))\\
        \sqrt{T_0}(\gammahat-\gammahatm)
    \end{bmatrix}.
\end{align*}

We define diagonal matrices by taking diagonal elements of ignoring off-diagonal covariance terms: 
\begin{equation*}
    \begin{aligned}
        \mathbf{V} &= \text{Diag}\left(\VY,{\rm diag}(\Vmu),{\rm diag}(\VSigma),{\rm diag}(\Vgamma)\right),\\
        {\rm \bf{Cov}} &= \text{Diag}\left(T_1\VY,{\rm diag}(T_1\Vmu),{\rm diag}(T_0\VSigma),{\rm diag}(T_0\Vgamma)\right),
    \end{aligned}
\end{equation*}
where $\text{Diag}(v)$ denotes the diagonal matrix with vector $v$ on its diagonal, and $\text{diag}(A)$ denotes the vector of diagonal elements of a matrix $A$.
Recall from Assumption~\ref{assumption: clt and cons} that we have
\begin{align*}
    &{\rm V}^{-1/2}_Y\left(\muhat_Y-\mu_Y\right)\dto\mathcal{N}(0,1),\quad \mathbf{V}_{\mu}^{-1/2} \left(\muhat-\mu\right)\dto\mathcal{N}(0,\mathbf{I}) \quad \text{as}\quad T_1\rightarrow \infty,\quad\text{and}\\
    &\mathbf{V}_{\Sigma}^{-1/2} \left(\vecl(\Sigmahat)-\vecl(\Sigma)\right)\dto\mathcal{N}(0,\mathbf{I}),\quad \mathbf{V}_{\gamma}^{-1/2}\left(\gammahat-\gamma\right)\dto\mathcal{N}(0,\mathbf{I}),\quad \text{as}\quad T_0\rightarrow \infty.
\end{align*}
By the union bound, we ensure
\begin{align}    \label{eq: union bound Zhat}\lim_{T_0,T_1\rightarrow\infty}\prob\left(\|{\rm \bf{Cov}}^{-1/2}\widehat{Z}\|_{\infty} \leq   z_{\alpha_0/(2p)} \right)\geq 1-\alpha_0,
\end{align}
for arbitrary $\alpha_0 \in (0,0.01]$. 

We define the event
$\mathcal{G}_0=\mathcal{G}_3\cap\mathcal{G}_4
\cap\mathcal{G}_5\cap\mathcal{G}_6$,
where $\mathcal{G}_3$ and $\mathcal{G}_4$ are defined in
\eqref{event: consistency}, and $\mathcal{G}_5$ and
$\mathcal{G}_6$ are defined as follows:
\begin{equation}
        \label{event: Covhat Zhat max}
    \begin{aligned}
         \mathcal{G}_5 &= \left\{\|{\rm \widehat{\bf{Cov}}}^{-1/2}\widehat{Z}\|_{\infty} \leq 1.05  z_{\alpha_0/(2p)}\right\},\\
          \mathcal{G}_6 &= \left\{\VY \leq 1.05^2\VhatY,\ \left\|\mathbf{V}_{\mu}-\Vhatmu\right\|_{\max} \leq d_{\mu},\ \left\|\mathbf{V}_\Sigma-\Vhat_\Sigma\right\|_{\max} \leq d_{\Sigma},\ \left\|\mathbf{V}_\gamma-\Vhat_\gamma\right\|_{\max} \leq d_{\gamma}\right\},
    \end{aligned}
\end{equation}
and ${\rm \widehat{\bf{Cov}}}$ is a diagonal matrix defined as follows:
\begin{equation}\label{eq: Covhat}
    {\rm \widehat{\bf{Cov}}}= \text{Diag}\left(T_1\VhatY,\text{diag}(T_1\Vhatmu+T_1d_{\mu}\mathbf{I}),\text{diag}(T_0\VhatSigma+T_0d_{\Sigma}\mathbf{I}),\text{diag}(T_0\Vhatgamma+T_0d_{\gamma}\mathbf{I})\right).
\end{equation}
with $d_{\mu}=\|\Vhatmu\|_{\max}$, $d_{\Sigma} = \|\VhatSigma\|_{\max}$, and $d_{\gamma}=\|\Vhatgamma\|_{\max}$. 
We introduce the following lemma to control the probability of $\mathcal{G}_6$, whose proof is provided in Section~\ref{pf: V and Vhat}.
\begin{Lemma}\label{lem: V and Vhat}
    Under Assumption~\ref{assumption: clt and cons}, the events $\mathcal{G}_6$ satisfies $\lim_{T_0,T_1\rightarrow\infty}\prob(\mathcal{G}_6)=1$.
\end{Lemma}

On $\mathcal{G}_6$ defined in \eqref{event: Covhat Zhat max}, 
\begin{equation*}
    \|{\rm \widehat{\bf{Cov}}}^{-1/2}{\rm \bf{Cov}}^{1/2}\|_{\max} = \max_{1\leq j \leq p}\sqrt{\frac{{\bf Cov}_{j,j}}{\widehat{\bf{Cov}}_{j,j}}}\leq 1.05.
\end{equation*}

This ensures 
\begin{align}
    \|{\rm \widehat{\bf{Cov}}}^{-1/2}\widehat{Z}\|_{\infty} \leq \|{\rm \widehat{\bf{Cov}}}^{-1/2}{\rm \bf{Cov}}^{1/2}{\rm \bf{Cov}}^{-1/2}\widehat{Z}\|_{\infty} \leq 1.05\|{\rm \bf{Cov}}^{-1/2}\widehat{Z}\|_{\infty}.    \label{eq: Covhat Cov comparison}
\end{align}
Let
\[
\mathcal A_0=
\left\{\|{\rm \bf Cov}^{-1/2}\widehat Z\|_\infty
\leq z_{\alpha_0/(2p)}\right\}.
\]
By \eqref{eq: Covhat Cov comparison},
$\mathcal A_0\cap\mathcal G_6\subseteq\mathcal G_5$.
Therefore,
\begin{align}
\liminf_{T_0,T_1\to\infty}\prob(\mathcal G_0)
&\geq
\liminf_{T_0,T_1\to\infty}\prob(\mathcal A_0)
-\limsup_{T_0,T_1\to\infty}
\sum_{j=3,4,6}\prob(\mathcal G_j^c) \geq 1-\alpha_0,
\label{eq: prob E0}
\end{align}
where the last inequality follows from
\eqref{eq: union bound Zhat},
Lemma~\ref{lem: cons of cov}, and
Lemma~\ref{lem: V and Vhat}.

In the following, we shall show 
\begin{align}\label{eq: E star and E}
\liminf_{T_0,T_1\rightarrow\infty}\prob\left(\mathcal{E}\right)\geq\liminf_{T_0,T_1\rightarrow\infty}\prob\left(\mathcal{G}_0\right),
\end{align}
where $\mathcal{E}$ is defined in \eqref{event: m star}.
Then by combining \eqref{eq: prob E0} and \eqref{eq: E star and E},  we establish
\begin{align*}
\liminf_{T_0,T_1\rightarrow\infty}\prob\left(\mathcal{E}\right)  \geq 1-\alpha_0,
\end{align*}
and this will conclude the proof.

{
We denote the observed data by $\mathcal O$. The conditional covariance matrix of $\widehat Z^{[m]}$ given $\mathcal O$ is the block-diagonal matrix
\begin{equation*}
\begin{aligned}
\widehat{\bf K}=\operatorname{blockdiag}\{&T_1\VhatY,\ T_1(\Vhatmu+d_\mu\mathbf I_N),\ T_0(\VhatSigma+d_\Sigma\mathbf I_q),\ T_0(\Vhatgamma+d_\gamma\mathbf I_N)\}.
\end{aligned}
\end{equation*}
where $q=N(N+1)/2$ denotes the dimension corresponding to $\Sigma$. The diagonal matrix $\widehat{\bf Cov}$ in \eqref{eq: Covhat}, used to normalize the filtering statistics, satisfies $\widehat{\bf Cov}=\operatorname{Diag}\{\operatorname{diag}(\widehat{\bf K})\}$. Let $f(\cdot\mid\mathcal O)$ denote the conditional density of $\widehat Z^{[m]}$. By the perturbation mechanism,
\begin{equation*}
f(\widehat Z^{[m]}=Z\mid\mathcal O)
=\frac{1}{\sqrt{(2\pi)^p\det(\widehat{\bf K})}}
\exp\left(-\frac{Z\tr\widehat{\bf K}^{-1}Z}{2}\right).
\end{equation*}
On $\mathcal G_0$, \eqref{eq: max cov}, \eqref{event: consistency}, and the definition of $\mathcal G_6$ imply
\begin{equation*}
\begin{aligned}
T_1\VhatY&\leq2c_1,\\
T_1\{(\Vhatmu)_{j,j}+d_\mu\}
&\leq T_1(\Vmu)_{j,j}+2T_1d_\mu\leq5c_1,\\
T_0\{(\VhatSigma)_{j,j}+d_\Sigma\}
&\leq T_0(\VSigma)_{j,j}+2T_0d_\Sigma\leq5c_1,\\
T_0\{(\Vhatgamma)_{j,j}+d_\gamma\}
&\leq T_0(\Vgamma)_{j,j}+2T_0d_\gamma\leq5c_1.
\end{aligned}
\end{equation*}
Assumption~\ref{assumption: clt and cons} also ensures that there exists a positive constant $c_2$ such that
\begin{equation}
\min\left\{T_1\VY,\min_{1\leq j\leq N}(T_1\Vmu)_{j,j},
\min_{1\leq j\leq q}(T_0\VSigma)_{j,j},
\min_{1\leq j\leq N}(T_0\Vgamma)_{j,j}\right\}\geq c_2.
\label{eq: min cov}
\end{equation}
Because the covariance estimators in the perturbation mechanism are positive semidefinite, each inflated block has smallest eigenvalue at least its diagonal inflation. On $\mathcal G_6$, the lower bound in \eqref{eq: min cov} and $\|\Vhatmu\|_{\max}=d_\mu$ imply
\begin{equation*}
c_2/T_1\leq(\Vmu)_{j,j}\leq |(\Vhatmu)_{j,j}|+d_\mu\leq2d_\mu,
\end{equation*}
for every $j$, and the same argument applies to the $\Sigma$ and $\gamma$ blocks. Together with $T_1\VhatY\geq T_1\VY/1.05^2$, this gives fixed constants $0<c_{\bf K}\leq C_{\bf K}<\infty$ such that
\begin{equation}
c_{\bf K}\leq\lambda_{\min}(\widehat{\bf K})
\leq\lambda_{\max}(\widehat{\bf K})\leq C_{\bf K},
\label{eq: K eigenvalue bounds}
\end{equation}
on $\mathcal G_0$ for fixed $N$. The upper bound follows from the preceding entrywise bounds and the fixed dimensions of the blocks.

Since $\mathcal G_0\subseteq\mathcal G_5$ and every diagonal element of $\widehat{\bf Cov}$ is at most $C_{\bf K}$,
\begin{equation*}
\begin{aligned}
\widehat Z\tr\widehat{\bf K}^{-1}\widehat Z
&\leq c_{\bf K}^{-1}\|\widehat Z\|_2^2
\leq c_{\bf K}^{-1}pC_{\bf K}(1.05z_{\alpha_0/(2p)})^2
\eqqcolon C_Q.
\end{aligned}
\end{equation*}
By \eqref{eq: K eigenvalue bounds}, $\det(\widehat{\bf K})\leq C_{\bf K}^p$. Define
\begin{equation}
\begin{aligned}
\kappa_N&=2^{-(p-q)}(4N)^{-q},\\
c_{\rm den}(\alpha_0)
&=(2\pi)^{-p/2}C_{\bf K}^{-p/2}\exp(-C_Q/2),\\
c^*(\alpha_0)&=\kappa_Nc_{\rm den}(\alpha_0).
\end{aligned}
\label{eq: c star alpha 0}
\end{equation}
The preceding bounds give
\begin{equation}
f(\widehat Z^{[m]}=\widehat Z\mid\mathcal O)
\I(\mathcal O\in\mathcal G_0)
\geq c_{\rm den}(\alpha_0)\I(\mathcal O\in\mathcal G_0).
\label{eq:term 1 bound}
\end{equation}
Both $c_{\rm den}(\alpha_0)$ and $c^*(\alpha_0)$ are fixed positive constants for fixed $N$.
Since $\nabla f(z)=-f(z)\widehat{\bf K}^{-1}z$, setting $u=z\tr\widehat{\bf K}^{-1}z$ and using \eqref{eq: K eigenvalue bounds} gives
\begin{equation*}
\begin{aligned}
\|\nabla f(z)\|_2^2
&=\frac{z\tr\widehat{\bf K}^{-2}z}{(2\pi)^p\det(\widehat{\bf K})}
\exp\left(-z\tr\widehat{\bf K}^{-1}z\right)\\
&\leq\frac{1}{(2\pi)^pc_{\bf K}^{p+1}}u\exp(-u).
\end{aligned}
\end{equation*}
Because $u\exp(-u)\leq e^{-1}$ for $u\geq0$, there exists a fixed constant $C>0$ such that $\|\nabla f(z)\|_1\leq C$ on $\mathcal G_0$. The mean value theorem therefore yields
\begin{equation}
\begin{aligned}
|f(\widehat Z^{[m]}=Z\mid\mathcal O)
-f(\widehat Z^{[m]}=\widehat Z\mid\mathcal O)|
\leq C\|Z-\widehat Z\|_\infty.
\end{aligned}
\label{eq: density local bound}
\end{equation}

We next construct a neighborhood that accounts for both filters defining $\mathbb M$ in \eqref{eq: index set}. For $d=\widehat Z-Z$, let $d_\Sigma\in\mathbb R^q$ denote the
subvector of $d$ corresponding to the third component of $\widehat Z$
in \eqref{eq: Z and Zm}, and let $D_\Sigma(d)$ be the unique symmetric
$N\times N$ matrix satisfying
$\vecl\{D_\Sigma(d)\}=d_\Sigma$. Define
\begin{equation*}
\begin{aligned}
\mathcal R_M(\widehat Z)
=\Bigg\{Z:\
&|d_j|\leq \err/2
\quad\text{for all coordinates corresponding to $\mu_Y$, $\mu$, and $\gamma$},\\
&\left\|D_\Sigma(d)-(\err/2)\mathbf I_N\right\|_{\max}
\leq \err/(4N)
\Bigg\}.
\end{aligned}
\end{equation*}
 The notation $\mathcal R_M(\widehat Z)$ indicates that $\widehat Z$ is its reference point. The neighborhood is symmetric around $\widehat Z$ in the coordinates corresponding to $\mu_Y$, $\mu$, $\gamma$, and the off-diagonal elements of $\Sigma$, but is shifted in the diagonal $\Sigma$ coordinates to preserve positive semi-definiteness.

For every $Z\in\mathcal R_M(\widehat Z)$, $\|Z-\widehat Z\|_\infty\leq\err$. Moreover,
\begin{equation*}
\lambda_{\min}\{D_\Sigma(d)\}\geq \err/2
-\left\|D_\Sigma(d)-(\err/2)\mathbf I_N\right\|_2\geq \err/4>0.
\end{equation*}
If $Z=\widehat Z^{[m]}$, then
\begin{equation*}
\widehat\Sigma^{[m]}=\Sigma+T_0^{-1/2}D_\Sigma(d),
\end{equation*}
so $\lambda_{\min}(\widehat\Sigma^{[m]})\geq0$. In addition, on $\mathcal G_0$, for all sufficiently large $M$,
\begin{equation*}
\begin{aligned}
|\widehat Z_j^{[m]}|
&\leq|\widehat Z_j|+\err
\leq1.05\sqrt{\widehat{\bf Cov}_{j,j}}z_{\alpha_0/(2p)}+\err\\
&\leq1.1\sqrt{\widehat{\bf Cov}_{j,j}}z_{\alpha_0/(2p)},
\end{aligned}
\end{equation*}
where the last inequality follows from \eqref{eq: min cov}, $\widehat{\bf Cov}_{j,j}\geq{\bf Cov}_{j,j}/1.05^2$, and $\err\to0$. Thus, $Z=\widehat Z^{[m]}\in\mathcal R_M(\widehat Z)$ implies $m\in\mathbb M$.

The set $\mathcal R_M(\widehat Z)$ is a $p$-dimensional hyperrectangle. Each of the $p-q$ coordinates corresponding to $\mu_Y$, $\mu$, and $\gamma$ ranges over an interval of width $\err$. For the remaining $q$ coordinates, the max-norm condition restricts each independent lower-triangular element of $D_\Sigma(d)-(\err/2)\mathbf I_N$ to an interval of width $\err/(2N)$. The diagonal shift $(\err/2)\mathbf I_N$ changes the location of these intervals but not their widths. Therefore, the volume of the neighborhood is
\begin{equation*}
\operatorname{Vol}\{\mathcal R_M(\widehat Z)\}=\err^{p-q}\left(\frac{\err}{2N}\right)^q=\kappa_N(2\err)^p.
\end{equation*}
By \eqref{eq:term 1 bound}, \eqref{eq: density local bound}, and $\err\to0$, for all sufficiently large $M$,
\begin{equation*}
\begin{aligned}
&\prob\{\widehat Z^{[m]}\in\mathcal R_M(\widehat Z)\mid\mathcal O\}
\I(\mathcal O\in\mathcal G_0)\\
&\quad\geq\frac{1}{2}c_{\rm den}(\alpha_0)
\operatorname{Vol}\{\mathcal R_M(\widehat Z)\}
\I(\mathcal O\in\mathcal G_0)\\
&\quad=\frac{1}{2}c^*(\alpha_0)(2\err)^p
\I(\mathcal O\in\mathcal G_0).
\end{aligned}
\end{equation*}

For fixed $\mathcal O$, define $\mathcal B_m=\{\widehat Z^{[m]}\in\mathcal R_M(\widehat Z)\}$ and $\pi_M=c^*(\alpha_0)(2\err)^p/2$. The preceding inequality gives $\prob(\mathcal B_m\mid\mathcal O)\geq\pi_M$ on $\mathcal G_0$. Moreover, because $\mathcal B_m$ implies both $m\in\mathbb M$ and $\|\widehat Z^{[m]}-\widehat Z\|_\infty\leq\err$, we have
\begin{equation*}
\bigcup_{m=1}^M\mathcal B_m
\subseteq
\left\{\min_{m\in\mathbb M}\|\widehat Z^{[m]}-\widehat Z\|_\infty\leq\err\right\}.
\end{equation*}
Conditional on $\mathcal O$, the events $\mathcal B_1,\ldots,\mathcal B_M$ are independent. Therefore,
\begin{equation*}
\begin{aligned}
&\prob\left(\min_{m\in\mathbb M}
\|\widehat Z^{[m]}-\widehat Z\|_\infty\leq\err
\given\mathcal O\right)\I(\mathcal O\in\mathcal G_0)\\
&\quad\geq\prob\left(\bigcup_{m=1}^M\mathcal B_m\given\mathcal O\right)
\I(\mathcal O\in\mathcal G_0)\\
&\quad=\left\{1-\prod_{m=1}^M
\left[1-\prob(\mathcal B_m\mid\mathcal O)\right]\right\}
\I(\mathcal O\in\mathcal G_0)\\
&\quad\geq\left[1-(1-\pi_M)^M\right]\I(\mathcal O\in\mathcal G_0)\\
&\quad\geq\left[1-\exp(-M\pi_M)\right]\I(\mathcal O\in\mathcal G_0)\\
&\quad=\left(1-\min\{T_0,T_1\}^{-1}\right)
\I(\mathcal O\in\mathcal G_0),
\end{aligned}
\end{equation*}
where the first inequality follows from the event inclusion above, the equality follows from conditional independence, the next inequality uses the one-perturbation bound $\prob(\mathcal B_m\mid\mathcal O)\geq\pi_M$, and the penultimate inequality follows from $1-x\leq\exp(-x)$. The final equality follows from \eqref{def: err}. Taking expectations with respect to $\mathcal O$ yields
\begin{equation*}
\prob(\mathcal E)\geq\left(1-\min\{T_0,T_1\}^{-1}\right)\prob(\mathcal G_0).
\end{equation*}
Taking $\liminf_{T_0,T_1\to\infty}\liminf_{M\to\infty}$ on both sides establishes \eqref{eq: E star and E} and concludes the proof.
}

\subsection{Proof of Theorem~\ref{thm: coverage lambda0}}\label{pf: coverage lambda0}
For the case where $\lambda=0$ and $\lambda_{\min}(\Sigma)>0$, the proof proceeds analogously to the proof of Theorem~\ref{thm: coverage} in Section~\ref{pf: coverage}, with $\lambda$ replaced by $\lambda_{\min}(\Sigma)/\sqrt{N}$. Then by Theorem~\ref{thm: negligible}, for $\alpha'=\alpha-\alpha_0$ with $\alpha\in(\alpha_0,1)$ and $\alpha_0\in(0,0.01]$, we establish that
\begin{equation*}
    \liminf_{T_0,T_1\rightarrow\infty}\liminf_{M\rightarrow\infty}\prob(\tau^*\in {\rm CI}_{\alpha}) \geq 1-\alpha' -\alpha_0=1-\alpha.
\end{equation*}

\subsection{Proof of Theorem~\ref{thm: precision lambda0}}\label{pf: precision lambda0}

Let $\lambda=0$ and suppose $\lambda_{\min}(\Sigma)>0$. We begin with the same length bound as in \eqref{eq: precision initial length}:
\begin{align}
    \operatorname{Length}({\rm CI}_{\alpha})
    \leq2\left[\max_{m\in\tilde{\mathbb M}}|\tauhatm-\tauhat|
    +z_{\alpha'/2}\VhatY^{1/2}\right],
    \label{eq: lambda0 precision initial}
\end{align}
where $\alpha'=\alpha-\alpha_0$. On $\mathcal G_3$, \eqref{eq: VhatY bound} gives
\begin{align}
    z_{\alpha'/2}\VhatY^{1/2}\lesssim\frac{1}{\sqrt{T_1}}.
    \label{eq: lambda0 precision half length}
\end{align}

We next control the distance between the candidate centers. From \eqref{eq: lambda0 sample weight distance}, on $\mathcal G_1$,
\begin{align}
    \|\betahat-\beta^{(0)}\|_2
    \leq\frac{\sqrt N(\rho+r_{T_0})}{\lambda_{\min}(\Sigma)}.
    \label{eq: lambda0 betahat distance}
\end{align}
For every $m\in\tilde{\mathbb M}$, the perturbed class $\Omegahatm(0)$ is non-empty and $\betahatm\in\Omegahatm(0)$; hence,
\begin{align*}
    \fhatm(\betahatm)\leq\rho_M.
\end{align*}
{On $\mathcal G_1\cap\mathcal G_4$, \eqref{eq: alternative uniform f} and \eqref{eq: fhatm fhat max m} give
\begin{equation}
    \|\Sigma(\betahatm-\beta^{(0)})\|_\infty
    =f(\betahatm)
    \leq\fhatm(\betahatm)+r_{T_0}+e_{T_0}
    \leq\rho_M+r_{T_0}+e_{T_0}.
    \label{eq: lambda0 perturbed discrepancy}
\end{equation}}
As in \eqref{eq: lambda0 sample weight distance}, positive definiteness of $\Sigma$ implies
\begin{align}
    \max_{m\in\tilde{\mathbb M}}\|\betahatm-\beta^{(0)}\|_2
    \leq\frac{\sqrt N(\rho_M+r_{T_0}+e_{T_0})}{\lambda_{\min}(\Sigma)}.
    \label{eq: lambda0 perturbed weight distance}
\end{align}

By the definitions of $\tauhatm$ and $\tauhat$ in \eqref{eq: m-th tauhat} and \eqref{eq: betahat tauhat}, respectively,
\begin{align*}
    |\tauhatm-\tauhat|
    &=|\muhat\tr\betahat-(\muhatm)\tr\betahatm|\\
    &\leq|(\muhat-\muhatm)\tr\betahatm|
    +|\muhat\tr(\betahat-\betahatm)|\\
    &\leq\|\muhatm-\muhat\|_\infty
    +\|\muhat\|_2\left(
    \|\betahat-\beta^{(0)}\|_2+
    \|\betahatm-\beta^{(0)}\|_2\right).
\end{align*}
{Assumption~\ref{assumption: post} implies
$\|\muhat-\mu\|_\infty=O_p(T_1^{-1/2})=o_p(1)$. Since $\mu$ is bounded and $N$ is fixed, $\|\muhat\|_2$ is bounded with probability tending to one. Combining \eqref{eq: muhatm bound on M}, \eqref{eq: lambda0 betahat distance}, and \eqref{eq: lambda0 perturbed weight distance}, on $\mathcal G_1\cap\mathcal G_3\cap\mathcal G_4\cap\mathcal E$ and this event we obtain
\begin{equation}
\begin{aligned}
    \max_{m\in\tilde{\mathbb M}}|\tauhatm-\tauhat|
    &\lesssim\frac{1}{\sqrt{T_1}}
    +\frac{\sqrt N}{\lambda_{\min}(\Sigma)}
    \left(\rho+\rho_M+2r_{T_0}+e_{T_0}\right)\\
    &\lesssim\frac{1}{\sqrt{T_1}}
    +\frac{\sqrt N}{\lambda_{\min}(\Sigma)}
    \frac{[\log(\max\{T_0,N\})]^{\frac{1+b}{2b}}}{\sqrt{T_0}}.
    \label{eq: lambda0 center sharp rate}
\end{aligned}
\end{equation}}
Combining \eqref{eq: lambda0 precision initial}, \eqref{eq: lambda0 precision half length}, and \eqref{eq: lambda0 center sharp rate}, we establish on the same event that
\begin{align}
    \operatorname{Length}({\rm CI}_{\alpha})
    \lesssim\frac{1}{\sqrt{T_1}}
    +\frac{\sqrt N}{\lambda_{\min}(\Sigma)}
    \frac{[\log(\max\{T_0,N\})]^{\frac{1+b}{2b}}}{\sqrt{T_0}}.
    \label{eq: lambda0 precision  rate}
\end{align}
{For a sufficiently large fixed constant $C$, the same probability argument as in the proof of Theorem~\ref{thm: precision} gives
\begin{equation*}
\begin{aligned}
&\liminf_{T_0,T_1\to\infty}\liminf_{M\to\infty}
\prob\left\{\operatorname{Length}({\rm CI}_{\alpha})
\leq C\left(\frac{1}{\sqrt{T_1}}
+\frac{\sqrt N}{\lambda_{\min}(\Sigma)}
\frac{[\log(\max\{T_0,N\})]^{\frac{1+b}{2b}}}{\sqrt{T_0}}\right)\right\}\\
&\quad\geq
\liminf_{T_0,T_1\to\infty}\liminf_{M\to\infty}
\prob\{\mathcal G_1\cap\mathcal G_3\cap\mathcal G_4
\cap\mathcal E\}\\
&\quad\geq1-\alpha_0.
\end{aligned}
\end{equation*}
This establishes the rate stated in Theorem~\ref{thm: precision lambda0} and concludes the proof.}

\subsection{Theoretical Justification of the Tuning Parameter \texorpdfstring{$\rho$}{rho}}\label{subsec: justify rho}
In this section, we justify our choice of tuning parameter $\rho$ in \eqref{eq: tuning from specifying} to construct $\Omegahat$ in \eqref{eq: estimator of omega} by assuming i.i.d.\ sub-gaussian data for fixed $N$. Throughout this section, we additionally assume the following:
\begin{itemize}
    \item[(A1)] There exist positive constants $c_1$ and $c_2$ such that $c_1\leq \lambda_{\min}(\Sigma)\leq \lambda_{\max}(\Sigma) \leq c_2$ where $\lambda_{\max}(\Sigma)$ denotes the maximum eigenvalue of $\Sigma=T_0^{-1}\sum_{t=1}^{T_0}\E X_tX_t\tr$.
    \item[(A2)] The subgaussian norm of $u_t^{(0)}$, which is defined as Definition 5.7 of \citet{vershynin2012introduction}, satisfies $\|u_t^{(0)}\|_{\psi_2}\leq \sigma_u$. Also, there exists a positive constant $c_3$ such that $\sigma^2 = \Var(u_t^{(0)})\leq c_3$ for all $t=1,\ldots,T_0$ where $u_t^{(0)}$ is defined in \eqref{eq: outcome model}.
    \item[(A3)] The pairs $\{(X_t,u_t^{(0)})\}_{t=1}^{T_0}$ are i.i.d.,
$u_t^{(0)}$ is independent of $X_t$, and there exists a positive constant
$c_4$ such that $\sigma_u\leq c_4\sigma$.
\end{itemize}

Now we provide the proof. Recall from \eqref{eq: def of f hat} and \eqref{eq: def of f} that for $\lambda \geq 0$,
\begin{align*}
    f(\beta) = \left\|\frac{1}{T_0}\sum_{t=1}^{T_0}\E\left[X_t\left(Y_{1,t}-X_t\tr\beta\right)\right]\right\|_{\infty}-\lambda,\quad \fhat(\beta) = \left\|\frac{1}{T_0}\sum_{t=1}^{T_0}X_t\left(Y_{1,t}-X_t\tr\beta\right)\right\|_{\infty}-\lambda,
\end{align*}
for $\beta\in\Delta^N$. In the following, we show that 
\begin{align*}
     \left|\fhat(\beta)-f(\beta)\right|
    &\lesssim \sqrt{\frac{\log N}{T_0}}\left(\lambda+\widehat{\sigma}\max_{2\leq j \leq N+1}\sqrt{\frac{1}{T_0}\sum_{t=1}^{T_0}Y_{j,t}^2}\right),
\end{align*}
for $\beta\in\Omega$ with high probability, thus, justify using $\rho$ as in \eqref{eq: tuning from specifying}.

For $\beta \in \Delta^{N}$, by \eqref{eq: diff f fhat}, we have
\begin{align*}
    \left|\fhat(\beta)-f(\beta)\right|&\leq\left\|\frac{1}{T_0}\sum_{t=1}^{T_0}\left[X_tu_t^{(0)}+(X_tX_t\tr-\E X_tX_t\tr)(\beta^{(0)}-\beta)\right]\right\|_{\infty}\\
    &\leq
    \left\|\frac{1}{T_0}\sum_{t=1}^{T_0}\left(X_tX_t\tr-\E X_tX_t\tr\right)(\beta^{(0)}-\beta)\right\|_{\infty}+\left\|\frac{1}{T_0}\sum_{t=1}^{T_0}X_tu_t^{(0)}\right\|_{\infty}.
\end{align*}
We define
\begin{align*}
    &\mathcal{F}_1(w,v,s) = \left\{\left|w\tr\left(\frac{1}{T_0}\sum_{t=1}^{T_0}X_tX_t\tr\right)v-w\tr\Sigma v\right|\lesssim \frac{s\|\Sigma^{1/2}w\|_2\|\Sigma^{1/2}v\|_2}{\sqrt{T_0}}\right\},\\
     &\mathcal{F}_2(w,\lambda,s) = \left\{\sup_{\beta \in \Omega}\left| w\tr\left(\frac{1}{T_0}\sum_{t=1}^{T_0}X_tX_t\tr-\Sigma\right)(\beta^{(0)}-\beta)\right| \lesssim \frac{s\|\Sigma^{1/2}w\|_2\sqrt{N}\lambda}{\sqrt{\lambda_{\min}(\Sigma)}\sqrt{T_0}}\right\}.
\end{align*}
By Lemma 11 of \citet{tony2020semisupervised}, $\mathbb{P}\left(\mathcal{F}_1(w,v,s)\right)\geq 1-2\exp(-c''s^2)$ for all $w,v\in\mathbb{R}^N$ for some positive constant $c''>0$. 

If $\lambda_{\min}(\Sigma)>0$, then $\Sigma^{-1}$ exists and
\begin{align*}
    \|\Sigma^{1/2}(\beta^{(0)}-\beta)\|_2 &=  \|\Sigma^{-1/2}\Sigma(\beta^{(0)}-\beta)\|_2\\
    &\leq \|\Sigma^{-1/2}\|_{2}\|\Sigma(\beta^{(0)}-\beta)\|_2\leq \sqrt{N} \|\Sigma^{-1/2}\|_{2}\|\Sigma(\beta^{(0)}-\beta)\|_{\infty}.
\end{align*}
Since $\|\Sigma(\beta^{(0)}-\beta)\|_{\infty}\leq \lambda$ for all $\beta\in\Omega$ and $\|\Sigma^{-1/2}\|_{2}=1/\sqrt{\lambda_{\min}(\Sigma)}$, we have
\begin{align*}
     \|\Sigma^{1/2}(\beta^{(0)}-\beta)\|_2\leq\frac{\sqrt{N}\lambda}{\sqrt{\lambda_{\min}(\Sigma)}}.
\end{align*}
Thus, on the event $\mathcal{F}_1(w,\beta^{(0)}-\beta,s))$ for all $\beta\in\Omega$, we have
\begin{align*}
    &\sup_{\beta \in \Omega}\left| w\tr\left(\frac{1}{T_0}\sum_{t=1}^{T_0}X_tX_t\tr-\Sigma\right)(\beta^{(0)}-\beta)\right|\lesssim \sup_{\beta \in \Omega}\frac{s\|\Sigma^{1/2}w\|_2\|\Sigma^{1/2}(\beta^{(0)}-\beta)\|_2}{\sqrt{T_0}} \lesssim \frac{s\|\Sigma^{1/2}w\|_2\sqrt{N}\lambda}{\sqrt{\lambda_{\min}(\Sigma)}\sqrt{T_0}}.
\end{align*}
Thus, $\mathbb{P}(\mathcal{F}_2(w,\lambda,s)) \geq \mathbb{P}(\mathcal{F}_1(w,\beta^{(0)}-\beta,s)) \geq 1-2\exp(-c''s^2)$ for all $w \in \mathbb{R}^N$, $\beta\in\Omega$, and $\lambda\geq0$. With the same logic above, we have that
\begin{align*}
    &\mathbb{P}\left(\sup_{\beta\in\Omega(\lambda)}\left\|\frac{1}{T_0}\sum_{t=1}^{T_0}\left(X_tX_t\tr-\E X_tX_t\tr\right)(\beta^{(0)}-\beta)\right\|_{\infty}\gtrsim \frac{s_0\sqrt{\lambda_{\max}(\Sigma)}\sqrt{N}\lambda}{\sqrt{\lambda_{\min}(\Sigma)}\sqrt{T_0}}\right)\\
    &\leq 2\exp(\log N -c''s_0^2)\leq 2\exp(\log (\max\{T_0,N\}) -c''s_0^2).
\end{align*}
Since $N$ is fixed, and $\lambda_{\min}(\Sigma)$ and $\lambda_{\max}(\Sigma)$ are constant, $\sqrt{N\lambda_{\max}(\Sigma)/\lambda_{\min}(\Sigma)}$ is also constant. Thus, by taking $s_0 \asymp s+\sqrt{\log (\max\{T_0,N\})}$, we establish
\begin{align}
    &\mathbb{P}\left(\sup_{\beta\in\Omega(\lambda)}\left\|\frac{1}{T_0}\sum_{t=1}^{T_0}\left(X_tX_t\tr-\E X_tX_t\tr\right)(\beta^{(0)}-\beta)\right\|_{\infty}\gtrsim 
    \frac{s+\sqrt{\log (\max\{T_0,N\})}}{\sqrt{T_0}}\cdot\lambda
    \right)\nonumber\\
    &\leq 2\exp(-c''s^2).\label{eq: conv of XX}
\end{align}

On the other hand, conditional on $\{X_t\}_{t=1}^{T_0}$, (A3) and Proposition 5.10 of \citet{vershynin2012introduction} imply that $T_0^{-1}\sum_{t=1}^{T_0}Y_{j,t}u_t^{(0)}$ is sub-gaussian for $j=2,\ldots,N+1$, with sub-gaussian norm satisfying
\begin{align*}
    \left\|\frac{1}{T_0}\sum_{t=1}^{T_0}Y_{j,t}u_t^{(0)}\right\|_{\psi_2} \lesssim \frac{\sigma_u}{T_0}\sqrt{\sum_{t=1}^{T_0}Y_{j,t}^2},
\end{align*}
By (A3), $\sigma_u\lesssim \sigma$, where $\sigma^2 = \Var(u_t^{(0)})$.
By a union bound, we establish
\begin{align*}
    \prob\left( \left\|\frac{1}{T_0}\sum_{t=1}^{T_0}X_tu_t^{(0)}\right\|_{\infty} \geq \delta\right) &\leq 2N\exp\left(-\frac{c\delta^2T_0^2}{\sigma^2\max_{2\leq j \leq N+1}\sum_{t=1}^{T_0}Y_{j,t}^2}\right)\\
    &\leq2\max\{T_0,N\}\exp\left(-\frac{c\delta^2T_0^2}{\sigma^2\max_{2\leq j \leq N+1}\sum_{t=1}^{T_0}Y_{j,t}^2}\right),
\end{align*}
for some $c>0$. By taking 
\begin{align*}
    \delta = \sigma\sqrt{\frac{\log (\max\{T_0,N\})}{T_0}}\max_{2\leq j \leq N+1}\sqrt{\frac{1}{T_0}\sum_{t=1}^{T_0}Y_{j,t}^2}+\frac{s}{\sqrt{T_0}},
\end{align*}
for some $s>0$, we establish
\begin{align}\label{eq: conv of Xu}
     \prob\left( \left\|\frac{1}{T_0}\sum_{t=1}^{T_0}X_tu_t^{(0)}\right\|_{\infty} \geq \sigma\sqrt{\frac{\log \max\{T_0,N\}}{T_0}}\max_{2\leq j \leq N+1}\sqrt{\frac{1}{T_0}\sum_{t=1}^{T_0}Y_{j,t}^2}+\frac{s}{\sqrt{T_0}}\right) \leq 2\exp(-c's^2).
\end{align}
for some $c'>0$.

For $\beta\in\Omega(\lambda)$, by combining \eqref{eq: conv of XX} and \eqref{eq: conv of Xu}, we have
\begin{align*}
     \left|\fhat(\beta)-f(\beta)\right|
    &\leq
    \left\|\frac{1}{T_0}\sum_{t=1}^{T_0}\left(X_tX_t\tr-\E X_tX_t\tr\right)(\beta^{(0)}-\beta)\right\|_{\infty}+\left\|\frac{1}{T_0}\sum_{t=1}^{T_0}X_tu_t^{(0)}\right\|_{\infty}\\
    &\lesssim \sqrt{\frac{\log \max\{T_0,N\}}{T_0}}\lambda+\sigma\sqrt{\frac{\log \max\{T_0,N\}}{T_0}}\max_{2\leq j \leq N+1}\sqrt{\frac{1}{T_0}\sum_{t=1}^{T_0}Y_{j,t}^2}\\
    &=\sqrt{\frac{\log \max\{T_0,N\}}{T_0}}\left(\lambda+\sigma\max_{2\leq j \leq N+1}\sqrt{\frac{1}{T_0}\sum_{t=1}^{T_0}Y_{j,t}^2}\right),
\end{align*}
with probability at least $1-2(\max\{T_0,N\})^{-c'}-2(\max\{T_0,N\})^{-c''}$. We estimate $\sigma^2$ with 
\begin{align*}
    \widehat{\sigma}^2 = \frac{1}{T_0}\sum_{t=1}^{T_0}\left(Y_{1,t}-X_t\tr\betahatSC\right)^2.
\end{align*}
By the definition of $\betahatSC$ in \eqref{eq: SC estimators}, we have
\begin{equation*}
\begin{aligned}
     \widehat{\sigma}^2&=\frac{1}{T_0}\sum_{t=1}^{T_0}\left(Y_{1,t}-X_t\tr\betahatSC\right)^2=\frac{1}{T_0}\sum_{t=1}^{T_0}\left[u_{t}^{(0)}-X_t\tr(\betahatSC-\beta^{(0)})\right]^2\\
     &=\frac{1}{T_0}\sum_{t=1}^{T_0}\left(u_t^{(0)}\right)^2+\frac{1}{T_0}\sum_{t=1}^{T_0}\left[X_t\tr(\betahatSC-\beta^{(0)})\right]^2-\frac{2}{T_0}\sum_{t=1}^{T_0}u_t^{(0)}X_t\tr(\betahatSC-\beta^{(0)}).
\end{aligned}
\end{equation*}
Since $\{u_t^{(0)}\}_{t=1}^{T_0}$ is a sequence of i.i.d.\ sub-gaussian random variables with sub-gaussian norm $\|u_{t}^{(0)}\|_{\psi_2}\leq \sigma_u$, Lemma 5.14 of \citet{vershynin2012introduction} shows that $\{(u_t^{(0)})^2\}_{t=1}^{T_0}$ is a sequence of i.i.d. sub-exponential random variables with sub-exponential norm $\|(u_t^{(0)})^2\|_{\psi_1}\leq 2 \sigma_u^2$, where $\|\cdot\|_{\psi_1}$ is defined in Definition 5.13 of \citet{vershynin2012introduction}. Since $\sigma_u\lesssim \sigma$, Corollary 5.17 of \citet{vershynin2012introduction} establishes that, for $t>0$ and some positive constant $c>0$,
\begin{equation*}
    \prob\left( \left| \frac{1}{T_0}\sum_{t=1}^{T_0}\left(u_t^{(0)}\right)^2 - \sigma^2 \right| \ge t \right) \le 2 \exp \left( -c T_0 \min \left\{ \frac{t^2}{\sigma^4}, \frac{t}{\sigma^2} \right\} \right).
\end{equation*}
By taking $t=\sigma^2\sqrt{\log \max\{T_0,N\}/T_0}$, we have
\begin{equation*}
    \prob\left( \left| \frac{1}{T_0}\sum_{t=1}^{T_0}\left(u_t^{(0)}\right)^2 - \sigma^2 \right| \le \sigma^2\sqrt{\frac{\log \max\{T_0,N\}}{T_0} }\right) \ge 1-2(\max\{T_0,N\})^{-c}.
\end{equation*}

Since we have
\begin{equation*}
\begin{aligned}
     \frac{1}{T_0}\sum_{t=1}^{T_0}\left(u_t^{(0)}\right)^2 &=  \widehat{\sigma}^2-\frac{1}{T_0}\sum_{t=1}^{T_0}\left[X_t\tr(\betahatSC-\beta^{(0)})\right]^2+\frac{2}{T_0}\sum_{t=1}^{T_0}u_t^{(0)}X_t\tr(\betahatSC-\beta^{(0)})\\
     &\leq \widehat{\sigma}^2+\frac{2}{T_0}\sum_{t=1}^{T_0}u_t^{(0)}X_t\tr(\betahatSC-\beta^{(0)}),
\end{aligned}
\end{equation*}
and H\"older inequality with $\|\beta^{(0)}-\betahatSC\|_{1}\leq2$ establishes that
\begin{equation*}
    \left|\frac{2}{T_0}\sum_{t=1}^{T_0}u_t^{(0)}X_t\tr(\beta^{(0)}-\betahatSC)\right| \lesssim\left\|\frac{1}{T_0}\sum_{t=1}^{T_0}X_tu_t^{(0)}\right\|_{\infty},
\end{equation*}
we ensure that with probability at least $1-2(\max\{T_0,N\})^{-c}$,
\begin{equation*}
      \sigma^2 \lesssim \widehat{\sigma}^2+\left\|\frac{1}{T_0}\sum_{t=1}^{T_0}X_tu_t^{(0)}\right\|_{\infty}+\sqrt{\frac{\log \max\{T_0,N\}}{T_0} },
\end{equation*}
since $\sigma^2\leq c_3$ for a positive constant $c_3>0$.

Finally, with probability at least $1-2(\max\{T_0,N\})^{-c}-2(\max\{T_0,N\})^{-c'}-2(\max\{T_0,N\})^{-c''}$, we establish that, for sufficiently large $T_0$,
\begin{equation*}
    \left|\fhat(\beta)-f(\beta)\right|\lesssim \left(\lambda+\widehat{\sigma}\max_{2\leq j \leq N+1}\sqrt{\frac{1}{T_0}\sum_{t=1}^{T_0}Y_{j,t}^2}\right)\sqrt{\frac{\log \max\{T_0,N\}}{T_0}}, 
\end{equation*}
since \eqref{eq: conv of Xu} establishes on the same event where the above holds that 
\begin{equation*}
    \left\|\frac{1}{T_0}\sum_{t=1}^{T_0}X_tu_t^{(0)}\right\|_{\infty} \lesssim \sqrt{\frac{\log \max\{T_0,N\}}{T_0} }.
\end{equation*}
Thus, it suggests to choose $\rho$ as
\begin{align*}
    \rho = C\left(\lambda+\widehat{\sigma}\max_{2\leq j \leq N+1}\sqrt{\frac{1}{T_0}\sum_{t=1}^{T_0}Y_{j,t}^2}\right)\sqrt{\frac{\log \max\{T_0,N\}}{T_0}},
\end{align*}
for some positive constant $C>0$. This concludes the proof.

\section{Proofs of Lemmas}\label{sec: pf lem}
In this section, we establish the proofs of lemmas required for the proofs in Section~\ref{sec: pf thm prop}.

\subsection{Proof of Lemma~\ref{lem: unif conv of f}}\label{pf: unif conv of f}
For $\beta \in \Delta^{N}$, under the model \eqref{eq: outcome model} we have 
\begin{align}
    &\left|\fhat(\beta)-f(\beta)\right| = \left|\left\|\frac{1}{T_0}\sum_{t=1}^{T_0}X_t\left(Y_{1,t}-X_t\tr\beta\right)\right\|_{\infty}-\lambda-\left\|\frac{1}{T_0}\sum_{t=1}^{T_0}\E\left[X_t\left(Y_{1,t}-X_t\tr\beta\right)\right]\right\|_{\infty}+\lambda\right|\nonumber\\
    &=\left|\left\|\frac{1}{T_0}\sum_{t=1}^{T_0}X_tX_t\tr(\beta^{(0)}-\beta)+\frac{1}{T_0}\sum_{t=1}^{T_0}X_tu_t^{(0)}\right\|_{\infty} -\left\|\frac{1}{T_0}\sum_{t=1}^{T_0}\E X_tX_t\tr(\beta^{(0)}-\beta)+\frac{1}{T_0}\sum_{t=1}^{T_0}\E X_tu_t^{(0)}\right\|_{\infty}\right|\nonumber\\
    &\leq\left\|\frac{1}{T_0}\sum_{t=1}^{T_0}X_tX_t\tr(\beta^{(0)}-\beta)+\frac{1}{T_0}\sum_{t=1}^{T_0}X_tu_t^{(0)}-\left(\frac{1}{T_0}\sum_{t=1}^{T_0}\E X_tX_t\tr(\beta^{(0)}-\beta)+\frac{1}{T_0}\sum_{t=1}^{T_0}\E X_tu_t^{(0)}\right)\right\|_{\infty}\nonumber\\
    &=\left\|\frac{1}{T_0}\sum_{t=1}^{T_0}\left[X_tu_t^{(0)}+(X_tX_t\tr-\E X_tX_t\tr)(\beta^{(0)}-\beta)\right]\right\|_{\infty},\label{eq: diff f fhat}
\end{align}
by the triangle inequality.
Thus, on the event $\mathcal{G}_1$ defined in \eqref{event: pre}, we can establish that
\begin{align*}
    \sup_{\beta\in\Delta^N}\left|\fhat(\beta)-f(\beta)\right|&\leq \sup_{\beta\in\Delta^N}\left\|\frac{1}{T_0}\sum_{t=1}^{T_0}\left[X_tu_t^{(0)}+(X_tX_t\tr-\E X_tX_t\tr)(\beta^{(0)}-\beta)\right]\right\|_{\infty}\\
    &\leq C_0\frac{\left(\log(\max\{T_0,N\})\right)^{\frac{1+b}{2b}}}{\sqrt{T_0}}
\end{align*}
for the positive constant $C_0>0$ in Assumption~\ref{assumption: pre}.
We conclude the proof.

\subsection{Proof of Lemma~\ref{lem: cons of cov}}\label{pf: cons of cov}
\noindent

By the triangle inequality and Assumption~\ref{assumption: clt and cons}, for the constant $c_1>0$ defined in \eqref{eq: max cov}, we have
\begin{equation*}
    \prob(T_1\VhatY>2c_1) \leq \prob(T_1\VY+|T_1(\VhatY-\VY)|> 2c_1)
         \leq\prob(T_1\VY>c_1)+\prob(|T_1(\VhatY-\VY)|> c_1) \to 0
\end{equation*}
as $T_1\to\infty$.  Also, we have
\begin{equation*}
    \begin{aligned}
        \prob(\|T_1\Vhatmu\|_{\max}>2c_1) &\leq \prob(\|T_1\Vmu\|_{\max}+\|T_1(\Vhatmu-\Vmu)\|_{\max}>2c_1)\\
        &\leq\prob(\|T_1\Vmu\|_{\max}>c_1)+\prob(\|T_1(\Vhatmu-\Vmu)\|_{\max}>c_1)\to 0
    \end{aligned}
\end{equation*}
as $T_1\to\infty$. Thus, $\lim_{T_1\to\infty}\prob(\mathcal{G}_3)=1$ by the union bound. Similarly, we can show that
\begin{equation*}
\prob(\|T_0\Vhatgamma\|_{\max}>2c_1) \to 0,\quad \prob(\|T_0\VhatSigma\|_{\max}>2c_1)\to 0,
\end{equation*}
as $T_0\to \infty$, so $\lim_{T_0\to\infty}\prob(\mathcal{G}_4)=1$ by the union bound.
This concludes the proof.

\subsection{Proof of Lemma~\ref{lem: fhatmstar f}}\label{pf: fhatmstar f}

For any $\beta\in\Delta^N$, by the triangle inequality,
\begin{align*}
    |\fhatm(\beta)-f(\beta)|&=\left|\|\gammahatm-\Sigmahatm\beta\|-\left\|\gamma-\Sigma\beta\right\|\right| \leq\left\|\gammahatm-\gamma-\left(\Sigmahatm-\Sigma\right)\beta\right\|_{\infty} \\
    &\leq \|\gammahatm-\gamma\|_{\infty}+ \|(\Sigmahatm-\Sigma)\beta\|_{\infty}.
\end{align*}
Using \eqref{eq: vecl norm}, we ensure that the last term above is bounded as
\begin{equation*}
    \begin{aligned}
         \|\gammahatm-\gamma\|_{\infty}+ \|(\Sigmahatm-\Sigma)\beta\|_{\infty}  &\leq \|\gammahatm-\gamma\|_{\infty}+\|\vecl(\Sigmahatm-\Sigma)\|_{\infty}\\
    &\leq 2\max\{\|\gammahatm-\gamma\|_{\infty},\|\vecl(\Sigmahatm-\Sigma)\|_{\infty}\}.
    \end{aligned}
\end{equation*}
Thus we establish
\begin{align*}
    \sup_{\beta\in\Delta^N}|\fhat^{[m]}(\beta)-f(\beta)|\leq 2\max\{\|\gammahatm-\gamma\|_{\infty},\|\vecl(\Sigmahatm-\Sigma)\|_{\infty}\}.
\end{align*}
On the event $\mathcal{E}$ defined in \eqref{event: m star}, there exists $m = m^*$ such that
\begin{align*}
    \max\{\|\gammahatmstar-\gamma\|_{\infty},\|\vecl(\Sigmahatmstar-\Sigma)\|_{\infty}\} \leq \frac{\err}{\sqrt{T_0}}.
\end{align*}
Thus, we establish \eqref{eq: fhatm conv}. By applying the above inequality, we establish that
\begin{equation*}
    \fhatmstar(\beta^{(0)})\leq f(\beta^{(0)}) +\frac{2\err}{\sqrt{T_0}} \leq -\lambda+\rho_M,
\end{equation*}
since $f(\beta^{(0)})=-\lambda$ and $\rho_M\geq 2\err/\sqrt{T_0}$. This concludes the proof.

\subsection{Proof of Lemma~\ref{lem: w conv}}\label{pf: w conv}

Let
\begin{align*}
    &w^* {\in} \argmin_{w \in \mathcal{H}} w\tr Aw - 2a\tr w, \quad \widehat{w} {\in} \argmin_{w \in \mathcal{H}} w\tr \widehat{A}w - 2\widehat{a}\tr w.
\end{align*}
By definition of $\widehat{w}$ and convexity of $\mathcal{H}$, for $t \in (0,1)$, we have
\begin{align*}
    \widehat{w}\tr\widehat{A}\widehat{w} -2\widehat{a}\tr\widehat{w}\leq \left(\what+t(w^*-\what)\right)\tr\widehat{A}\left(\what+t(w^*-\what)\right)-2\widehat{a}\tr(\what+t(w^*-\what)).
\end{align*}
Dividing both terms by $t$ further leads to 
\begin{align*}
    2(w^*-\what)\tr\widehat{A}\what+t(w^*-\what)\tr\widehat{A}(w^*-\what)-2\widehat{a}\tr(w^*-\what) \geq 0,
\end{align*}
which is equivalent to
\begin{align*}
    (t-2)(w^*-\what)\tr\widehat{A}(w^*-\what)+2(w^*-\what)\widehat{A}w^*-2\widehat{a}\tr(w^*-\what)\geq 0. 
\end{align*}
By taking $\underset{t\rightarrow 0+}{\lim}$ both sides, we have
\begin{align*}
    (w^*-\what)\tr\widehat{A}(w^*-\what) \leq (w^*-\what)\tr\widehat{A}w^* -\widehat{a}\tr(w^*-\what),
\end{align*}
which also leads to
\begin{align*}
    (w^*-\what)\tr\widehat{A}\what \geq \widehat{a}\tr(w^*-\what).
\end{align*}
Since the definitions of $w^*$ and $\what$ are symmetric, by exchanging the roles of $\{A,a,w^*\}$ and $\{\widehat{A},\widehat{a},\what\}$, we have
\begin{align*}
     (\what-w^*)\tr Aw^* \geq a\tr(\what-w^*).
\end{align*}
Using the above inequality, we further establish that
\begin{align*}
    (w^*-\what)\tr\widehat{A}w^* -\widehat{a}\tr(w^*-\what) &= (w^*-\what)\tr Aw^*+(w^*-\what)\tr(\widehat{A}-A)w^*-\widehat{a}\tr(w^*-\what)\\
    &\leq \left(a-\widehat{a}+(\widehat{A}-A)w^*\right)\tr(w^*-\what),
\end{align*}
which further leads to
\begin{align*}
    (w^*-\what)\tr\widehat{A}(w^*-\what) \leq \left(a-\widehat{a}+(\widehat{A}-A)w^*\right)\tr(w^*-\what)\leq \lVert (\widehat{a}-a) +(A-\widehat{A})w^*\rVert_{\infty}\lVert \what-w^*\rVert_1.
\end{align*}
In the last inequality, we use the H\"older's inequality. By switching the roles of $\{A,a,w^*\}$ and $\{\widehat{A},\widehat{a},\what\}$ in the above, we establish
\begin{align*}
    (w^*-\what)\tr A(w^*-\what) \leq \lVert (a-\widehat{a}) +(\widehat{A}-A)\what\rVert_{\infty}\lVert w^*-\what\rVert_1.
\end{align*}
This concludes the proof.

\subsection{Proof of Lemma~\ref{lem: g eq}}\label{pf: g eq}
Since $\tilde{w}\tr A\tilde{w}\geq 0$ for a positive semi-definite matrix $A$,
\begin{align*}
    h(\bar{w})-h(\tilde{w}) &= \bar{w}\tr A\bar{w}-2a\tr\bar{w} - \left(\tilde{w}\tr A\tilde{w}-2a\tr\tilde{w}\right)\\
    &= (\bar{w}-\tilde{w})\tr A(\bar{w}-\tilde{w})-2\tilde{w}\tr A\tilde{w}+2\tilde{w}\tr A\bar{w}-2a\tr\left(\bar{w}-\tilde{w}\right)\\
     &=(\bar{w}-\tilde{w})\tr A(\bar{w}-\tilde{w})+2\tilde{w}\tr A\left(\bar{w}-\tilde{w}\right)-2a\tr\left(\bar{w}-\tilde{w}\right).
\end{align*}
Since $\tilde{w},\bar{w}\in \mathcal{H}$ and $\mathcal{H}$ is a convex set, we ensure for given $t\in (0,1)$, $(1-t)\tilde{w}+t\bar{w}\in\mathcal{H}$. This leads to $h(\tilde{w}) = \min_{w\in\mathcal{H}}h(w)\leq h\left((1-t)\tilde{w}+t\bar{w}\right)$. This implies that
\begin{align*}
     h(\tilde{w}) = \tilde{w}\tr A\tilde{w}-2a\tr\tilde{w}&\leq h\left((1-t)\tilde{w}+t\bar{w}\right)\\
     &=\left((1-t)\tilde{w}+t\bar{w}\right)\tr A\left((1-t)\tilde{w}+t\bar{w}\right)-2a\tr\left((1-t)\tilde{w}+t\bar{w}\right).
\end{align*}

The above inequality is equivalent to
\begin{align*}
    t^2(\bar{w}-\tilde{w})\tr A(\bar{w}-\tilde{w})+2t\tilde{w}\tr A(\bar{w}-\tilde{w})-2ta\tr(\bar{w}-\tilde{w})\geq 0.
\end{align*}
By taking $\underset{t\rightarrow 0}{\lim}$ after dividing by $t$ on both sides, we have $\tilde{w}\tr A(\bar{w}-\tilde{w})-a\tr(\bar{w}-\tilde{w}) \geq 0$. 
Then the lower bound of $h(\bar{w})-h(\tilde{w})$ is further bounded as
\begin{align*}
    h(\bar{w})-h(\tilde{w})&\geq(\bar{w}-\tilde{w})\tr A(\bar{w}-\tilde{w})+2\tilde{w}\tr A\left(\bar{w}-\tilde{w}\right)-2a\tr\left(\bar{w}-\tilde{w}\right)\geq(\bar{w}-\tilde{w})\tr A(\bar{w}-\tilde{w}).
\end{align*}
This concludes the proof.

\subsection{Proof of Lemma~\ref{lem: control of obj}}\label{pf: control of obj}

If $e'(\tilde{w})\leq c'$, then $\tilde{w}\in \mathcal{H} \cap \mathcal{H}'$, so $h(\tilde{w}) = h(\bar{w})$ and the claim holds. Thus, we only consider the case $e'(\tilde{w})>c'$.

We define $\tilde{w}^0 = \xi\tilde{w}+(1-\xi)w^0$ for some $w^0\in\Delta^{N}$ where $e'(w^0)<c'$ and $\xi = \frac{e'(w^0)-c'}{e'(w^0)-e'(\tilde{w})}\in (0,1)$.
Since $e'$ is a convex function, direct calculation shows that $$e'(\tilde{w}^0)\leq \xi e'(\tilde{w})+(1-\xi)e'(w^0) = c'.$$ Also, by the convexity of the function $e$, we have
$$e(\tilde{w}^0) \leq \xi e(\tilde{w})+(1-\xi)e(w^0).$$
Since $\tilde{w}\in\mathcal{H}$, we have $e(\tilde{w})\leq c$ and $e(w^0)\leq c$ by definition. Thus, $e(\tilde{w}^0)\leq c$.
Combining these facts, we have $\tilde{w}^0\in\mathcal{H}\cap\mathcal{H}'.$ This ensures
\begin{align*}
    h(\tilde{w}^0)\geq h(\bar{w})\geq h(\tilde{w}).
\end{align*}
Using the above inequalities, we have
\begin{align*}
    h(\bar{w})-h(\tilde{w})\leq h(\tilde{w}^0)-h(\tilde{w})\leq \nabla h(\tilde{w}^0)\tr(\tilde{w}^0-\tilde{w})
\end{align*}
by the supporting hyperplane theorem since $h$ is a convex and differentiable function \citep[][Equation~(3.2)]{boyd2004convex}. Since $\nabla h(w) = 2Aw-2a$ and $\|w\|_{\infty}\leq \|w\|_{1}\leq 1$ for any $w\in\Delta^N$, we have
\begin{align*}
    \|\nabla h(w)\|_{\infty} \leq 2(\|A\|_{\infty}+\|a\|_{\infty}).
\end{align*}
By H\"older's inequality, we can further bound
\begin{align*}
    \nabla h(\tilde{w}^0)\tr(\tilde{w}^0-\tilde{w}) \leq  \|\nabla h(\tilde{w}^0)\|_{\infty}\|\tilde{w}^0-\tilde{w}\|_{1} &\leq 2(\|A\|_{\infty}+\|a\|_{\infty})\|\tilde{w}^0-\tilde{w}\|_{1}.
\end{align*}
Since $\tilde{w}^0-\tilde{w}=(1-\xi)(w^0-\tilde{w})$ and $\|w^0-\tilde{w}\|_1\leq \|w^0\|_1+\|\tilde{w}\|_1\leq 2$,
\begin{align*}
    2(\|A\|_{\infty}+\|a\|_{\infty})\|\tilde{w}^0-\tilde{w}\|_{1}\leq 4(\|A\|_{\infty}+\|a\|_{\infty})(1-\xi)
\end{align*}

On the other hand, since $e'(\tilde{w})>c'$ and $e'(w^0)< c'$, we have
\begin{align*}
    1-\xi = \frac{e'(\tilde{w})-c'}{e'(\tilde{w})-e'(w^0)}< \frac{e'(\tilde{w})-c'}{c'-e'(w^0)}.
\end{align*}
If $c\leq c'$,  $e(\tilde{w})\leq c \leq c'$ and thus $e'(\tilde{w})-c' \leq e'(\tilde{w})-e(\tilde{w})$. %
This ensures that
\begin{align*}
    1-\xi \leq \frac{e'(\tilde{w})-e(\tilde{w})}{c'-e'(w^0)}.
\end{align*}
To summarize, we establish that when $c\leq c'$,
\begin{align*}
    h(\bar{w})-h(\tilde{w}) \leq 4(\|A\|_{\infty}+\|a\|_{\infty})\frac{e'(\tilde{w})-e(\tilde{w})}{c'-e'(w^0)}.
\end{align*}

If $c'<c$ and $e(\tilde{w})\leq c'$, then
$$e'(\tilde{w})-c'\leq e'(\tilde{w})-e(\tilde{w}),$$
so the claim above still holds.
If $c'<c$ and $c'<e(\tilde{w})\leq c$, then $0<e(\tilde{w})-c'\leq c-c'$, so
\begin{align*}
    e'(\tilde{w})-c'\leq e'(\tilde{w})-e(\tilde{w})+c-c'.
\end{align*}
That is, we proved that
$$1-\xi \leq \frac{e'(\tilde{w})-e(\tilde{w})+\max\{c-c',0\}}{c'-e'(w^0)} .$$
Since $e'(\tilde{w})-e(\tilde{w}) \leq \sup_{w}|e'(w)-e(w)|$, this concludes the proof.

\subsection{Proof of Lemma~\ref{lem: ghatm fhatm}}\label{pf: ghatm fhatm}

On the event $\mathcal{E}$ defined in \eqref{event: m star}, we can take $m=m^*$ such that \eqref{eq: m star} holds, that is, 
\begin{align*}
    &\max \left[\sqrt{T_0}\|\vecl(\Sigmahatmstar-\Sigma)\|_{\infty},\sqrt{T_0}\|\gammahatmstar-\gamma\|_{\infty},\right]\leq \err,\\
    &\max \left[\sqrt{T_1}|\muhatmstar_Y-\mu_Y|,\sqrt{T_1}\|\muhatmstar-\mu\|_{\infty}\right]\leq \err.
\end{align*}

 We denote
 \begin{align*}
     \beta \in \Omega \iff \{f(\beta)\leq 0\},\quad \beta\in\Omegahatm  \iff \{\fhatm(\beta)\leq \rho_M\}.
 \end{align*}
 By the definition of $f$ in \eqref{eq: def of f}, $f(\beta^{(0)})=\|\gamma-\Sigma\beta^{(0)}\|_{\infty}-\lambda=-\lambda<0$ for $\lambda>0$. Since $\rho_M = C_1[\log(\min\{T_0,T_1\})/M]^{1/p}/\sqrt{T_0}$ and $C_1 \geq (2/c^*(\alpha_0))^{1/p}$, it follows that
 \begin{align*}
     \rho_M \geq \frac{1}{\sqrt{T_0}}\left(\frac{2\log(\min\{T_0,T_1\})}{M\cdot c^*(\alpha_0)}\right)^{\frac{1}{p}} =\frac{2\err}{\sqrt{T_0}}.
 \end{align*}
 On the event $\mathcal{E}$, such $\beta^{(0)}$ satisfies $\fhatmstar(\beta^{(0)})\leq -\lambda+\rho_M<\rho_M$ for $\rho_M\geq 2\err/\sqrt{T_0}$ by Lemma~\ref{lem: fhatmstar f}. 
 By definition of $\tildebetamstar$ and $\barbetamstar$, $\tildebetamstar \in \Omega$ and $\barbetamstar \in \Omega\cap\Omegahatmstar$ where $\Omegahatmstar$ is defined in \eqref{eq: m-th omega}.
 By Lemma~\ref{lem: control of obj}, on the event $\mathcal{E}$, we establish that
 \begin{align*}
     \ghatmstar(\barbetamstar)-\ghatmstar(\tildebetamstar)\lesssim \left(\|\Gammahatmstar\|_{\infty}+\|\deltahatmstar\|_{\infty}\right)\frac{\sup_{\beta\in\Delta^N}\left|\fhatmstar(\beta)-f(\beta)\right|}{\rho_M-\fhatmstar(\beta^{(0)})}.
 \end{align*}
Since $\|\Gammahatmstar\|_{\infty} \lesssim \|\Gamma\|_{\infty}+\err$ and $\|\deltahatmstar\|_{\infty} \lesssim \|\delta\|_{\infty}+\err$ for sufficiently large $M$, by Lemma~\ref{lem: fhatmstar f}, we ensure
\begin{align*}
    \ghatmstar(\barbetamstar)-\ghatmstar(\tildebetamstar)\lesssim \frac{\err}{\lambda\sqrt{T_0}} \asymp \frac{\left(\log(\min\{T_0,T_1\})/{M}\right)^{1/p}}{\lambda\sqrt{T_0}},
\end{align*}
since $\err \asymp [\log(\min\{T_0,T_1\})/M]^{1/p}$.
Similarly by Lemma~\ref{lem: control of obj}, on the event $\mathcal{E}$, we establish that
\begin{align*}
    \ghatmstar(\barbetamstar)-\ghatmstar(\betahatmstar)\lesssim \frac{1}{\lambda}\left(\frac{\err}{\sqrt{T_0}}+\rho_M\right) \asymp \frac{\left(\log(\min\{T_0,T_1\})/{M}\right)^{1/p}}{\lambda\sqrt{T_0}},
\end{align*}
since $\rho_M \asymp [\log(\min\{T_0,T_1\})/M]^{1/p}/\sqrt{T_0}$.
 This concludes the proof.

\subsection{Proof of Lemma~\ref{lem: control lambda0}}\label{pf: control lambda0}
Since $h:\Delta^N\to\mathbb{R}$ is convex and differentiable and $\beta^{(0)}\in\Delta^N$, we ensure $h(\beta^{(0)})-h(\beta)\leq \nabla h(\beta^{(0)})\tr(\beta^{(0)}-\beta)$ for $\beta\in\Delta^N$ \citep[][Equation~(3.2)]{boyd2004convex}. By H\"older inequality, we can further bound $h(\beta^{(0)})-h(\beta)$ as
\begin{align}
    h(\beta^{(0)})-h(\beta)\leq \nabla h(\beta^{(0)})\tr(\beta^{(0)}-\beta)\leq \|\nabla h(\beta^{(0)})\|_{1}\|\beta^{(0)}-\beta\|_{\infty}.\label{eq: h holder}
\end{align}
Since $\lambda=0$, the function $f(\beta)$ defined in \eqref{eq: def of f} can be simplified as
\begin{align*}
    f(\beta)=\left\|\frac{1}{T_0}\sum_{t=1}^{T_0}\E\left[X_t\left(X_t\tr\beta^{(0)}+u_t^{(0)}-X_t\tr\beta\right)\right]\right\|_{\infty} = \left\|\Sigma(\beta-\beta^{(0)})\right\|_{\infty}.
\end{align*}
Thus, by \eqref{eq: beta and beta0}, we have $$  f(\beta) \geq \frac{\lambda_{\min}(\Sigma)}{\sqrt{N}}\|\beta-\beta^{(0)}\|_{\infty},$$ and since $\lambda_{\min}(\Sigma)>0$, we establish
\begin{align}
    \|\beta-\beta^{(0)}\|_{\infty}\leq \frac{\sqrt{N}f(\beta)}{\lambda_{\min}(\Sigma)}.\label{eq: beta norm in f}
\end{align}
Combining \eqref{eq: h holder} and \eqref{eq: beta norm in f}, we establish
\begin{align*}
    h(\beta^{(0)})-h(\beta) \leq \|\nabla h(\beta^{(0)})\|_{1}\frac{\sqrt{N}f(\beta)}{\lambda_{\min}(\Sigma)}.
\end{align*}
This concludes the proof.

\subsection{Proof of Lemma~\ref{lem: V and Vhat}}\label{pf: V and Vhat}
\noindent

By the definition of $\mathcal{G}_6$ defined in \eqref{event: Covhat Zhat max}, we remark that
\begin{align}
    \limsup_{T_0,T_1\to\infty}\prob(\mathcal{G}_6^c) &\leq \limsup_{T_0,T_1\to\infty}\prob(\VY> 1.05^2\VhatY)+\limsup_{T_0,T_1\to\infty}\prob(\|\mathbf{V}_{\mu}-\Vhatmu\|_{\max} > d_{\mu})\nonumber\\
    &+\limsup_{T_0,T_1\to\infty}\prob(\|\mathbf{V}_{\gamma}-\Vhatgamma\|_{\max} > d_{\gamma})+\limsup_{T_0,T_1\to\infty}\prob(\|\mathbf{V}_{\Sigma}-\VhatSigma\|_{\max} > d_{\Sigma}).\label{eq: E1c upper}
\end{align}
We define 
\begin{align*}
     \sigma_Y^2= {T_1}\VY,\quad \bm{\Cov}_{\mu} = {T_1}\mathbf{V}_{\mu},\quad  \bm{\Cov}_{\Sigma}= {T_0}\mathbf{V}_{\Sigma}, \quad \bm{\Cov}_{\gamma}= {T_0}\mathbf{V}_{\gamma}.
\end{align*}
We observe that
\begin{equation*}
\begin{aligned}
    \prob( \VY> 1.05^2\VhatY) &= \prob (1.05^2T_1(\VY-\VhatY)> (1.05^2-1)T_1\VY) \\
        &  \leq \prob (1.05^2T_1(\VY-\VhatY)> (1.05^2-1)c_2)+\prob(T_1\VY> c_1),
\end{aligned}
\end{equation*}
where $c_1$ and $c_2$ are positive constants defined in \eqref{eq: max cov} and \eqref{eq: min cov}, respectively.
Since $T_1\VY \geq c_2$ and the probability of the event $\{T_1\VY \leq c_1\}$ is 1 for positive constants $c_1,c_2>0$ by \eqref{eq: max cov} and \eqref{eq: min cov}, we have
\begin{equation*}
\lim_{T_1\to\infty}\prob (1.05^2T_1(\VY-\VhatY)> (1.05^2-1)c_2)+\lim_{T_1\to\infty}\prob(T_1\VY> c_1) = 0.
\end{equation*}
Thus, we have $\limsup_{T_1\to\infty}\prob( \VY> 1.05^2\VhatY)=0$.
Recall that $d_{\mu} = \|\Vhatmu\|_{\max}$, $d_{\Sigma} = \|\VhatSigma\|_{\max}$, $d_{\gamma} = \|\Vhatgamma\|_{\max}$.
For positive constant $c_2$ in \eqref{eq: min cov}, we have
\begin{align*}
    &\prob\left(\|\Vmu-\Vhatmu\|_{\max} > \|\Vhatmu\|_{\max}\right)\leq \prob\left(\|\Vmu-\Vhatmu\|_{\max} > \|\Vmu\|_{\max}-\|\Vmu-\Vhatmu\|_{\max}\right)\\
    &\leq \prob\left(\|T_1(\Vmu-\Vhatmu)\|_{\max}>\|T_1\Vmu\|_{\max}/2\right)\\
    &\leq  \prob\left(\|T_1(\Vmu-\Vhatmu)\|_{\max}>c_2/2\right).
\end{align*}
Since $\prob(\|T_1(\Vmu-\Vhatmu)\|_{\max}>c_2/2)\to 0$ as $T_1\to\infty$ by Assumption~\ref{assumption: clt and cons}, we establish
\begin{align*}
    \lim_{T_1\to\infty}\prob(\|\mathbf{V}_{\mu}-\Vhatmu\|_{\max} > \|\Vhatmu\|_{\max}) = 0
\end{align*}
In a similar way, we have
\begin{align*}
    \lim_{T_0\to\infty}\prob(\|\mathbf{V}_{\gamma}-\Vhatgamma\|_{\max} > \|\Vhatgamma\|_{\max}) =\lim_{T_0\to\infty}\prob(\|\mathbf{V}_{\Sigma}-\VhatSigma\|_{\max} > \|\VhatSigma\|_{\max})= 0
\end{align*}
Thus, the right-hand side terms in \eqref{eq: E1c upper} vanish, and we proved that $\limsup_{T_0,T_1\to\infty}\prob(\mathcal{G}_6^c)=0$.

\section{Detailed Procedure of Semi-real Data Analysis}\label{sec: semi-real simulation}
In this section, we present the full procedures and results of the semi-real data analyses based on the Basque Country data \citep{abadie2003economic}, illustrating the challenges of applying the SC method discussed in Section~\ref{subsec: challenges}.
We first examine weight instability by adding random noise to the pre-treatment data.
For $c \in \{0, 0.05, 0.1, 0.15\}$, which denotes noise level, we generate perturbed pre-treatment outcomes for both the treated and control units, denoted $\tilde{Y}_{1,t}$ and $\tilde{X}_t$, by adding small random noise to the original outcomes. The precise formulation of these perturbations is given below:
\begin{align*} &\tilde{Y}_{1,t} = Y_{1,t} + Z_t, \quad Z_t \overset{i.i.d.}{\sim} \mathcal{N}(0, (c\sigma_Y)^2), \quad \text{for } t = 1, \dots, T_0, \\ &\tilde{X}_t = X_t + W_t, \quad W_t \overset{i.i.d.}{\sim} \mathcal{N}(\mathbf{0}_N, (c\sigma_{X})^2 \mathbf{I}_N), \quad \text{for } t = 1, \dots, T_0, \end{align*}
where we compute the sample variance of $\{Y_{1,t}\}_{t=1}^{T_0}$ as
$$\sigma_Y^2 = \frac{1}{T_0-1}\sum_{t=1}^{T_0}\left(Y_{1,t}-\frac{1}{T_0}\sum_{t=1}^{T_0}Y_{1,t}\right)^2,$$ and the minimum diagonal element of the sample covariance matrix of $\{X_t\}_{t=1}^{T_0}$ as
\begin{align}
    \sigma_X^2=\min\left\{{\rm diag}\left[\frac{1}{T_0-1}\sum_{t=1}^{T_0}\left(X_t-\frac{1}{T_0}\sum_{t=1}^{T_0}X_t\right)\left(X_t-\frac{1}{T_0}\sum_{t=1}^{T_0}X_t\right)\tr\right]\right\},\label{eq: sigma X}
\end{align}
which are computed by assuming i.i.d.\ pre-treatment data.

\begin{figure}[ht]
    \centering
    \includegraphics[width=\linewidth]{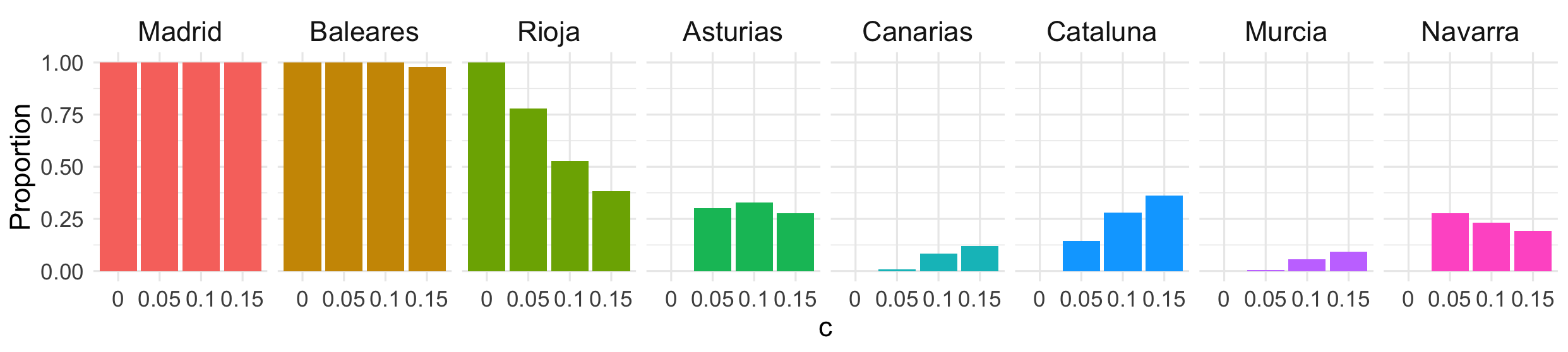}
    \caption{ 
    The proportion of the control units being selected by the SC method out of 1000 perturbed data sets. 
    The variable \texttt{c} indicates the noise level applied to the dataset to generate the pre-treatment data.}
    \label{fig: basque supp adding error full}
\end{figure}

With the newly generated pre-treatment outcomes $\{\tilde{Y}_{1,t}, \tilde{X}_t\}_{t=1}^{T_0}$, we apply the SC method, by estimating $\taubar$ using $\tauhatSC$ in \eqref{eq: SC estimators}. Figure~\ref{fig: basque supp adding error full} reports the full results of the proportion of times each control unit being selected by
SC across 1000 simulated pre-treatment data.
Madrid and Baleares remain consistently selected across noise levels, whereas Rioja is selected less frequently and the similar regions Asturias and Cataluna are selected more often as the noise level increases. This substitution among highly correlated controls illustrates the weight instability caused by failure of (E1): small perturbations of the pre-treatment data can produce different, nearly equivalent synthetic weights.

\begin{figure}
\centering
\raisebox{-10px}{
\begin{minipage}[t]{.15\linewidth}
\vspace{0pt}
 \centering
  \small
   \begin{tabular}[b]{c|cc}
Region &$\beta^{(0)}$ &$\beta^{(1)}$\\ \hline
 Madrid    & 0.483     & 0.483 \\
 Baleares       & 0.311        & $(1-\kappa)\cdot0.311$\\
Rioja        & 0.206          & $(1-\kappa)\cdot0.206$\\  
Cataluna & 0  & $\kappa\cdot0.311$\\ 
Asturias & 0  & $\kappa\cdot0.206$ \\ \hline
\end{tabular}%
\end{minipage}
}
\hfill
\begin{minipage}[t]{.6\linewidth}
\vspace{1pt}
    \centering
    \includegraphics[width=.87\linewidth]{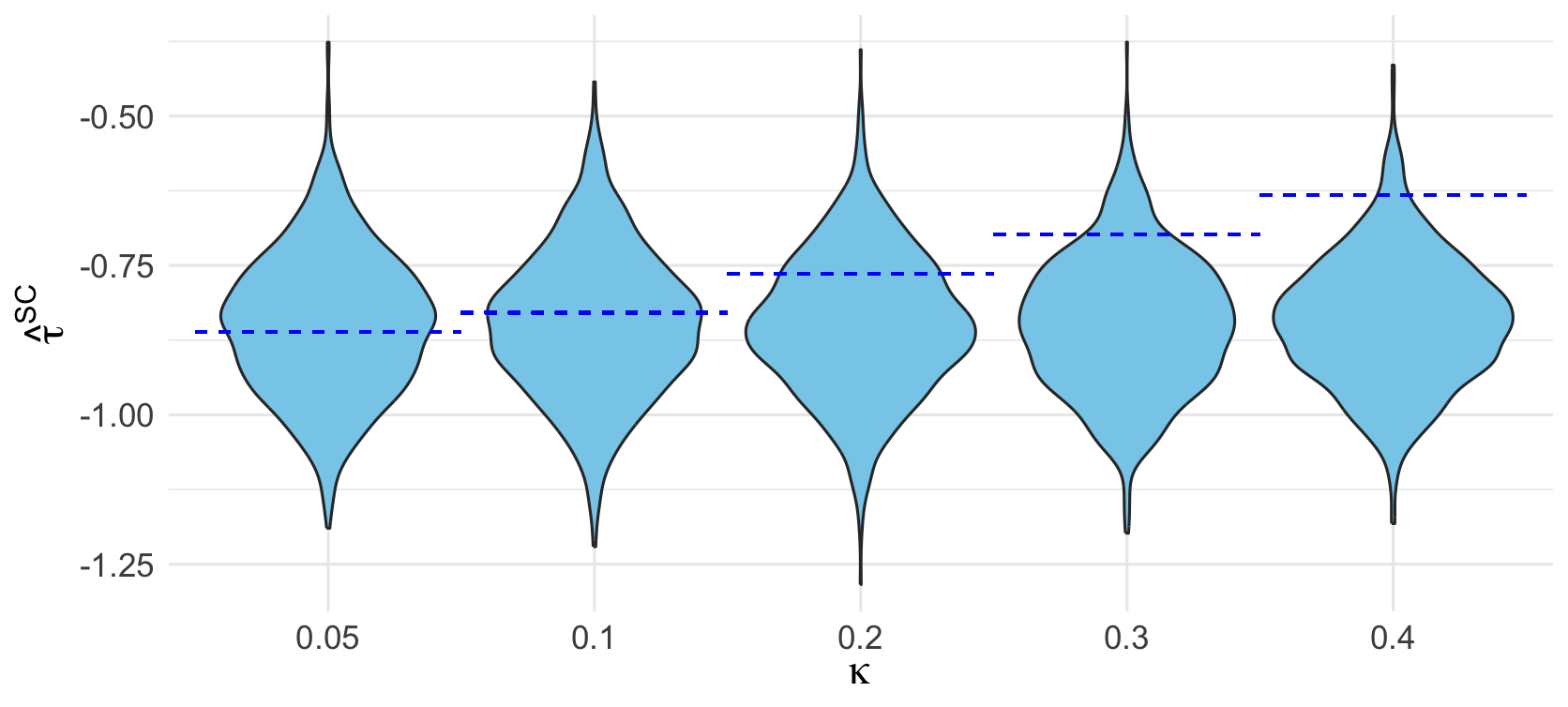}
\end{minipage}
    \caption{The left table reports the pre- and post-treatment weights used to generate the model \eqref{eq: outcome model}. The right panel shows violin plots of $\tauhatSC$ in \eqref{eq: SC estimators} across 1000 simulations for each $\kappa \in \{0.05, 0.1, 0.2,0.3, 0.4\}$; the blue dashed line denotes the true $\taubar$.}
    \label{fig: basque weight shift}
\end{figure}

Next, we describe the simulation procedure for weight shifts. To investigate the impact of weight shifts, we fix all parameters based on the dataset. Specifically, using the original data, we set $\beta^{(0)}$ in \eqref{eq: outcome model} based on the estimated weights from the SC method, as shown in the left table of Figure~\ref{fig: basque weight shift}. With this weight $\beta^{(0)}$, we define $\tau_t = Y_{1,t} - X_t\tr \beta^{(1)}$ and $\taubar = T_1^{-1}\sum_{t=T_0+1}^{T} \tau_t$. To model a post-treatment weight shift, we introduce a new weight $\beta^{(1)}$ for the post-treatment period, as shown in the left table of Figure~\ref{fig: basque weight shift}.

For each simulation, we generate control units by adding random noise to the pre-treatment outcomes:
 $$\tilde{X_t} = X_t +W_t,\quad W_t\overset{i.i.d.}{\sim}\mathcal{N}(\mathbf{0}_N,\sigma_{X}^2\mathbf{I}_N),\quad t=1,...,T, $$ 
 where $\sigma_{X}^2$ is defined in \eqref{eq: sigma X}.
We then generate potential outcomes for the treated unit under control, denoted by $\tilde{Y}_{1,t}^{(0)}$, according to \eqref{eq: outcome model}, using $\tilde{Y}_{1,t}^{(0)}$ along with $\{\tilde{X}_t\}_{t=1}^{T}$, $\beta^{(0)}$, and $\beta^{(1)}$.
The error terms $\{u_t\}_{t=1}^{T}$ in \eqref{eq: outcome model} are independently drawn as $u_t \sim \mathcal{N}(0, \sigma_{u}^2)$ for $t = 1, \dots, T$, where $\sigma_{u}^2 = T_0^{-1} \sum_{t=1}^{T_0} (Y_{1,t} - X_t\tr \beta^{(0)})^2$.
We generate $\tilde{Y}_{1,t}^{(1)}$, perturbed potential outcomes for the treated unit under treatment for the post-treatment period, as follows:
\begin{align*}
    &\tilde{Y}_{1,t}^{(1)} = \tilde{Y}_{1,t}^{(0)}+\tau_t+v_t, \quad v_t\overset{i.i.d.}{\sim}\mathcal{N}(0,\sigma_{v}^2), \text{ for } t=T_0+1,...,T,
\end{align*}
where $\sigma_{v}^2 = T_1^{-1} \sum_{t=T_0+1}^{T} (\tau_t - \taubar)^2$. With the newly generated outcomes $\{\tilde{Y}_{1,t}, \tilde{X}_t\}_{t=1}^{T}$, we compute the SC and WRoTE estimators as described in \eqref{eq: SC estimators} and Algorithm~\ref{alg:DRoSC}, respectively. For our proposal, we set $\lambda = \|\Sigma(\beta^{(0)} - \beta^{(1)})\|_\infty$ with $\Sigma = T_0^{-1} \sum_{t=1}^{T_0} X_t X_t\tr + \sigma_{X}^2 \mathbf{I}_N$.

Next, we examine how weight shifts affect the SC estimator, illustrating the failure of (E2) discussed in Section~\ref{subsec: challenges} of the main paper. We set $\beta^{(0)}=\betahatSC$ and construct $\beta^{(1)}$ by shifting weights away from $\beta^{(0)}$ across pairs of similar regions (Baleares--Cataluna and Rioja--Asturias), with the shift magnitude controlled by $\kappa\in[0.05,0.4]$, as specified in the left panel of Figure~\ref{fig: basque weight shift}. For each $\kappa$, we generate 1,000 semi-real datasets by adding independent noise to the control units and computing the treated unit's outcomes from \eqref{eq: outcome model} with $\beta^{(0)}$ and $\beta^{(1)}$, and we compute $\tauhatSC$ in \eqref{eq: SC estimators} and $\tauhat$ in \eqref{eq: betahat tauhat} for each dataset.

\begin{figure}[ht]
    \centering
    \includegraphics[width=.9\textwidth]{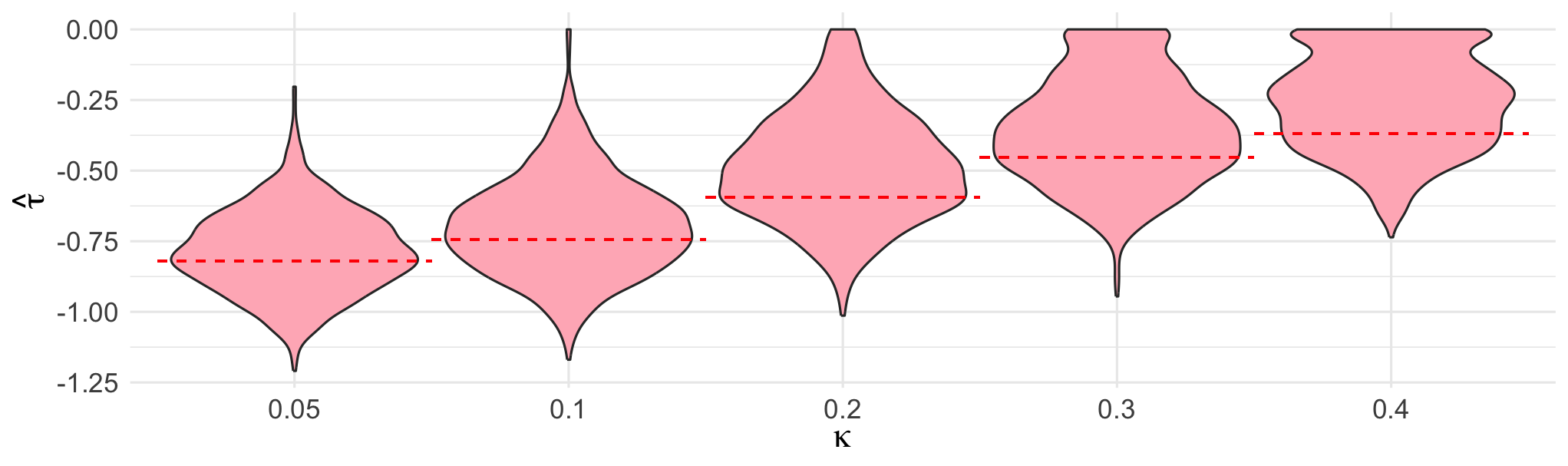}
    \caption{The violin plots of $\tauhat$, defined in \eqref{eq: betahat tauhat}, across 1000 simulations using newly generated data for each $\kappa \in \{0.05, 0.1, 0.2,0.3, 0.4\}$. The red dashed line marks the true value of $\tau^*$.}
    \label{fig: basque weight shift DRoSC}
\end{figure}

We present the results for the SC and proposed estimators in Figures~\ref{fig: basque weight shift} and \ref{fig: basque weight shift DRoSC}, respectively.
The right panel of Figure~\ref{fig: basque weight shift} shows that $\tauhatSC$ fails to accurately estimate $\taubar$ as $\kappa$ grows, reflecting bias induced by weight shifts and thus a violation of (E2).
In contrast, our proposed estimator $\tauhat$ consistently recovers the estimand $\tau^*$, as shown by the alignment of the violin plot centers with the red dashed line in Figure~\ref{fig: basque weight shift DRoSC}. However, as $\kappa$ increases, the violin plots become increasingly irregular in shape, implying possible non-regularity of the estimator.

\section{Additional Simulation Results}\label{sec: additional simulation}
In this section, we present the exact data-generating models and additional simulation results omitted from the main paper. Section~\ref{subsec: dgp} details the data-generating models for control units' outcomes. Section~\ref{subsec: simulation negligible} demonstrates the effectiveness of the perturbation-based inference procedure. 
Section~\ref{subsec: sens to prop} illustrates how the proportion of feasible solutions affects the resulting CI. 
Section~\ref{subsec:inconsistency} provides simulation results for inference using the covariance estimators in \eqref{eq: Vhat} under an AR(1) process.
 Section~\ref{subsec: additional inference} presents additional inference results.
 Finally, Sections~\ref{subsec: sim est} present estimation results, and Section~\ref{subsec: sens rho} examines sensitivity to the choice of $\rho$ for estimation, respectively.

\subsection{Data-generating Models}\label{subsec: dgp}
We specify the data-generating model of control units' outcomes $X_t$ for $t=1,\ldots,T$. In the pre-treatment period, we generate $\{X_t\}_{t=1}^{T_0}$ as follows:
\begin{align}\label{eq: equi-correlation X}
    X_t-\mu_0 = \phi(X_{t-1}-\mu_0)+\nu_t,\quad\nu_t\overset{i.i.d.}{\sim}\mathcal{N} (0,(1-\rho_0)\mathbf{I}_N+\rho_0\mathbf{1}_N\mathbf{1}_N\tr)\quad\text{for}\quad t=1,...,T_0,
\end{align} 
where $X_0$ is set to $\mu_0$ to ensure \eqref{eq: equi-correlation X} is well-defined, and $\phi \in [0,1)$ denotes the AR(1) coefficient. The covariance structure of $\nu_t$ induces an equicorrelation of $\rho_0$ between each pair of control units in the pre-treatment period.
In the post-treatment period, the control units' outcomes follow a similar model as \eqref{eq: equi-correlation X} with different mean and covariance:
\begin{align*}
   X_t-\mu = \phi(X_{t-1}-\mu)\I(t >T_0+1)+\nu_t,\quad\nu_t\overset{i.i.d.}{\sim}\mathcal{N} (0,\mathbf{I}_N)\quad\text{for}\quad t=T_0+1,...,T.
\end{align*}

We generate the setting (S1) with Conditions (E1) and (E2) satisfied. 
\begin{enumerate}
\item[(S1)] $\mu = \mu_0$, $\rho_0 = 0.25$, and $\beta^{(1)} = \beta^{(0)}$.
\end{enumerate}
In (S1), the control units are not highly correlated and no weight shift occurs, so $\taubar$ is identifiable by the SC method. In the additional two settings, we introduce violations of (E1) or (E2). (S2) is presented in Section~\ref{sec: simulation studies} in the main paper. (S3) is defined as follows:
\begin{enumerate}
\item[(S3)] We consider a mean shift with $\mu_0 = \mu = \mathbf{1}_N + N^{-1}(1, \dots, N)\tr$, a large weight shift via $\beta^{(1)} = \beta^{(0)} + 0.2 \cdot (-\mathbf{1}_3\tr, \mathbf{0}_{N-6}\tr, \mathbf{1}_3\tr)\tr$, and  low %
correlations by setting $\rho_0 = 0.25$.
\end{enumerate}

\subsection{Effectiveness of the Perturbation-based Inference}\label{subsec: simulation negligible}
In this section, we provide the numerical results to justify our perturbation-based inference method. Recall the decomposition of the estimation error $\tauhatm-\tau^*$ in \eqref{eq: decomposition of tauhatm}:
$$\tauhatm-\tau^* = \muhat_Y-\mu_Y-\left[(\muhatm)\tr\betahatm-\mu\tr\beta^*\right].$$
Theorem~\ref{thm: negligible} shows that with high probability, we ensure the existence of an index $m^*$ which makes $(\muhatmstar)\tr\betahatmstar \approx \mu\tr\beta^*$ for sufficiently large $M$. For such an $m^*$, we only need to quantify the asymptotically normal component $\muhat_Y-\mu_Y$. 

\begin{figure}[ht]
    \centering
    \includegraphics[width=.9\textwidth]{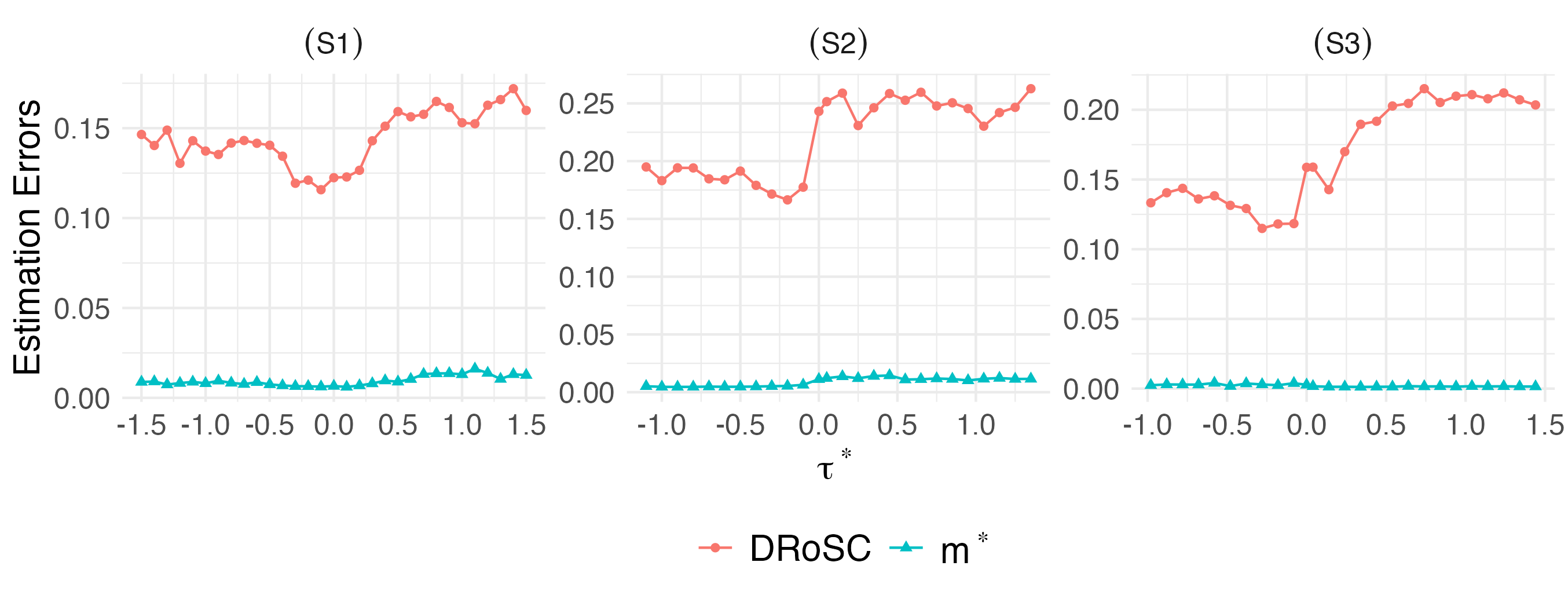}
    \caption{Empirical average of estimation errors under (S1), (S2), and (S3) from Section~\ref{sec: simulation studies}, {with respect to a range of $\tau^*$ values specified in the $x$-axis.} {\texttt{Proposed} and \texttt{m$^*$}} correspond to $|\muhat\tr\betahat - \mu\tr \beta^*|$ and $\min_{m\in\mathbb{M}}|(\muhatm)\tr\betahatm - \mu\tr \beta^*|$ where $\mathbb{M}$ is later defined in \eqref{eq: index set}, respectively. }
    \label{fig: est err}
\end{figure}

We provide numerical results for such an $m^*$ existing in Figure~\ref{fig: est err} with $T_0=T_1=25$ and $\phi=0$ and omit the results for other settings since they are similar to those with $T_0=T_1=25$ and $\phi=0$. Figure~\ref{fig: est err} illustrates that  $(\muhatmstar)\tr\betahatmstar - \mu\tr\beta^*$ is much smaller than $\muhat\tr\betahat-\mu\tr\beta^*$, having the values near zero.
This observation supports the intuition behind the decomposition of $\tauhatm-\tau^*$ in \eqref{eq: decomposition of tauhatm}.

\subsection{Sensitivity to the Proportion of Feasible Solutions}\label{subsec: sens to prop}

Our perturbation-based inference method involves a tuning parameter, defined as the proportion of feasible solutions $\betahatm$ obtained from the perturbed optimization problem \eqref{eq: m-th betahat} across $M$ perturbations. We set the default threshold to 10\%. Since this choice may influence the coverage and precision of the confidence interval in \eqref{eq: aggregated CI}, we empirically examine the sensitivity of the results to this proportion. The default threshold requires that at least 10\% of the $M$ perturbations yield feasible solutions, but we also consider alternative thresholds of 20\% and 30\% in this section. Consistent with the previous inference simulation in Section~\ref{sec: simulation studies}, we set $M = 500$. For simplicity, we present results with $T_0=25$ and $T_1=25$, since the other results with the different values of $T_0$ and $T_1$ are similar to those with $T_0=T_1=25$.

\begin{figure}[ht]
   \centering
   \includegraphics[width=.9\textwidth]{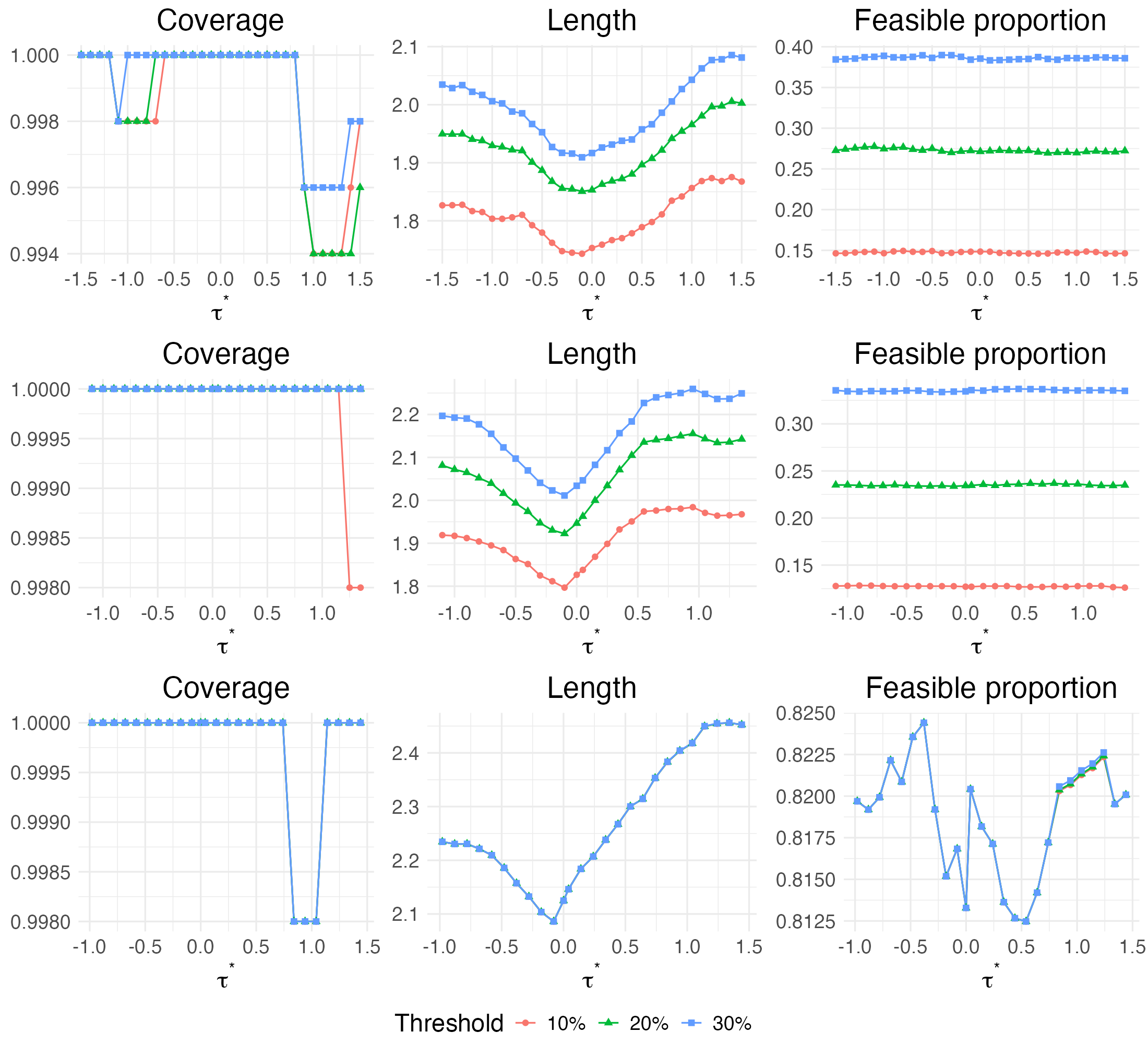}
  \caption{Sensitivity of coverage and average length of the proposed CI to varying proportions of feasible solutions. The leftmost and middle panels denote the empirical coverages and means of lengths of confidence intervals. The rightmost panel displays the empirical proportion of feasible solutions for each threshold. $x$-axes in all panels denote $\tau^*$. \texttt{Prop} denotes the threshold for the proportion of feasible solutions among the $M=500$ perturbations.}
  \label{fig: sensitivity to prop}
\end{figure}

As shown in Figure~\ref{fig: sensitivity to prop}, the average length of the proposed CI varies slightly with the choice of threshold in settings (S1) and (S2), while the empirical coverage remains relatively stable. In contrast, both coverage and length remain largely unchanged in (S3). This difference arises due to the value of $\lambda$ used in each simulation. Specifically, we recall that we set $\lambda = \|\Sigma(\beta^{(1)} - \beta^{(0)})\|_{\infty}$ in (S2) and (S3). Since the weight shifts are small in (S2), the corresponding $\lambda$ is also small and it is even zero in (S1). In contrast, (S3) involves large weight shifts, resulting in a larger $\lambda$. This larger $\lambda$ results in a greater number of feasible solutions even with a small tuning parameter $\rho_M$, thereby making the effect of the threshold choice less pronounced in this setting. This pattern is evident in the rightmost panel of Figure~\ref{fig: sensitivity to prop}: the proportion of feasible solutions increases with the threshold in (S1) and (S2), but shows little change in (S3).

\subsection{Inference Results with Inconsistent Covariance Estimators}\label{subsec:inconsistency}

{In this section, we report empirical coverage and interval lengths for the oracle and proposed CIs with $M=1000$. Although $\phi=0.5$, we deliberately use the i.i.d.-based covariance estimators in \eqref{eq: Vhat}, which are inconsistent under dependence; this probes the sensitivity of the procedure to misspecification of the perturbation covariances.} For simplicity, we only report the results with $T_0=25$ since the results with $T_0=50$ are similar to those with $T_0=25$.

\begin{figure}[ht]
    \centering
     \begin{subfigure}{.7\textwidth}
    \centering
    \includegraphics[width=\textwidth]{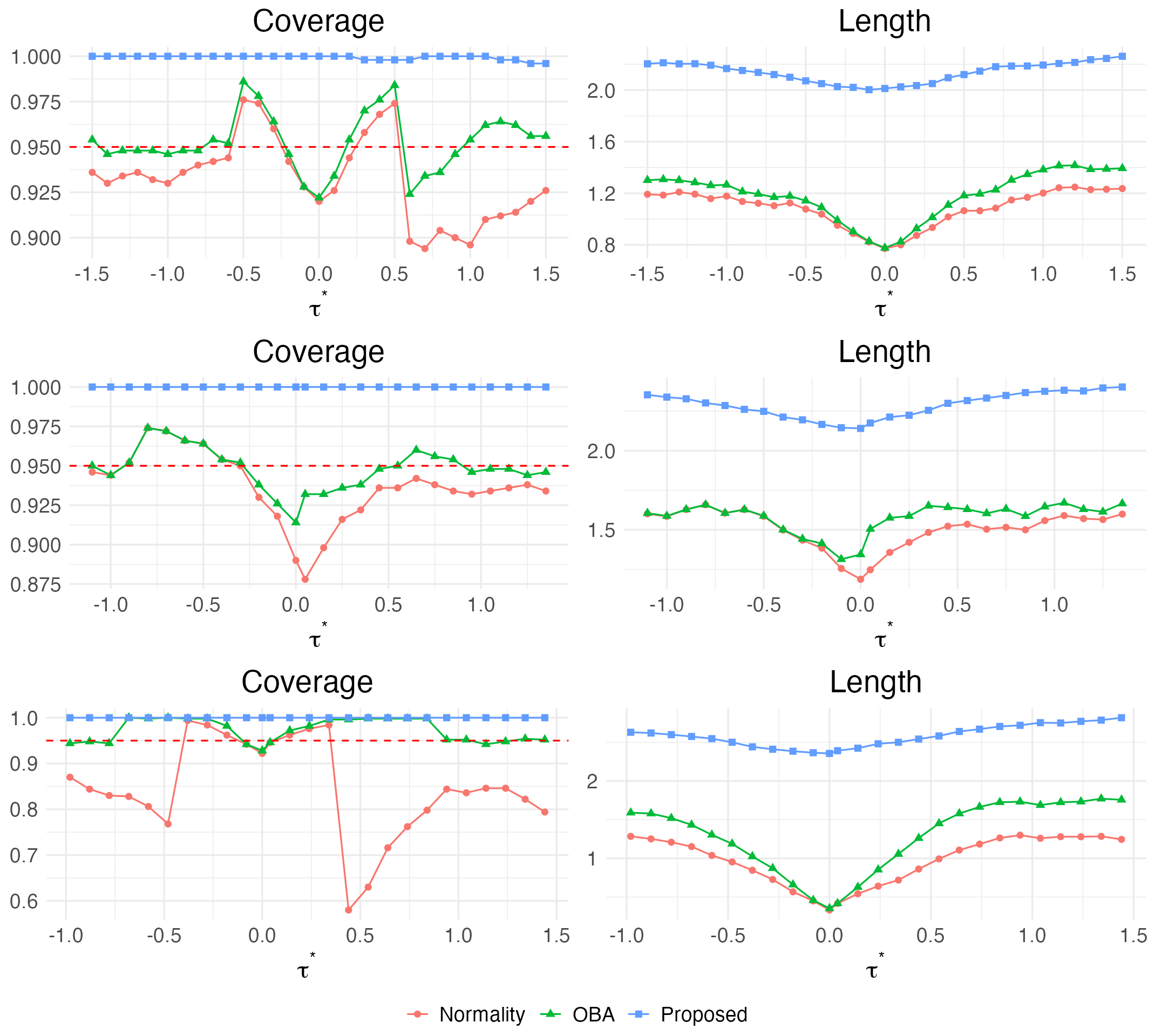}
  \end{subfigure}
   \begin{subfigure}{.7\textwidth}
    \centering
    \includegraphics[width=\textwidth]{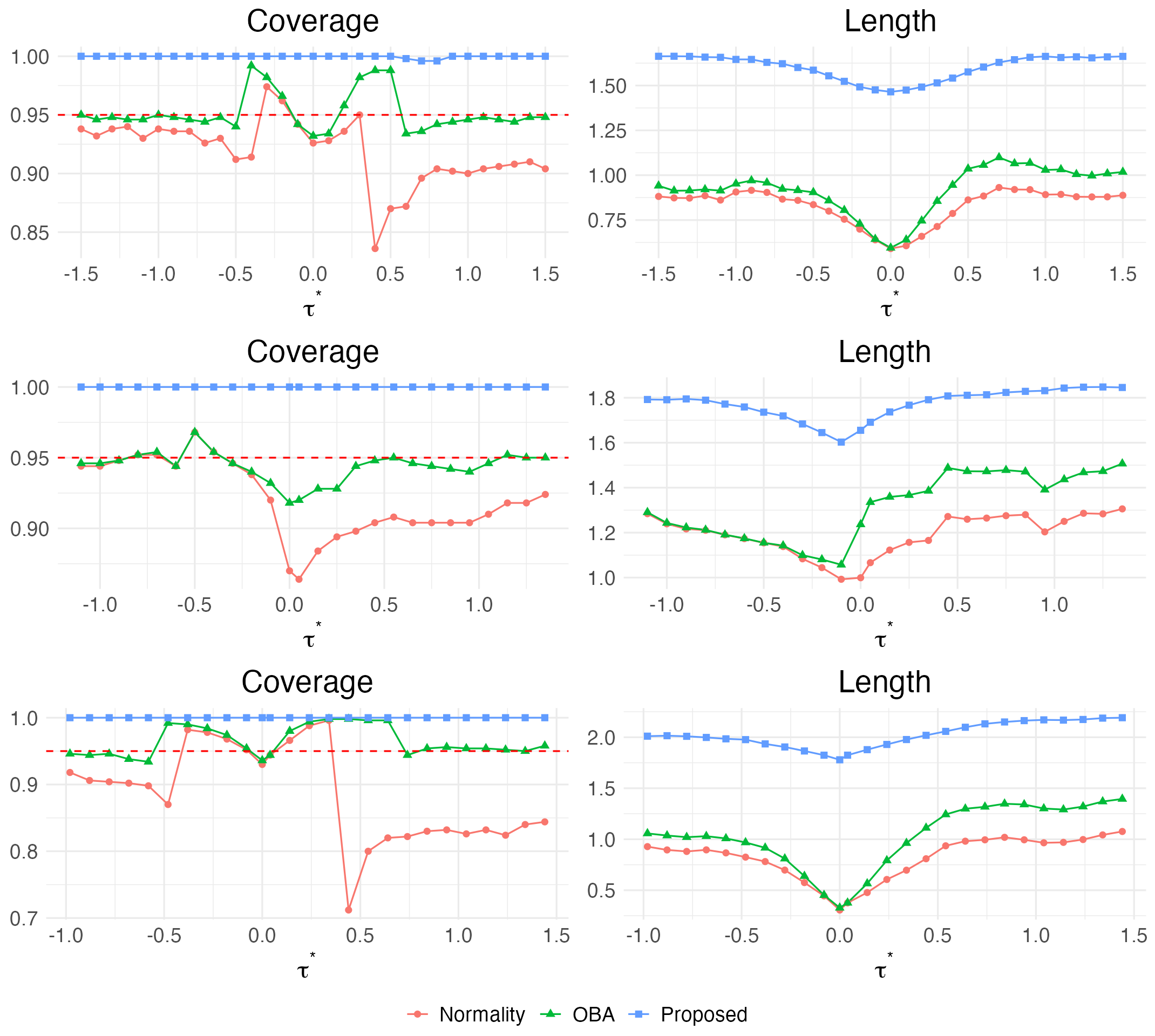}
  \end{subfigure}
    \caption{Empirical coverage and interval lengths under settings (S1), (S2), and (S3) described in Section~\ref{sec: simulation studies} with $\phi = 0.5$.  From top to bottom, Panels 1–3 correspond to $T_0=25, T_1=25$ with (S1), (S2), and (S3), respectively, while Panels 4–6 correspond to $T_0=25, T_1=50$ with (S1), (S2), and (S3), respectively. The $x$-axis denotes $\tau^*$.
    \texttt{Normality} refers to the oracle normality-based CI, \texttt{OBA} denotes the oracle bias-aware CI, and \texttt{Proposed} corresponds to the proposed CI in \eqref{eq: aggregated CI}.}
    \label{fig:infer phi0.5 pre25}
\end{figure}

The results are presented in Figures~\ref{fig:infer phi0.5 pre25}. We emphasize that while we use inconsistent covariance estimator, our proposed CI does not suffer from under-coverage issue severely comparing to the oracle normality-based CI, and its length is comparable to and consistently longer than that of the oracle bias-aware CI. 

\subsection{Additional Inference Results}\label{subsec: additional inference}
\begin{figure}[ht]
    \centering
     \begin{subfigure}{.9\textwidth}
    \centering
    \includegraphics[width=\textwidth]{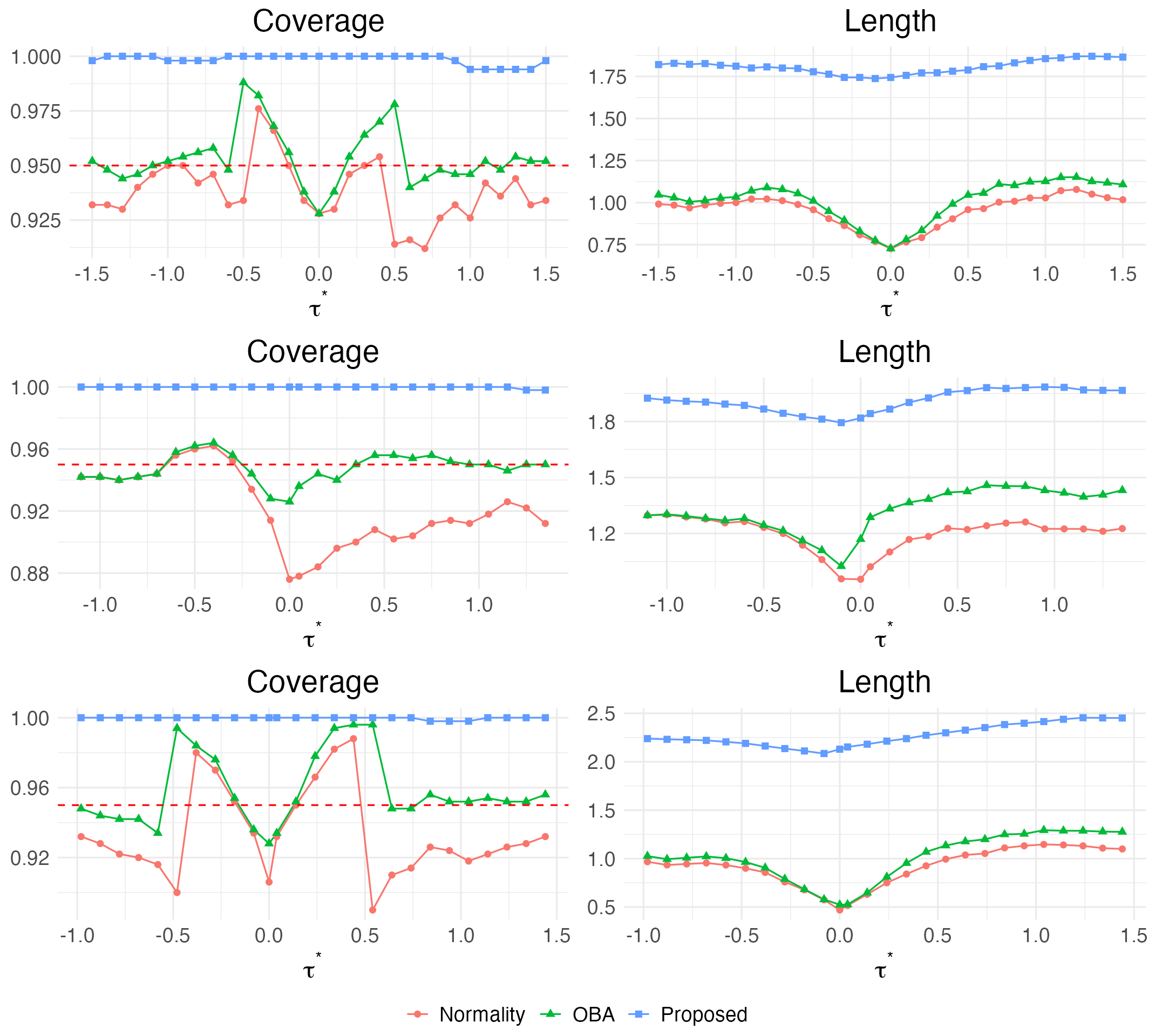}
  \end{subfigure}
    \caption{Empirical coverage and interval lengths under settings (S1), (S2), and (S3) described in Section~\ref{sec: simulation studies} with $\phi = 0$.  From top to bottom, Panels 1–3 correspond to $T_0=25, T_1=25$ with (S1), (S2), and (S3), respectively. The $x$-axis denotes $\tau^*$.
    \texttt{Normality} refers to the oracle normality-based CI, \texttt{OBA} denotes the oracle bias-aware CI, and \texttt{Proposed} corresponds to the proposed CI in \eqref{eq: aggregated CI}.}
    \label{fig:infer phi0 post25}
\end{figure}

\begin{figure}[ht]
    \centering
     \begin{subfigure}{.7\textwidth}
    \centering
    \includegraphics[width=\textwidth]{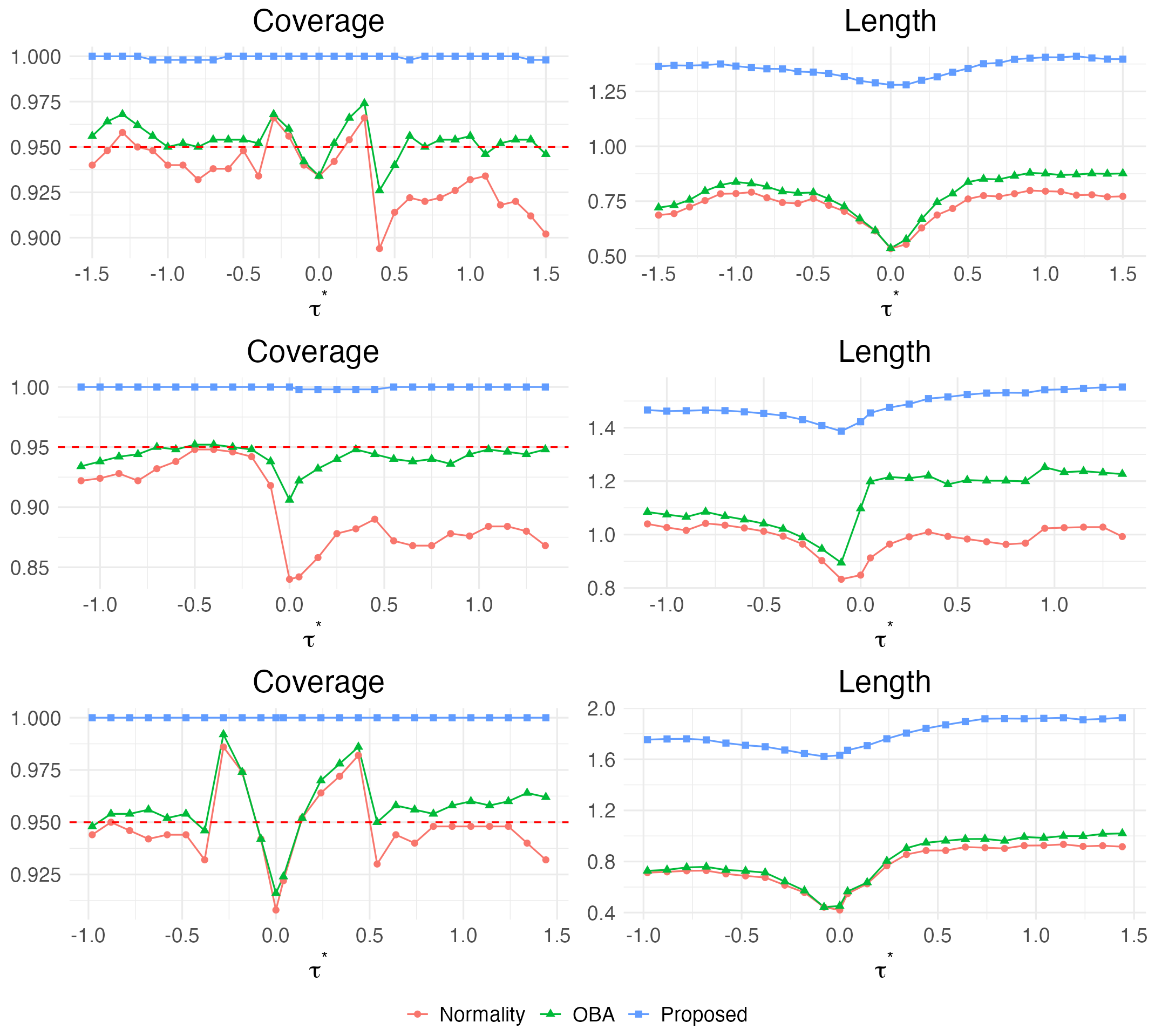}
  \end{subfigure}
   \begin{subfigure}{.7\textwidth}
    \centering
    \includegraphics[width=\textwidth]{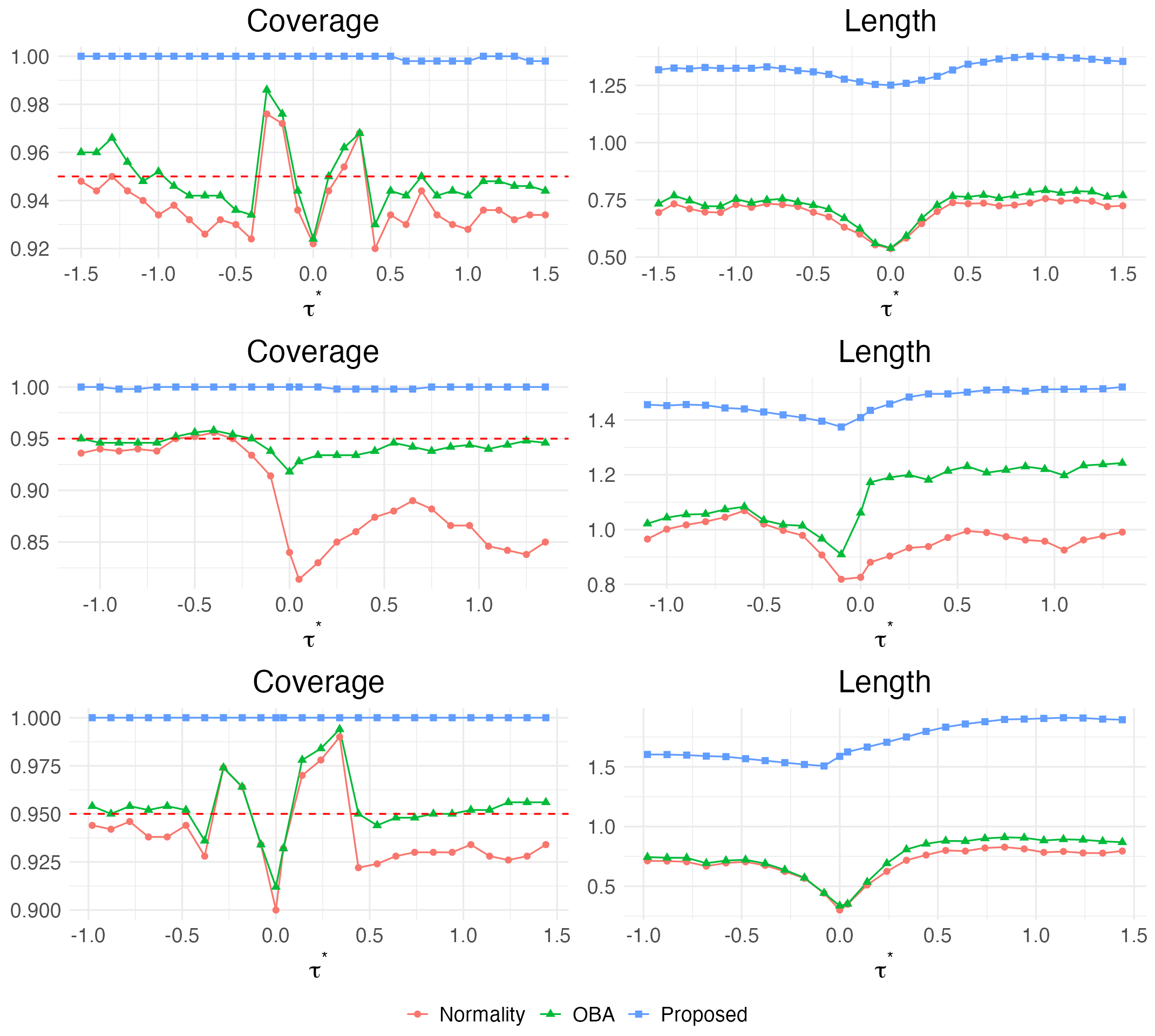}
  \end{subfigure}
    \caption{Empirical coverage and interval lengths under settings (S1), (S2), and (S3) described in Section~\ref{sec: simulation studies} with $\phi = 0$.  From top to bottom, Panels 1–3 correspond to $T_0=25, T_1=50$ with (S1), (S2), and (S3), respectively, while Panels 4–6 correspond to $T_0=50, T_1=50$ with (S1), (S2), and (S3), respectively. The $x$-axis denotes $\tau^*$.
    \texttt{Normality} refers to the oracle normality-based CI, \texttt{OBA} denotes the oracle bias-aware CI, and \texttt{Proposed} corresponds to the proposed CI in \eqref{eq: aggregated CI}.}
    \label{fig:infer phi0 post50}
\end{figure}

In this section, we provide empirical coverage and interval lengths for the oracle and proposed CIs. We present the setting presented in the main paper ($\phi=0$, $T_0=T_1=25$) under (S1) and (S3). Furthermore, we additionally report the results with $T_1 = 50$ and $\phi=0$ for simplicity since the result with $T_1=25$ is similar. The results are presented in Figures~\ref{fig:infer phi0 post25} and \ref{fig:infer phi0 post50}. 
Overall, the patterns are similar to those in Figure~\ref{fig:infer pre25 post25} of the main paper: our proposed CI does not suffer from under-coverage issue severely comparing to the oracle normality-based CI, and its length is slightly longer but comparable to that of the oracle bias-aware CI. 

\subsection{Estimation Results}\label{subsec: sim est}
In this section, we evaluate the numerical performance of our estimator $\tauhat$ for $\tau^*$, defined in \eqref{eq: betahat tauhat}, under settings (S1)–(S3) in Section~\ref{sec: simulation studies}. We implement the estimation procedure detailed in Section~\ref{sec: estimation of DRoSC}, using $\lambda=0$ for (S1) and $\lambda=\|\Sigma(\beta^{(1)}-\beta^{(0)})\|_{\infty}$ for (S2) and (S3), where $\Sigma=\Sigma_0+\mu_0\mu_0\tr$. Each setting is simulated 500 times, and we compare performance with the SC estimator $\tauhatSC$ for $\taubar$ from \eqref{eq: SC estimators}.

\begin{figure}[ht]
    \centering
    \begin{subfigure}{.85\textwidth}
    \centering
    \includegraphics[width=\textwidth]{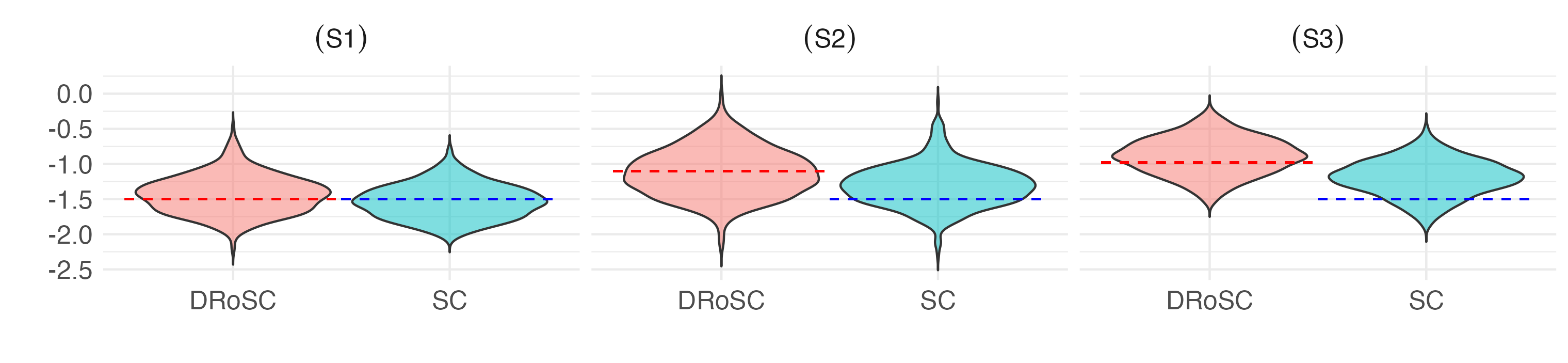}
  \end{subfigure}
  \begin{subfigure}{.85\textwidth}
    \centering
    \includegraphics[width=\textwidth]{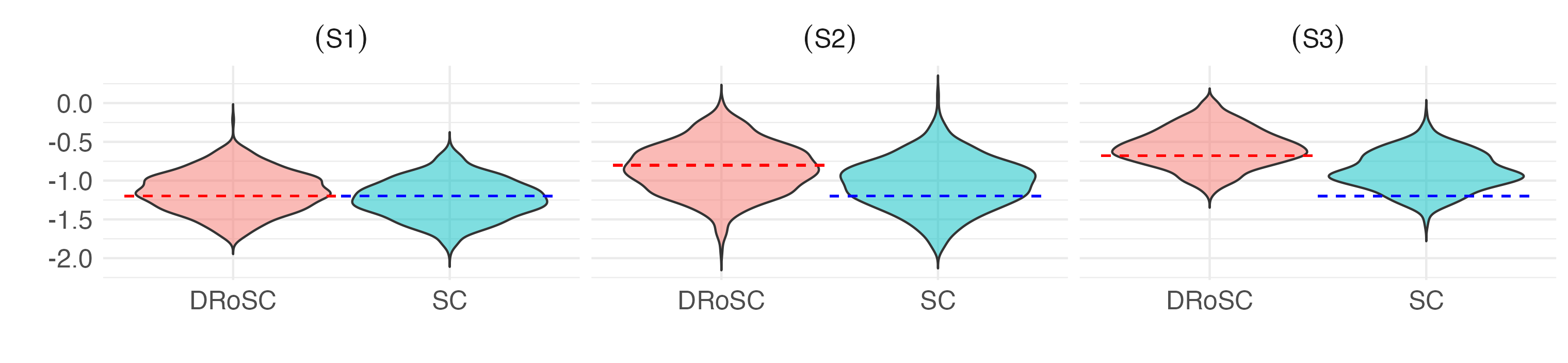}
  \end{subfigure}
  \begin{subfigure}{.9\textwidth}
    \centering
    \includegraphics[width=\textwidth]{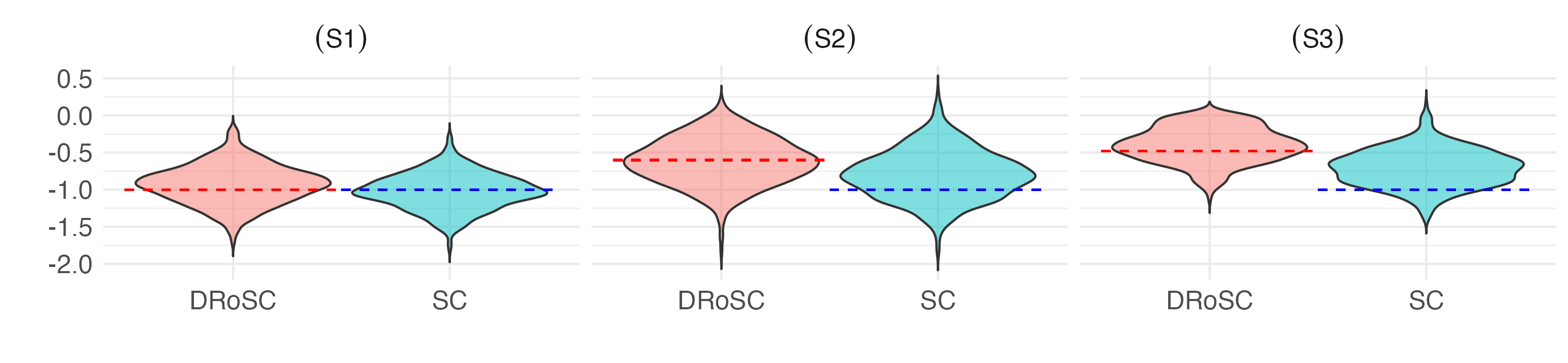}
  \end{subfigure}

    \caption{ {Violin plots of the proposed and SC estimators} from settings (S1), (S2), and (S3) when $\taubar = -1.5$, $-1.2$, and $-1$.
    The figures, from top to bottom, correspond to $\taubar=-1.5$, $-1.2$, and $-1$ respectively. 
    \texttt{DRoSC} and \texttt{SC} in $x$-axis denote the estimators \eqref{eq: betahat tauhat} and \eqref{eq: SC estimators} from our method and the SC method respectively. The red and blue dashed lines denote $\tau^*$ and $\taubar$ respectively. }
    \label{fig:violin pre25 post25}
\end{figure}

We report results for $\taubar=-1.5,-1.2,-1$ for (S1)–(S3) in Figure~\ref{fig:violin pre25 post25}. 
 In (S1), $\taubar$ is identifiable and $\taubar=\tau^*$.
Both $\tauhatSC$ and $\tauhat$ target the same quantity. Thus, in the left panel of Figure~\ref{fig:violin pre25 post25}, the blue ($\taubar$) and red dashed lines ($\tau^*$) coincide, and both estimators are consistent. In (S2) and (S3) where (E1) and (E2) are violated, $\taubar$ is unidentifiable and $\tau^*$ becomes a conservative proxy for $\taubar$. 
In such settings, $\tauhatSC$ is no longer a consistent estimator of $\taubar$. 
In contrast, our proposed estimator remains consistent for $\tau^*$ (red dashed lines in the middle and right panels of Figure~\ref{fig:violin pre25 post25}), even when (E1) and (E2) are violated.

\begin{figure}[ht]
    \centering
    \begin{subfigure}{.9\textwidth}
    \centering
    \includegraphics[width=\textwidth]{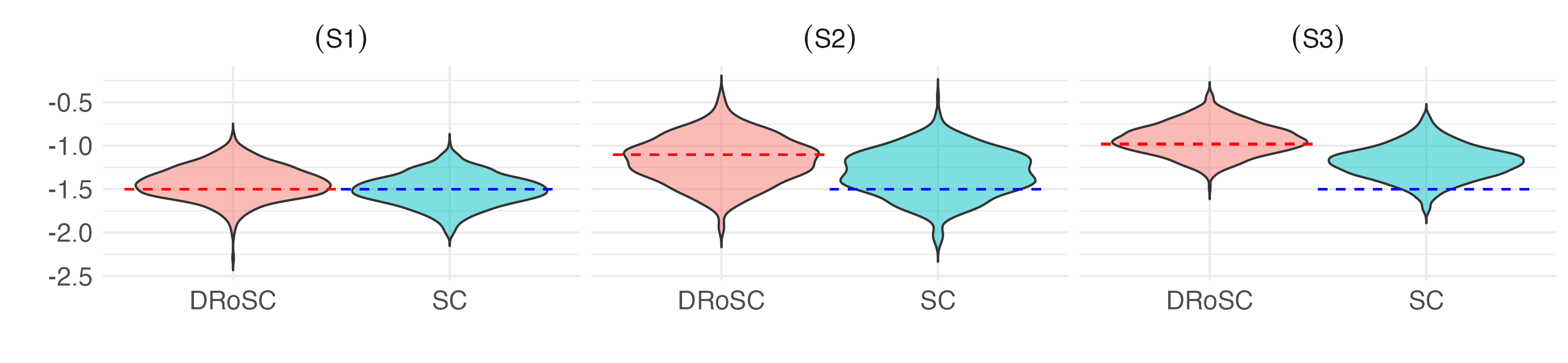}
  \end{subfigure}
  \begin{subfigure}{.9\textwidth}
    \centering
    \includegraphics[width=\textwidth]{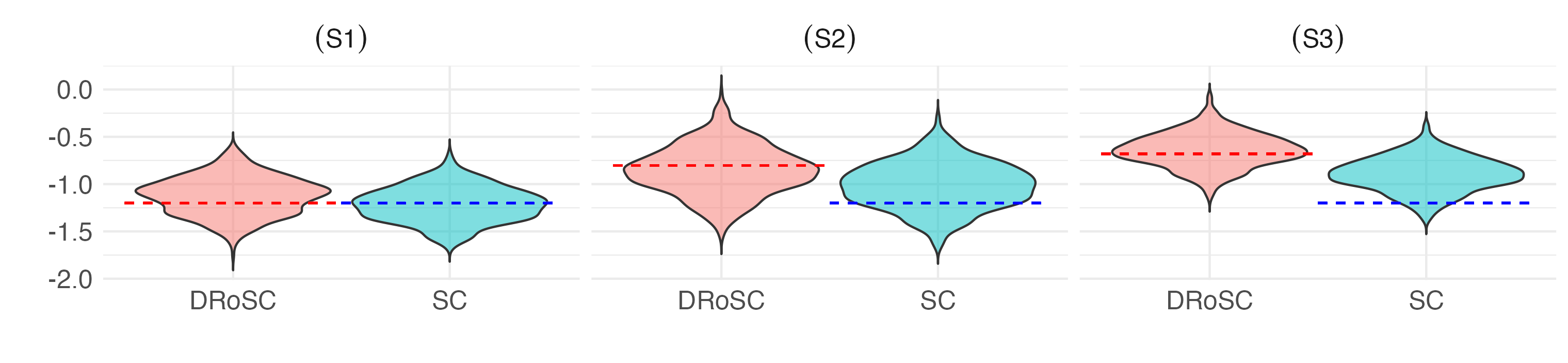}
  \end{subfigure}
  \begin{subfigure}{.9\textwidth}
    \centering
    \includegraphics[width=\textwidth]{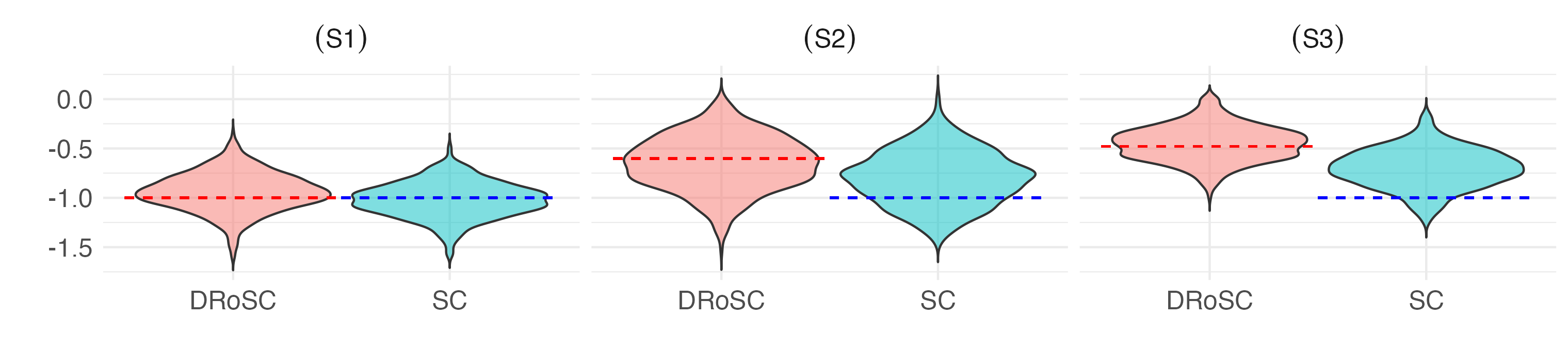}
  \end{subfigure}

    \caption{Simulation studies from settings (S1), (S2), and (S3) with $\phi = 0$, $T_0=25$, and $T_1=50$. The figures, from top to bottom, correspond to $\taubar=-1.5$, $-1.2$, and $-1$ respectively.  \texttt{DRoSC} and \texttt{SC} in $x$-axis denote the estimators \eqref{eq: betahat tauhat} and \eqref{eq: SC estimators} from our method and the SC method respectively. The red and blue dashed lines denote $\tau^*$ and $\taubar$ respectively. }
    \label{fig:violin pre25 post50}
\end{figure}

\begin{figure}[ht]
    \centering
    \begin{subfigure}{.8\textwidth}
    \centering
    \includegraphics[width=\textwidth]{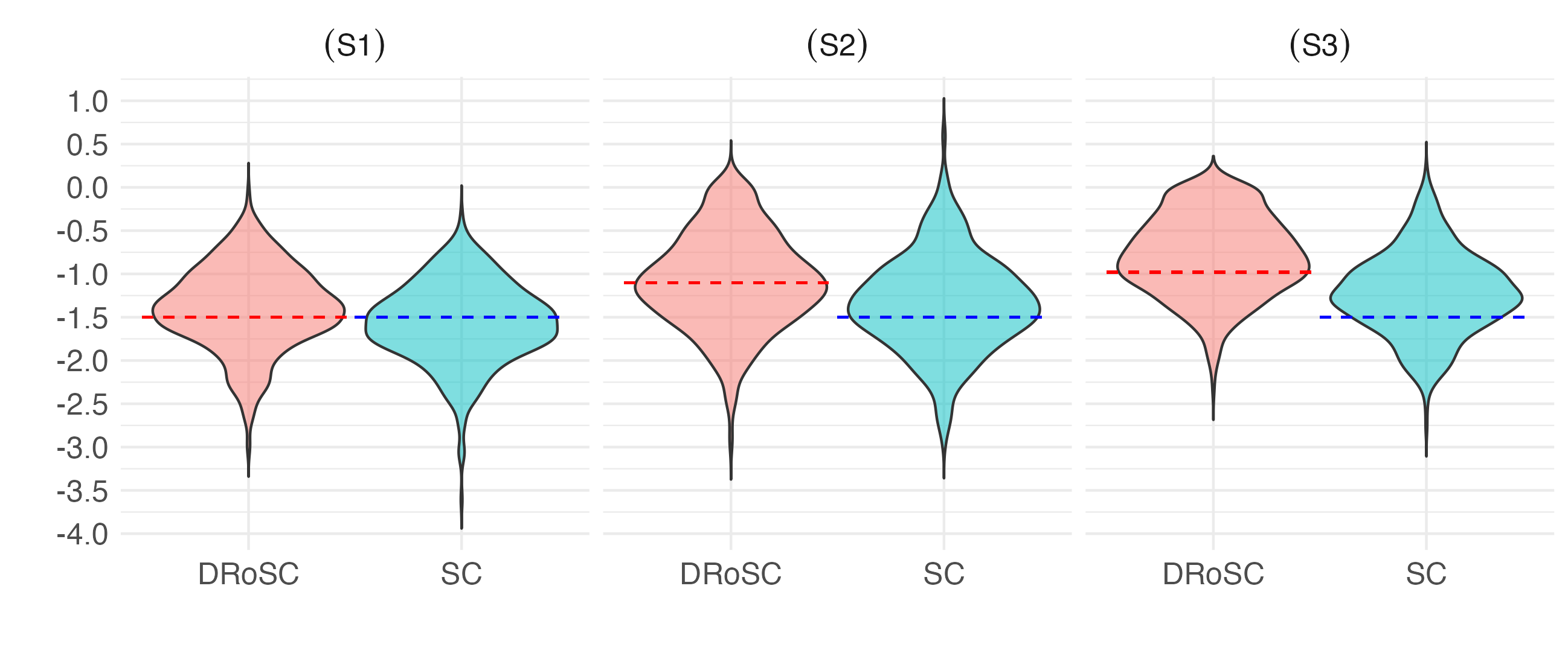}
  \end{subfigure}
  \begin{subfigure}{.8\textwidth}
    \centering
    \includegraphics[width=\textwidth]{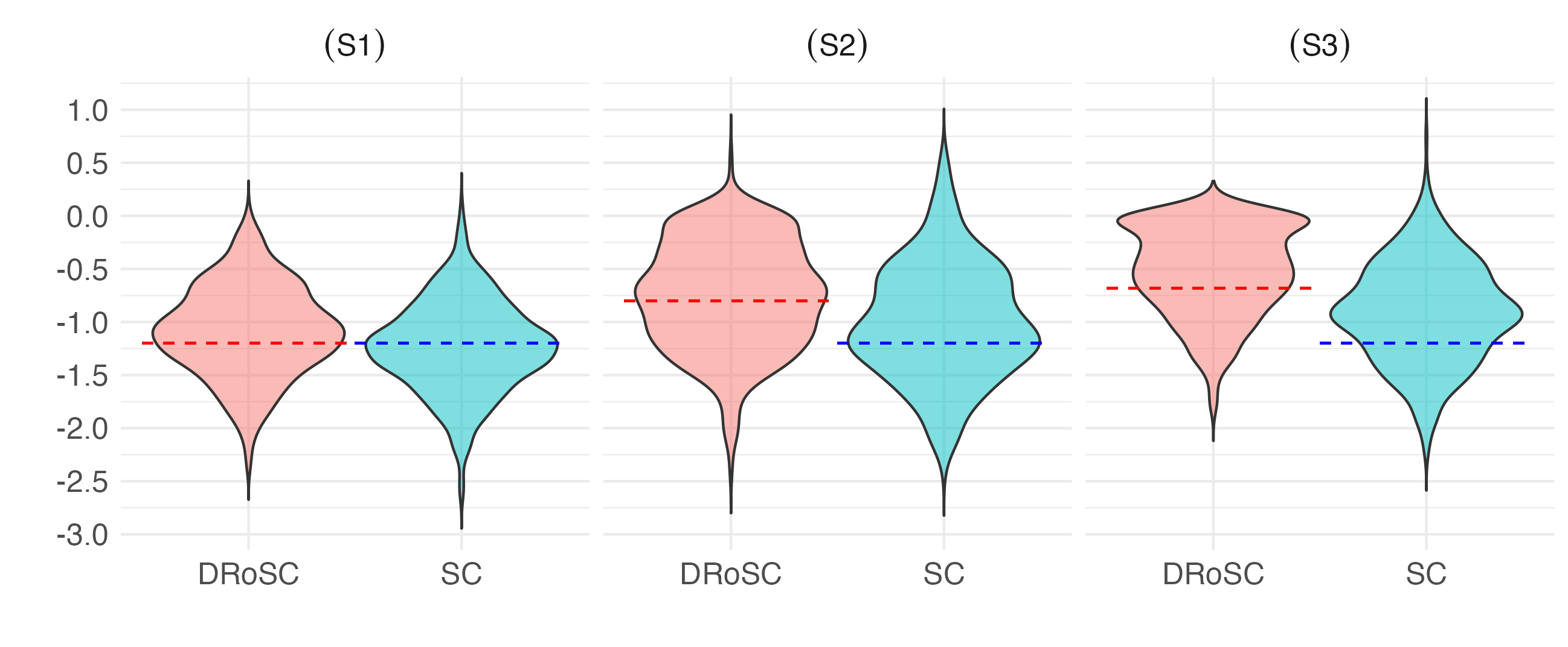}
  \end{subfigure}
  \begin{subfigure}{.8\textwidth}
    \centering
    \includegraphics[width=\textwidth]{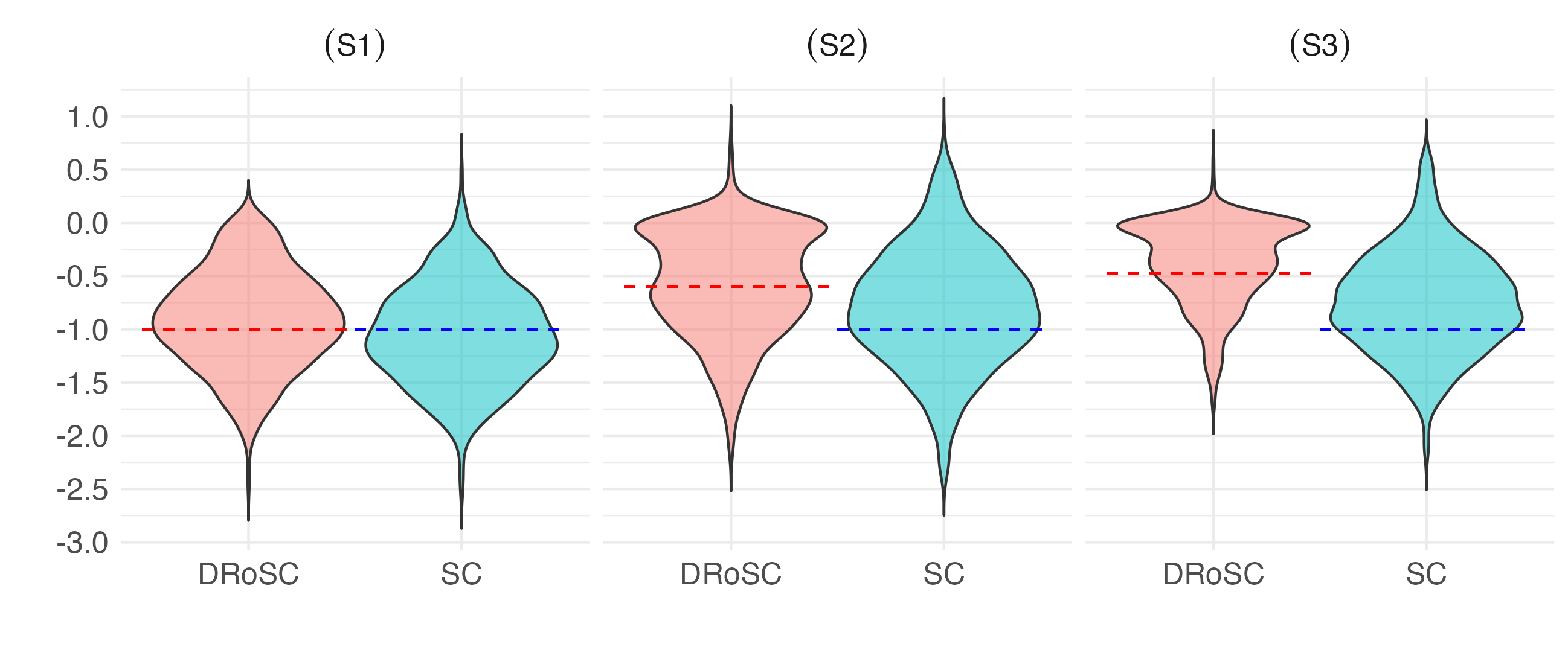}
  \end{subfigure}

    \caption{Simulation studies from settings (S1), (S2), and (S3) with $\phi = 0.5$, $T_0=25$, and $T_1=25$. The figures, from top to bottom, correspond to $\taubar=-1.5$, $-1.2$, and $-1$ respectively. \texttt{DRoSC} and \texttt{SC} in $x$-axis denote the estimators \eqref{eq: betahat tauhat} and \eqref{eq: SC estimators} from our method and the SC method respectively. The red and blue dashed lines denote $\tau^*$ and $\taubar$ respectively. }
    \label{fig: violin phi0.5 pre25 post25}
\end{figure}

\begin{figure}[ht]
    \centering
    \begin{subfigure}{.8\textwidth}
    \centering
    \includegraphics[width=\textwidth]{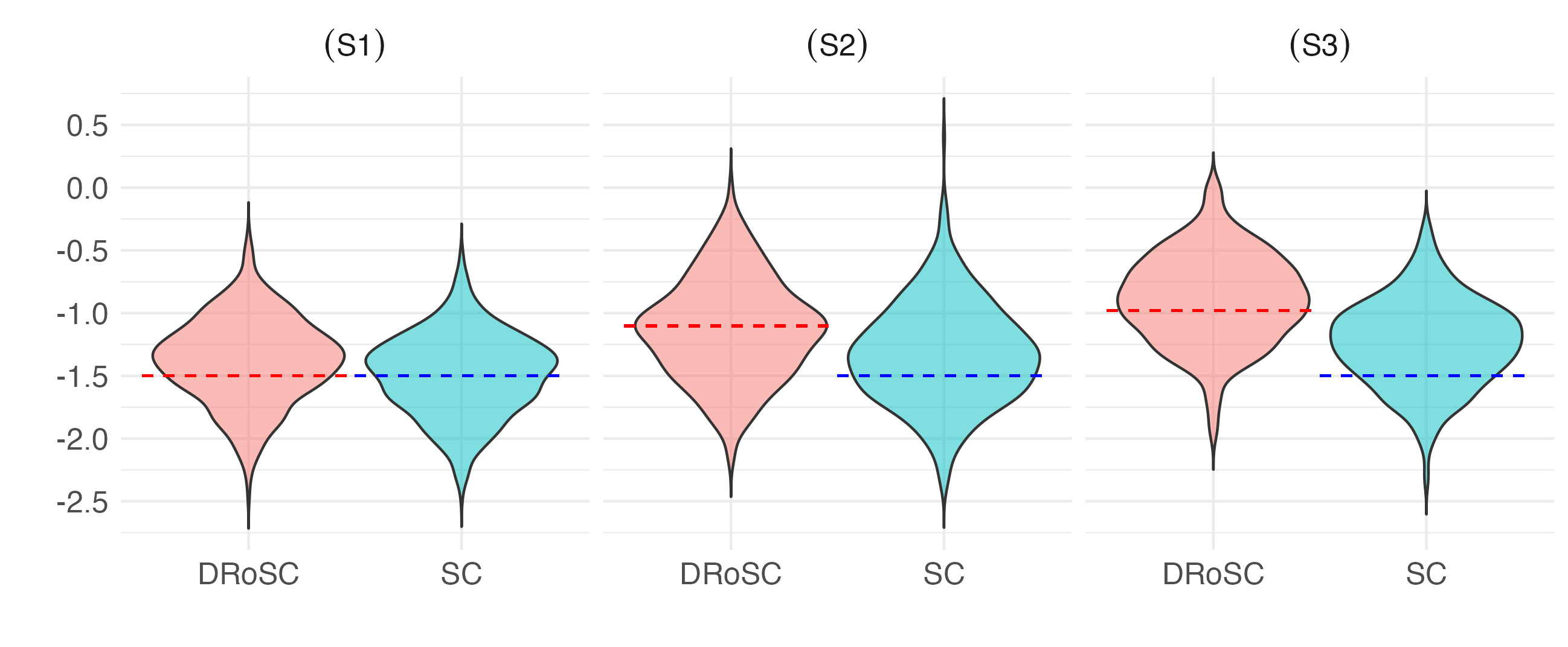}
  \end{subfigure}
  \begin{subfigure}{.8\textwidth}
    \centering
    \includegraphics[width=\textwidth]{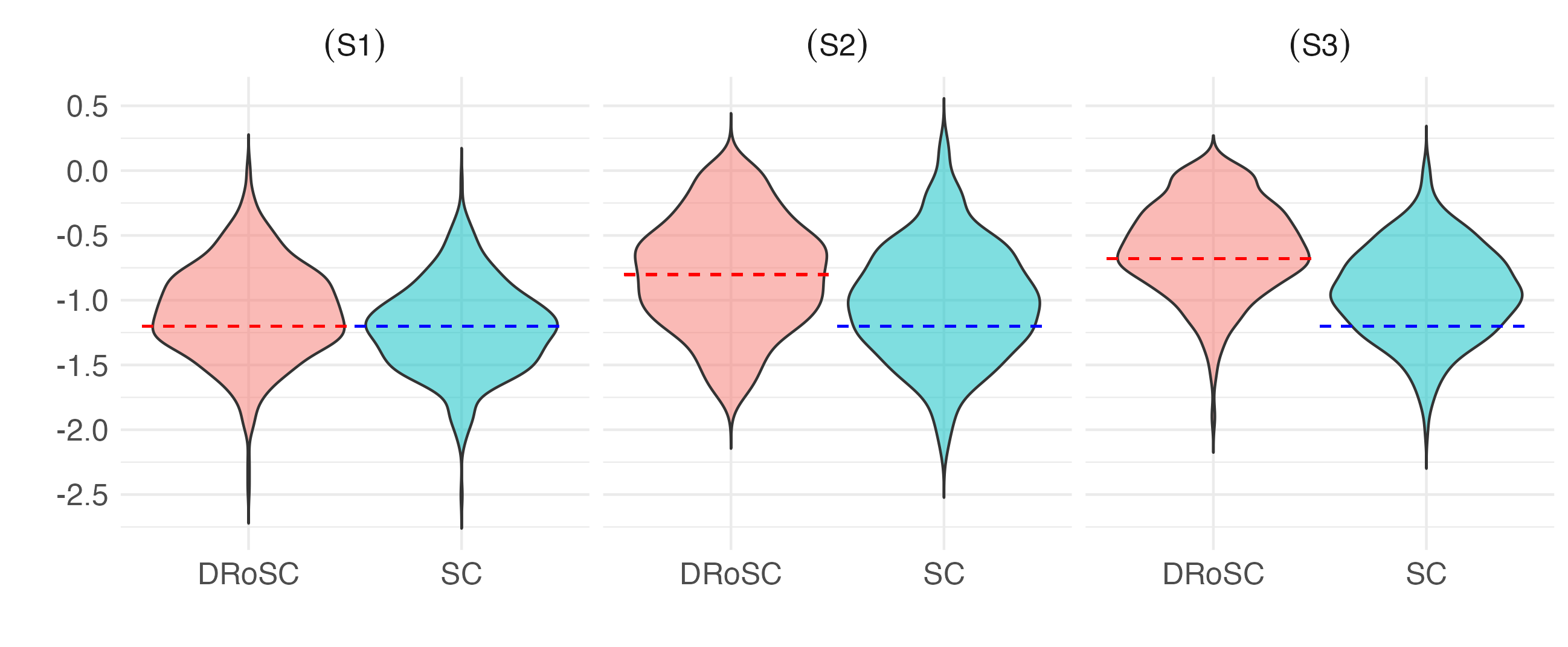}
  \end{subfigure}
  \begin{subfigure}{.8\textwidth}
    \centering
    \includegraphics[width=\textwidth]{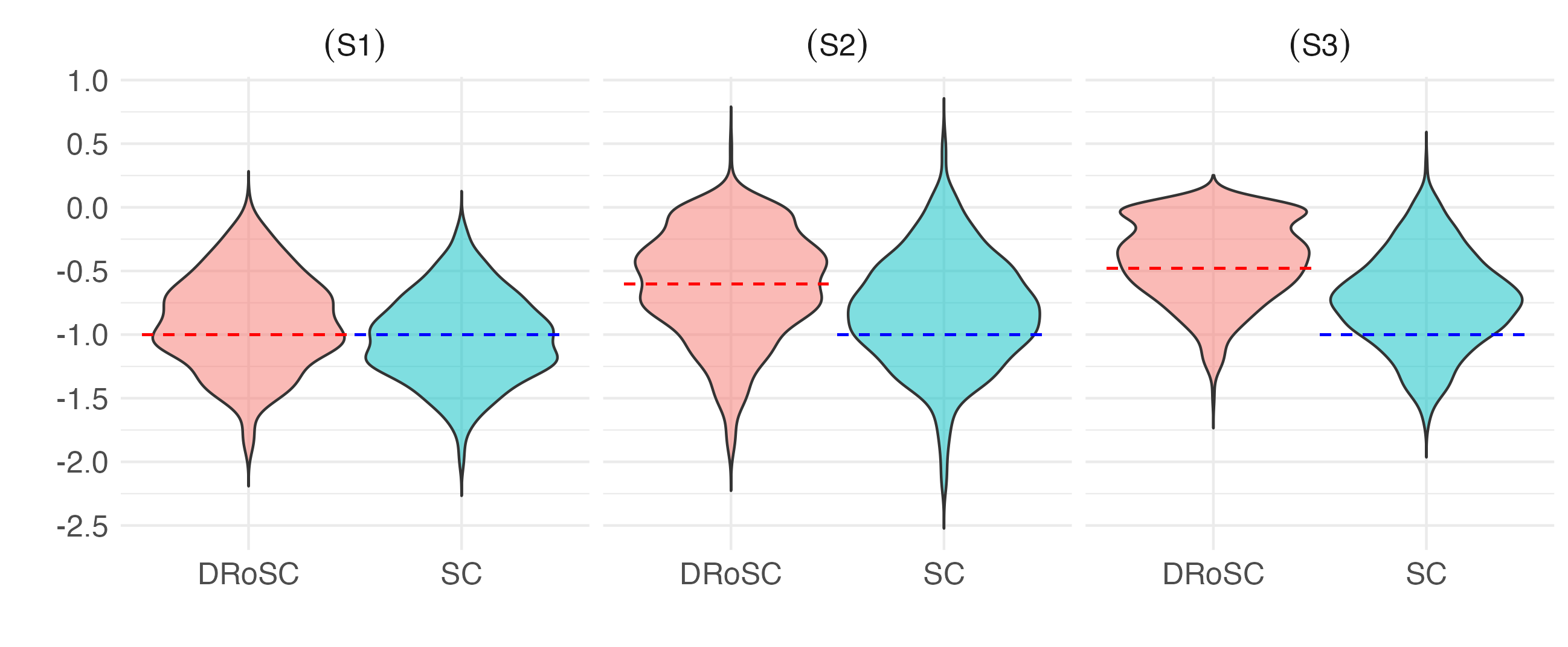}
  \end{subfigure}

    \caption{Simulation studies from settings (S1), (S2), and (S3) with $\phi = 0.5$, $T_0=25$, and $T_1=50$. The figures, from top to bottom, correspond to $\taubar=-1.5$, $-1.2$, and $-1$ respectively. \texttt{DRoSC} and \texttt{SC} in $x$-axis denote the estimators \eqref{eq: betahat tauhat} and \eqref{eq: SC estimators} from our method and the SC method respectively. The red and blue dashed lines denote $\tau^*$ and $\taubar$ respectively. }
    \label{fig: violin phi0.5 pre25 post50}
\end{figure}

Additional estimation results with $T_1=50$ and $\phi=0.5$ are presented in Figures~\ref{fig:violin pre25 post50}, \ref{fig: violin phi0.5 pre25 post25}, and \ref{fig: violin phi0.5 pre25 post50}. Overall, the patterns mirror those in Figure~\ref{fig:violin pre25 post25}: $\tauhat$ remains consistent, whereas the SC estimator $\tauhatSC$ does not.

\subsection{Sensitivity to the Order of \texorpdfstring{$\rho$}{rho}}\label{subsec: sens rho}
In this section, we study the sensitivity of the proposed estimator $\tauhat$ in \eqref{eq: betahat tauhat} to the choice of the tuning parameter $\rho$. The specification in \eqref{eq: tuning from specifying} sets $\rho$ of order $[\log(\max\{T_0,N\})]^{1/2}/\sqrt{T_0}$, whereas Assumption~\ref{assumption: pre} suggests the more general order depending on the dependence structure of the pre-treatment data:
\begin{equation}\label{eq: general rho}
    [\log(\max\{T_0,N\})]^{a}/\sqrt{T_0}
\end{equation} for $a=(1+b)/(2b)$ where $b>0$ is from Assumption~\ref{assumption: pre}. Since the true exponent $a$ and the multiplicative constant are typically unknown, we adopt a practical procedure that increases a multiplicative constant $C$ in $\rho$ in \eqref{eq: tuning from specifying} from a small constant value (e.g. $C=0.01$) until a feasible solution of \eqref{eq: betahat tauhat} is obtained. To assess robustness, we conduct simulations varying the exponent $a \in \{0.5,1,2\}$. The case $a=0.5$ coincides with \eqref{eq: tuning from specifying}, while larger values represent departures from that baseline.

\begin{figure}[ht]
    \centering
    \begin{subfigure}{.8\textwidth}
    \centering
    \includegraphics[width=\textwidth]{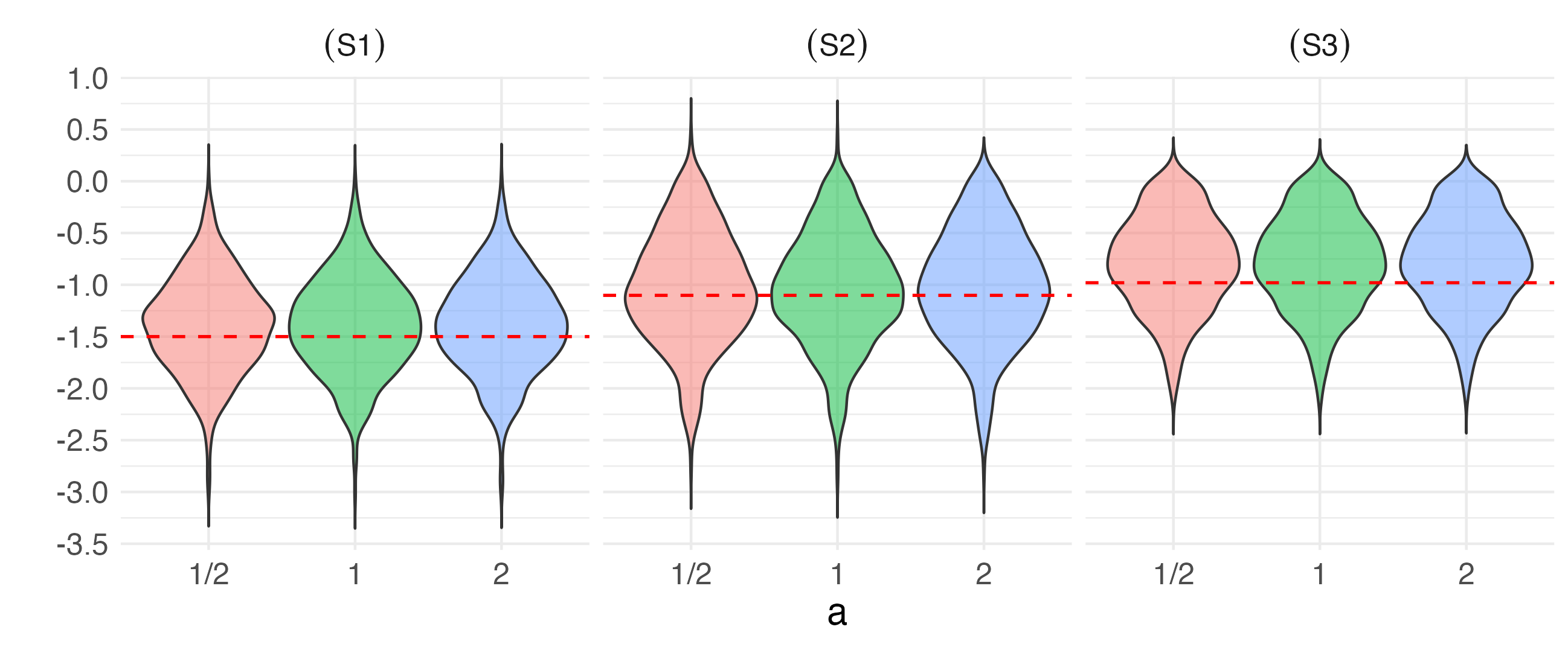}
  \end{subfigure}
  \begin{subfigure}{.8\textwidth}
    \centering
    \includegraphics[width=\textwidth]{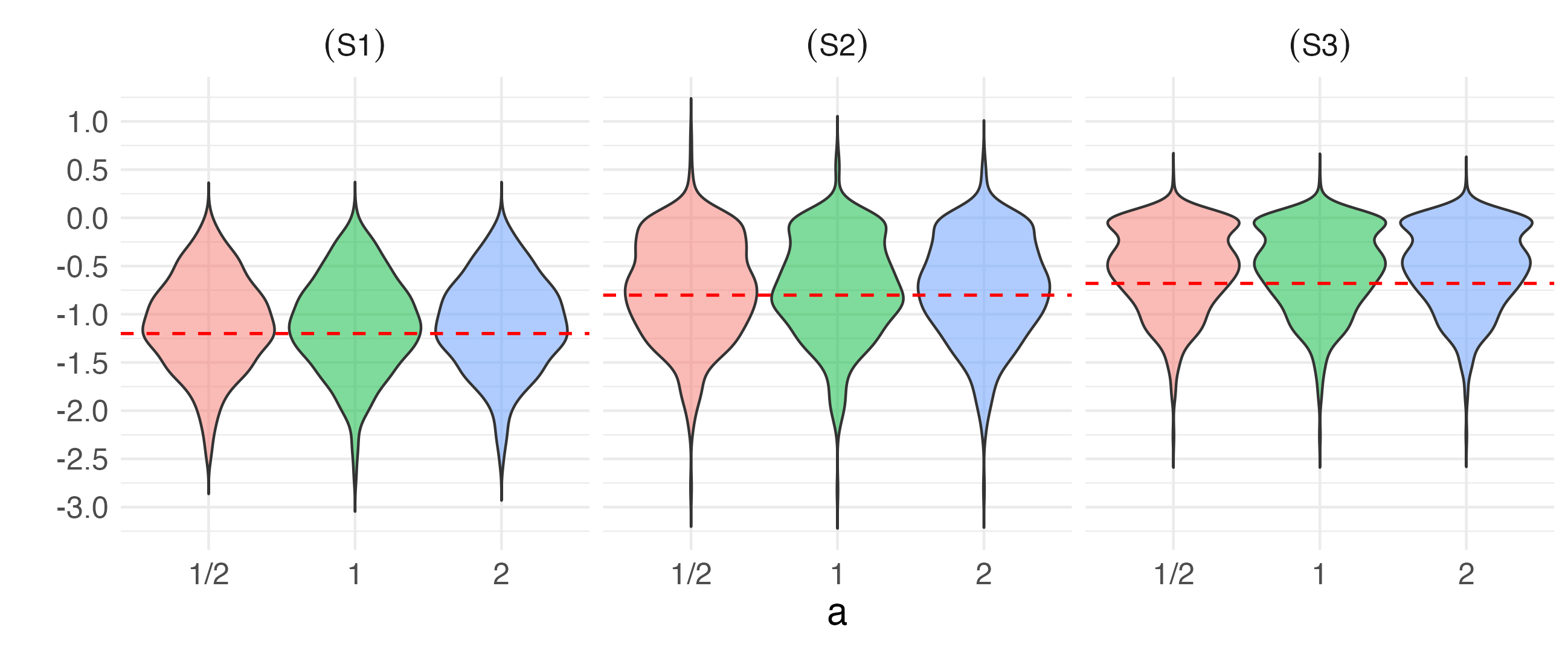}
  \end{subfigure}
  \begin{subfigure}{.8\textwidth}
    \centering
    \includegraphics[width=\textwidth]{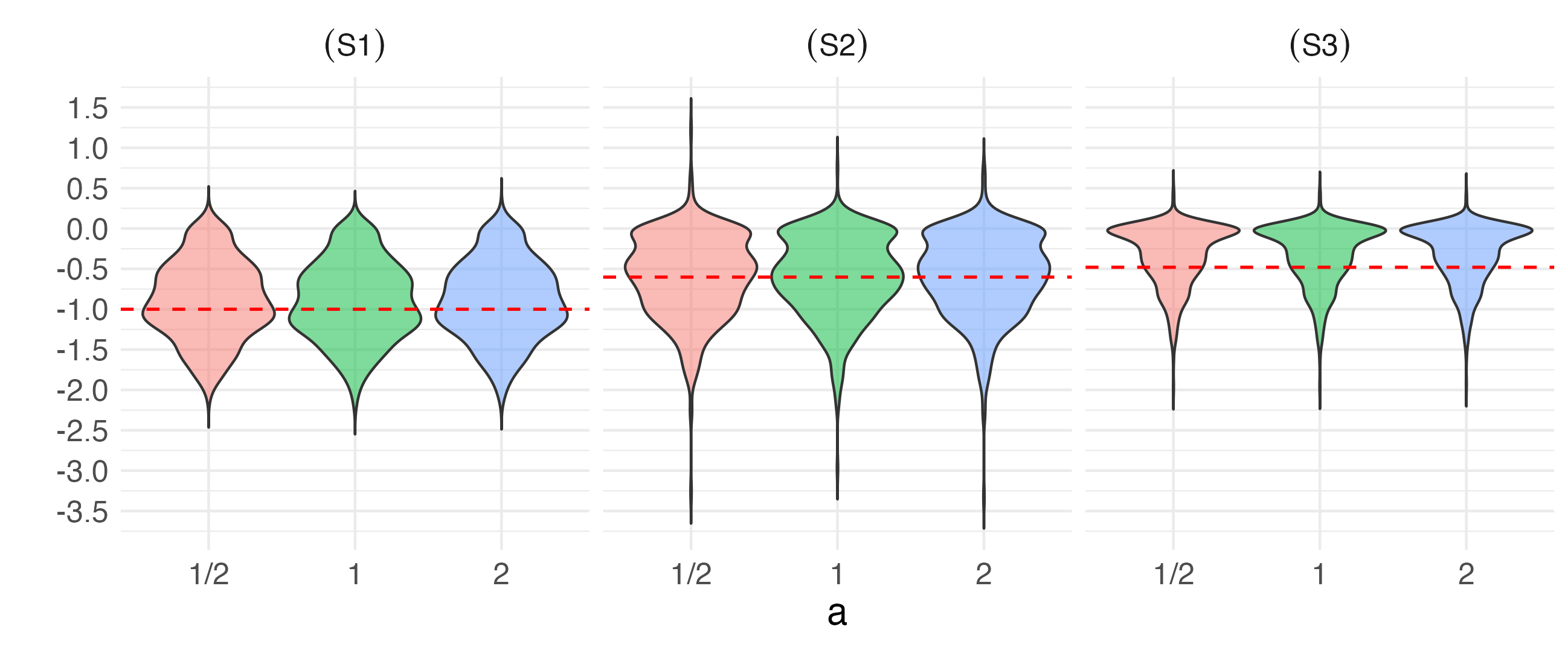}
  \end{subfigure}

    \caption{Simulation studies from settings (S1), (S2), and (S3). The figures, from top to bottom, correspond to $\taubar=-1.5$, $-1.2$, and $-1$ respectively. \texttt{a} in $x$-axis denote the exponent $a$ in \eqref{eq: general rho}. The red dashed line denotes $\tau^*$. }
    \label{fig:a value violin pre25 post25}
\end{figure}

\begin{figure}[ht]
    \centering
    \begin{subfigure}{.8\textwidth}
    \centering
    \includegraphics[width=\textwidth]{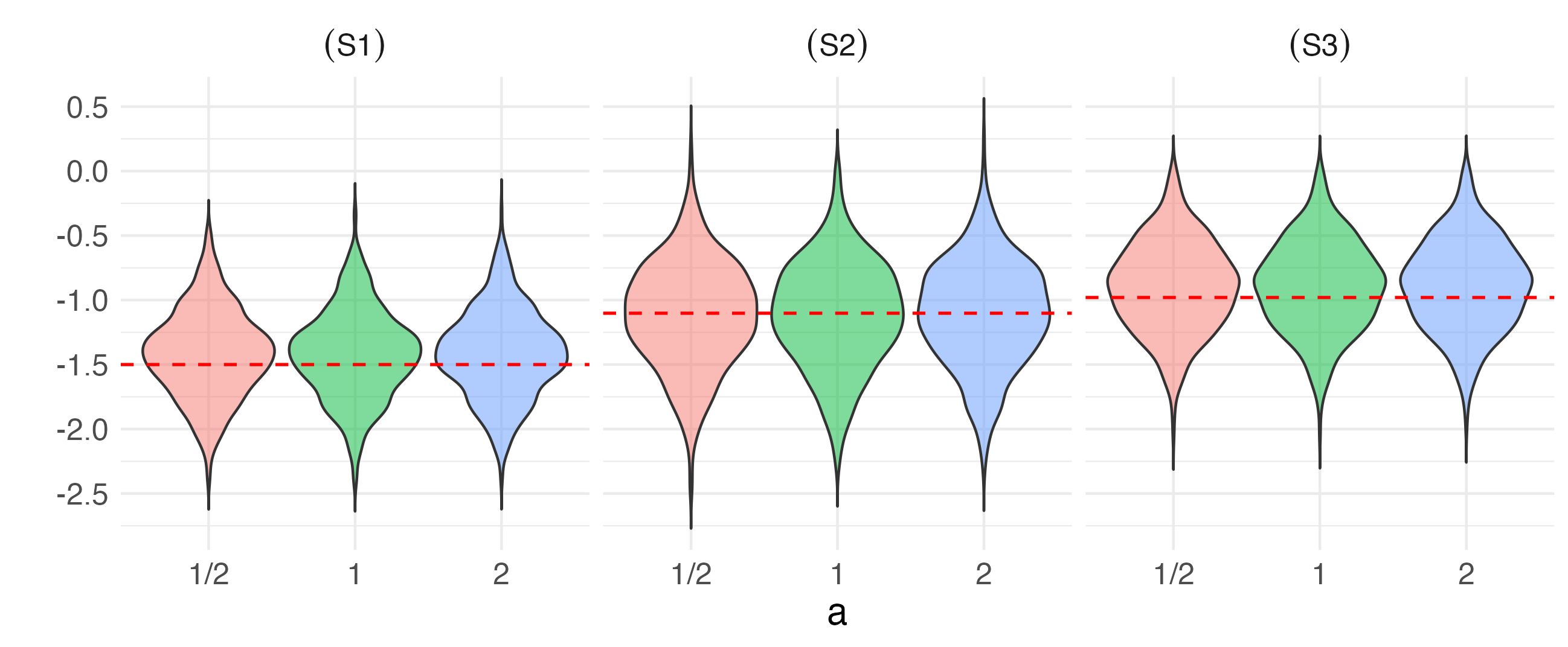}
  \end{subfigure}
  \begin{subfigure}{.8\textwidth}
    \centering
    \includegraphics[width=\textwidth]{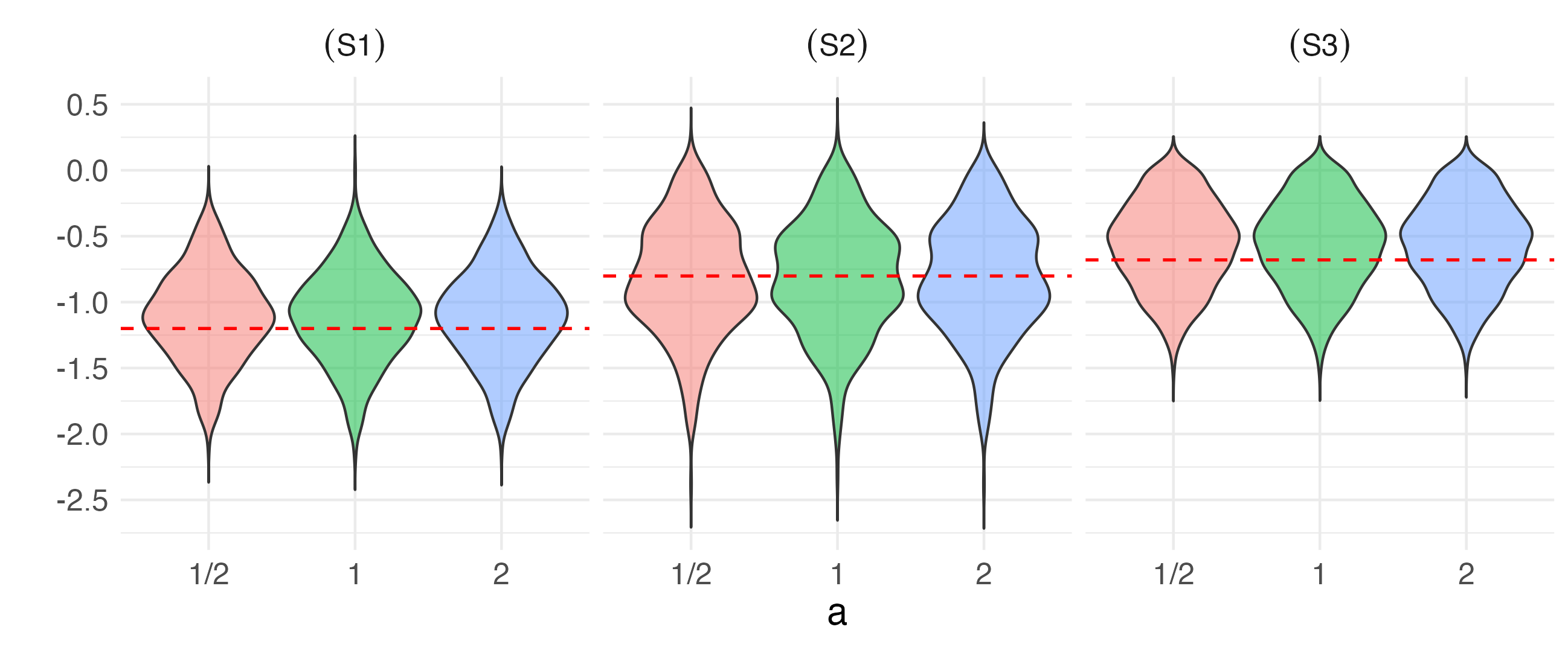}
  \end{subfigure}
  \begin{subfigure}{.8\textwidth}
    \centering
    \includegraphics[width=\textwidth]{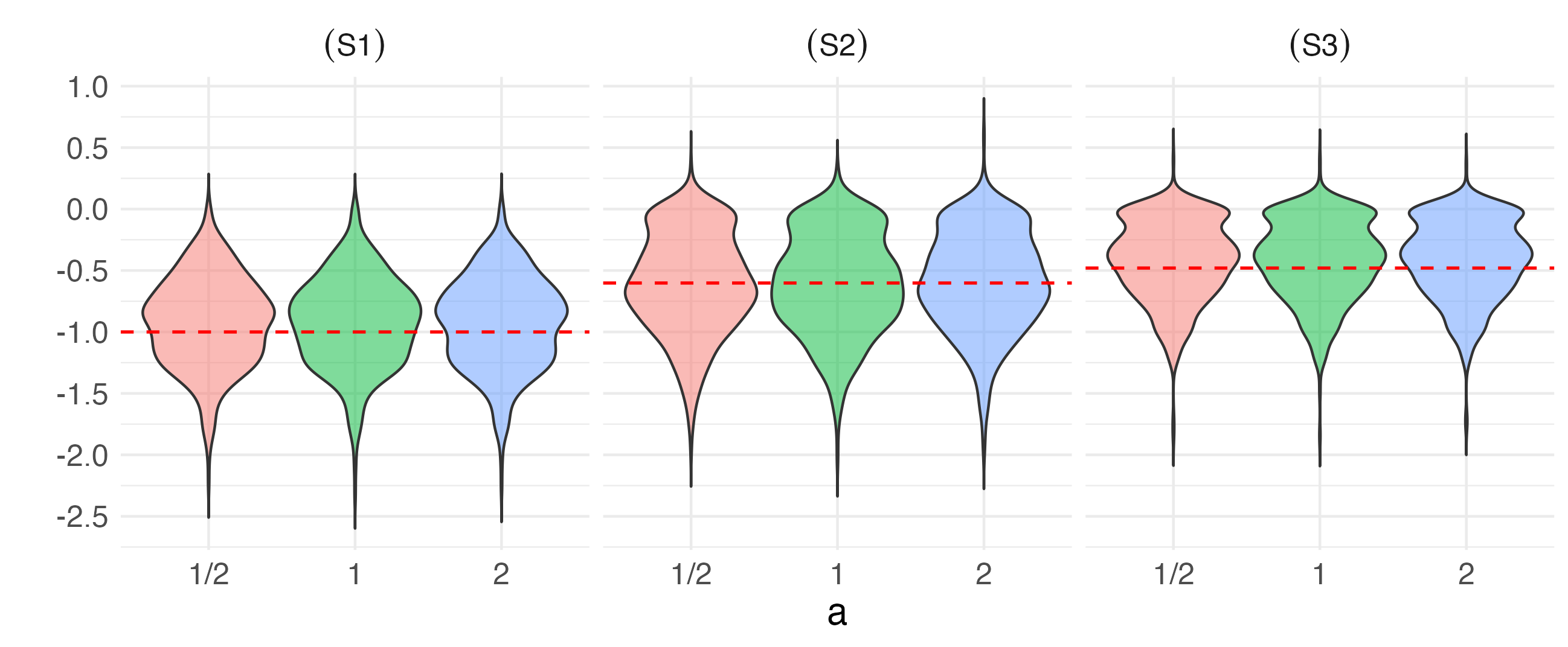}
  \end{subfigure}

    \caption{Simulation studies from settings (S1), (S2), and (S3). The figures, from top to bottom, correspond to $\taubar=-1.5$, $-1.2$, and $-1$ respectively. \texttt{a} in $x$-axis denote the exponent $a$ in \eqref{eq: general rho}. The red dashed line denotes $\tau^*$.}
    \label{fig:a value violin pre25 post50}
\end{figure}

For simplicity, we only present results with $T_0=T_1=25$ and $\phi=0.5$, using the same values of $\taubar$ as in Section~\ref{subsec: sim est}, since the results with different values of $T_0$ and $T_1$ are similar to these results. Note that $a=0.5$ is not theoretically valid here, as it assumes i.i.d.\ data. However, the results, shown in Figures~\ref{fig:a value violin pre25 post25} and \ref{fig:a value violin pre25 post50}, reveal only minor differences across choices of $a$, with violin plots remaining largely similar and consistently demonstrating stable performance of the estimator.

\end{document}